\documentclass{report}
 \usepackage{amsbsy,amssymb,amscd,amsfonts,latexsym,amstext,delarray,
 amsmath,graphicx}
 \input{amssymb.sty}

 \usepackage[dvips]{epsfig}
\def\taille{\long\def\epsfsize##1##2{\facteur\textwidth}}
\def\fig#1#2{\long\def\facteur{#1}\taille\epsffile{#2}}

 \input xypic

\newtheorem{thm}{Theorem}[chapter]
\newtheorem{prop}[thm]{Proposition}
\newtheorem{cor}[thm]{Corollary}
\newtheorem{lem}[thm]{Lemma}

\newtheorem{defn}[thm]{Definition}

\numberwithin{equation}{chapter}

\def\C{{\mathbb C}}
\def\Cb{{\mathbb C}}

\newcommand{\bG}{\mathbb{G}}

\def\K{{\mathbb K}}

\def\N{{\mathbb N}}

\renewcommand{\P}{{\mathbb P}}
\def\bP{{\mathbb P}}

\def\Q{{\mathbb Q}}
\def\R{{\mathbb R}}
\def\Rb{{\mathbb R}}

\def\T{{\mathbb T}}

\def\Z{{\mathbb Z}}

\newcommand{\scr}{\mathcal}

\def\sC{{\mathcal C}}
\def\Ec{{\mathcal E}}

\newcommand{\cF}{\scr{F}}
\def\cG{{\mathcal G}}

\def\cH{{\mathcal H}}
\def\Hc{{\mathcal H}}

\def\Lc{{\mathcal L}}

\def\cL{{\mathcal L}}

\def\cM{{\mathcal M}}
\def\Nc{{\mathcal N}}
\def\Oc{{\mathcal O}}
\newcommand{\fO}{{\mathcal{O}}}
\renewcommand{\O}{{\mathcal O}}

\def\Qc{{\mathcal Q}}

\def\cS{{\mathcal S}}
\def\Tc{{\mathcal T}}
\def\sT{{\mathcal T}}

\newcommand{\cU}{\scr{U}}
\def\cV{{\mathcal V}}

\newcommand{\fg}{{\mathfrak{g}}}

\newcommand{\fh}{{\mathfrak{h}}}

\def\a{\alpha}
\def\b{\beta}
\def\g{\gamma}
\def\G{\Gamma}
\def\d{\delta}
\def\D{\Delta}
\def\ve{\varepsilon}

\def\t{\theta}

\def\s{\sigma}

\def\vp{\varphi}

\def\qq{{\,,\quad \forall}}

\newcommand{\wt}{\widetilde}

\def\fl{\forall}
\def\ify{\infty}
\def\lgl{\langle}

\def\op{\oplus}
\def\ot{\otimes}

\def\part{\partial}
\def\rgl{\rangle}
\def\sbs{\subset}
\def\semi{>\!\!\!\lhd}
\def\sm{\simeq}
\def\ts{\times}

\def\ra{\rightarrow}

\def\text{\hbox}

\def\Aut{\mathop{\rm Aut}\nolimits}

\def\End{\mathop{\rm End}\nolimits}
\def\Ext{{\rm Ext}}
\def\Gal{{\rm Gal}}
\def\GL{{\rm GL}}
\def\Gr{{\rm Gr}} 
\def\Hom{\mathop{\rm Hom}\nolimits}

\def\Ind{{\rm Ind}}

\def\Ker{\mathop{\rm Ker}\nolimits}
\def\Lie{{\rm Lie}}

\def\SL{{\rm SL}}

\def\Sp{{\rm Spec}}

\def\Tr{{\rm Tr}}

\def\diff{\mathop{\rm diff}\nolimits}

\newcommand{\ie}{{\it i.e.\/}\ }
\newcommand{\eg}{{\it e.g.\/}\ }
\newcommand{\cf}{{\it cf.\/}\ }

 \title
 {From Physics to Number theory via Noncommutative Geometry, II}
 \author{Alain Connes and Matilde Marcolli}

\begin{document}
\maketitle

\setcounter{chapter}{1}

\chapter{Renormalization, the Riemann--Hilbert correspondence,
and motivic Galois theory}

\tableofcontents

\section{Introduction}\label{Sintro}

We give here a comprehensive treatment of the mathematical theory
of perturbative renormalization (in the minimal subtraction scheme with
dimensional regularization), in the framework of the Riemann--Hilbert
correspondence and motivic Galois theory. We give a detailed
overview of the work of Connes--Kreimer \cite{3CK}, \cite{3cknew}. We
also cover some background material on affine group schemes, Tannakian
categories, the Riemann--Hilbert problem in the regular singular and
irregular case, and a brief introduction to motives and
motivic Galois theory. We then give a complete account of our results
on renormalization and motivic Galois theory announced in
\cite{cmln}.

\smallskip

Our main goal is to show how the divergences of quantum field theory,
which may at first appear as the undesired effect of a mathematically
ill-formulated theory, in fact reveal the presence of a very rich
deeper mathematical structure, which manifests itself through the
action of a hidden ``cosmic Galois group''\footnote{The idea of a
``cosmic Galois group'' underlying perturbative renormalization was
proposed by Cartier in \cite{3Cart1}.}, which is of an arithmetic
nature, related to motivic Galois theory.

\medskip

Historically,  perturbative renormalization has always appeared as one
of the most elaborate recipes created by modern physics, capable of
producing numerical quantities of great physical relevance out of 
a priori meaningless mathematical expressions.
 In this respect, it is fascinating for mathematicians and
 physicists alike. The depth of its origin in quantum field
 theory and the precision with which it is confirmed by experiments
 undoubtedly make it into one of the jewels of modern theoretical physics.

 \smallskip

  For a mathematician in quest of ``meaning" rather than
 heavy formalism, the attempts to cast the perturbative
 renormalization technique in a conceptual framework
 were so far falling short of accounting for the
 main computational aspects, used for instance in QED. These
 have to do with the subtleties involved in the subtraction of
 infinities in the evaluation of Feynman graphs and do not fall under
 the range of ``asymptotically free theories'' for which constructive
 quantum field theory can provide a mathematically satisfactory
 formulation.

 \smallskip

  The situation recently changed through the work of Connes--Kreimer
 (\cite{3CK1}, \cite{3ck}, \cite{3CK}, \cite{3cknew}), where
 the conceptual meaning of the detailed computational devices used in
 perturbative renormalization is analysed. Their work shows that the
 recursive procedure used by physicists is in fact identical to a
 mathematical method of extraction of finite values known as the
 Birkhoff decomposition, applied to a loop $\g(z)$
 with values in a complex pro-unipotent Lie group $G$. 

\smallskip

This result, and the close relation between the Birkhoff factorization
of loops and the Riemann--Hilbert problem, suggested the existence of
a geometric interpretation of perturbative renormalization in terms of the  
Riemann--Hilbert correspondence. Our main result in this paper
is to identify explicitly the Riemann--Hilbert correspondence underlying
perturbative renormalization in the minimal subtraction scheme
with dimensional regularization. 

 \smallskip

   Performing the Birkhoff (or Wiener-Hopf) decomposition of
 a loop $\g(z)\in G$ consists of describing it as a product
 \begin{equation}
 \g \, (z) = \g_- (z)^{-1} \, \g_+ (z) \qquad z \in C \,
 ,\label{renorm15}
 \end{equation}
 of boundary values of holomorphic maps (which we still
 denote by the same symbol)
 \begin{equation}
 \g_{\pm} : C_{\pm} \ra G \, . \label{gpm}
 \end{equation}
 defined on the connected components $C_{\pm}$ of the complement of the
 curve $C$ in the Riemann sphere $\P^1(\C)$.

\begin{figure}
\begin{center}
\includegraphics[scale=1.2]{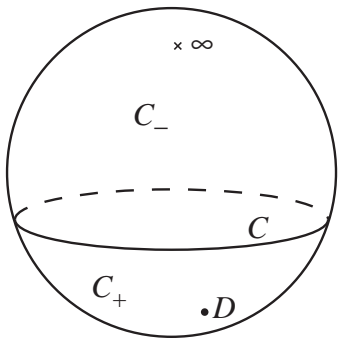}
\caption{\label{FigBirkhoff} Birkhoff Decomposition }
\end{center}
\end{figure}

 \medskip

  The geometric meaning of this decomposition, for instance when
 $G={\rm GL}_n (\Cb)$, comes directly from the theory of holomorphic
 bundles with structure group $G$ on the Riemann sphere $\P^1(\C)$. The
 loop $\g(z)$ describes the clutching data to construct the bundle from
 its local trivialization and the Birkhoff decomposition provides a
 global trivialization of this bundle. While in the case of ${\rm
GL}_n (\Cb)$ the existence of a Birkhoff decomposition may be
obstructed by the non-triviality of the bundle, in the case of a
pro-unipotent complex Lie group $G$, as considered in the CK theory of
renormalization, it is always possible to obtain a factorization
\eqref{renorm15}.

 \smallskip

  In perturbative renormalization the points of $\P^1(\C)$ are
 ``complex dimensions'', among which the dimension $D$ of the relevant
 space-time is a preferred point. The little devil that conspires to
 make things interesting makes it impossible to just evaluate the
 relevant physical observables at the point $D$, by letting them
 diverge precisely at that point. One can nevertheless
 encode all the evaluations at points $z\neq D$ in the form of a loop
 $\g(z)$ with values in the group $G$. The  perturbative
 renormalization technique then acquires the following general
 meaning: while  $\g(D)$ is meaningless, the physical quantities
 are in fact obtained by evaluating
 $\g_+ (D)$, where $\g_+$ is the term 
that is holomorphic at $D$
for the Birkhoff decomposition relative to an  
infinitesimal circle with center $D$.

 \smallskip

  Thus, renormalization appears as a special case of a general
 principle of extraction of finite results from divergent
 expressions based on the Birkhoff decomposition.

 \smallskip

 The nature of the group $G$ involved in perturbative renormalization
 was clarified in several steps in the work of Connes--Kreimer
 (CK). The first was Kreimer's discovery \cite{3dhopf} of a Hopf
 algebra structure underlying the recursive formulae of \cite{3BP},
 \cite{3hepp}, \cite{3zim}. The resulting Hopf algebra of rooted trees
depends on the physical theory $\Tc$ through the use of suitably
decorated trees. The next important ingredient was the
 similarity between the Hopf algebra of rooted trees of \cite{3dhopf} and
 the Hopf algebra governing the symmetry of transverse geometry in
 codimension one of \cite{3cm1}, which was observed already in \cite{3CK1}.
 The particular features of a given physical theory were then
 better encoded by a Hopf algebra defined in \cite{3CK} directly in terms of
 Feynman graphs. This Hopf algebra of Feynman graphs depends on the theory
$\Tc$ by construction. It determines $G$ as
the associated affine group scheme, which is referred to as {\em
diffeographisms} of the theory, 
$G={\rm Difg}(\Tc)$. Through the Milnor-Moore theorem \cite{3MM}, the
Hopf algebra of Feynman graphs determines a Lie algebra, whose
corresponding infinite dimensional pro-unipotent Lie group is given
by the complex points $G(\C)$ of the affine group scheme of diffeographisms.

\smallskip

This group is related to the formal group of Taylor expansions of
diffeomorphisms. 
 It is this infinitesimal feature of the expansion that accounts
 for the ``perturbative'' aspects inherent to
 the computations of Quantum Field Theory.
 The next step in the CK theory of renormalization is
 the construction of an action of ${\rm Difg}(\Tc)$
 on the coupling constants of the physical theory,
 which shows a close relation between ${\rm Difg}(\Tc)$
 and the group of diffeomorphisms of the
 space of Lagrangians.

 \smallskip

  In particular, this allows one to lift the renormalization group
 to a one parameter subgroup of ${\rm Difg}$, defined intrinsically
 from the independence of the term $\g_- (z)$ in the Birkhoff
 decomposition from the choice of an additional mass scale $\mu$.
 It also shows  that the polar expansions of the
 divergences are entirely determined by their residues (a strong form
 of the 't Hooft relations), through
 the scattering formula of \cite{3cknew}
  \begin{equation} \g_- (z) = \lim_{t \to \infty}
e^{-t \left( \frac{\b}{z} + Z_0 \right)} \, e^{t Z_0} \, .
\label{scattering}
 \end{equation}

\medskip

After a brief review of perturbative renormalization in QFT (\S
\ref{Srenorm}), we give in Sections \ref{Sgraphs},
\ref{SLie}, \ref{SBirk}, \ref{difgdiff}, and in part of Section
\ref{Rensect}, a detailed 
account of the main results mentioned above of the CK theory of
perturbative renormalization and its formulation in terms of Birkhoff
decomposition. This overview of the work of Connes--Kreimer is 
partly based on an English translation of \cite{CoGal1} \cite{CoGal2}.  

\smallskip

The starting point for our interpretation of renormalization as
a Riemann--Hilbert correspondence is presented in Sections
\ref{expansional} and \ref{Rensect}. It consists of rewriting the scattering
formula \eqref{scattering} in terms of the time ordered
exponential of physicists (also known as {\em expansional} in
mathematical terminology), as
\begin{equation} 
\g_- (z) ={\bf{\rm T}e^{-\frac{1}{z}\,\int_0^\infty\,\t_{-t}(\beta)\,dt}},
\label{expscattering}
 \end{equation}
where $\theta_t$ is the one-parameter group of automorphisms
implementing the grading by loop number on the Hopf algebra of Feynman
graphs. We exploit the more suggestive form \eqref{expscattering} to
clarify the relation between the Birkhoff decomposition used in
\cite{3CK} and a form of the Riemann-Hilbert correspondence.

\smallskip

In general terms, as we recall briefly in Section \ref{RHsect},
the Riemann--Hilbert correspondence is an
equivalence between a class of singular differential systems and
representation theoretic data. The classical example is that of
regular singular differential systems and their monodromy
representation.

\smallskip

In our case, the geometric problem underlying perturbative
renormalization consists of the classification of {\em
``equisingular"} $G$-valued flat connections on the total space
$B$ of a principal $\bG_m$-bundle over an infinitesimal punctured
disk $\Delta^*$. An equisingular connection is a $\bG_m$-invariant
$G$-valued connection, singular on the fiber over zero, and
satisfying the following property: the equivalence class of the
singularity of the pullback of the connection by a section of the
principal $\bG_m$-bundle only depends on the value of the section
at the origin.

\smallskip

The physical significance of this geometric setting is the following.
The expression \eqref{expscattering} in expansional
form can be recognized as the solution of a differential system 
\begin{equation}\label{diffsystG}
 \gamma^{-1}\,d\gamma =\omega. 
\end{equation}
This identifies a class of connections naturally associated to the
differential of the regularized quantum field theory, viewed as a
function of the complexified dimension. The base $\Delta^*$ is the
space of complexified dimensions around the critical dimension $D$. 
The fibers of the principal $\bG_m$-bundle $B$
describe the arbitrariness in the normalization
of integration in complexified dimension $z\in \Delta^*$,
in the Dim-Reg regularization procedure.
The  $\bG_m$-action corresponds to the rescaling of the normalization
factor of integration in complexified dimension $z$, which can be
described in terms of the scaling 
$\hbar\,\, \partial /\partial \hbar$ on the expansion in powers of $\hbar$.  
The group defining $G$-valued connections
is $G={\rm Difg}(\Tc)$.
The physics input that the
counterterms are independent of the additional choice of a unit of
mass translates, in geometric terms, into the notion of
equisingularity for the connections associated to the differential
systems \eqref{diffsystG}.

\smallskip

On the other side of our Riemann--Hilbert correspondence, the
representation theoretic setting equivalent to the classification
of equisingular flat connections is provided by finite dimensional
linear representations of a universal group $U^*$, unambiguously
defined independently of the physical theory. Our main result is the explicit
description of $U^*$ as the semi-direct product by its grading of
the graded pro-unipotent Lie group $U$ whose Lie algebra is the
free graded Lie algebra $$\cF(1,2,3,\cdots)_{\bullet}$$ generated
by elements $e_{-n}$ of degree $n$, $n>0$.
As an affine group scheme, $U^*$ is identified uniquely via the
formalism of Tannakian categories. Namely, equisingular
flat connections on finite dimensional vector bundles can be
can be organized into a Tannakian category with a natural fiber
functor to the category of vector spaces. This category is 
equivalent to the category of finite dimensional representations 
of the affine group scheme $U^*$. These main results are presented
in detail in Sections \ref{Sconnections}, \ref{classif}, and
\ref{SGalois}. 

\smallskip

This identifies a new level at which Hopf algebra structures
enter the theory of perturbative renormalization, after 
Kreimer's Hopf algebra of rooted trees and the CK Hopf algebra of Feynman
graphs. Namely, the Hopf algebra associated to the affine group scheme
$U^*$ is universal with respect to the set of physical theories.
The ``motivic Galois group'' $U$ acts on the set of dimensionless
coupling constants of physical theories, through the map $U^* \to {\rm
Difg}^*$ to the group of diffeographisms of a given theory, which in
turns maps to formal diffeomorphisms as shown in \cite{3cknew}. Here
${\rm Difg}^*$ is the semi-direct product of ${\rm Difg}$ by the
action of the grading $\t_t$, as in \cite{3cknew}. 

\smallskip

We then construct in Section \ref{Sunivframe} a specific universal
singular frame on principal $U$-bundles over $B$. We show that, when
using in this frame the 
dimensional regularization technique of QFT, all divergences
disappear and one obtains a finite theory which only depends upon
the choice of a local trivialization for the principal
$\bG_m$-bundle $B$ and produces the physical theory in the
minimal subtraction scheme.

\smallskip

The coefficients of the universal singular frame, written out in
the expansional form, are the same as those appearing in the local index
formula of Connes--Moscovici \cite{cmindex}. This leads to the very
interesting question of the explicit relation to noncommutative
geometry and the local index formula.

\smallskip

In particular, the coefficients of the universal singular frame are
rational numbers. This means that we can view equisingular 
flat connections on finite dimensional vector bundles as endowed
with arithmetic structure. Thus, the Tannakian category of flat
equisingular bundles can be defined over any field of characteristic
zero. Its properties are very reminiscent of the formalism of mixed
Tate motives (which we recall briefly in Section \ref{Smotives}).

\smallskip

In fact, group schemes closely related to $U^*$ appear in motivic Galois
theory. For instance, $U^*$ is abstractly (but non-canonically)
isomorphic to the motivic Galois group $ G_{\cM_T}(\fO) $ (\cite{dg},
\cite{3sasha4}) of
 the scheme $S_4={\rm Spec}(\fO) $
of $4$-cyclotomic integers, $\fO=\,\Z[i][\frac{1}{2}]$.

\smallskip

The existence of a universal pro-unipotent group $U$ underlying the
theory of perturbative renormalization, canonically defined and
independent of the physical theory, confirms a suggestion made by
Cartier in \cite{3Cart1}, 
that in the Connes--Kreimer theory of perturbative renormalization
one should find a hidden ``cosmic Galois group'' closely related
in structure to the Grothendieck--Teichm\"uller group. The
question of relations between the work of Connes--Kreimer, motivic
Galois theory, and deformation quantization was further emphasized
by Kontsevich in \cite{Kont}, as well as the conjecture of an action
of a motivic Galois group on the coupling constants of physical
theories. At the level of the Hopf algebra of rooted trees,
relations between renormalization and motivic Galois theory were
also investigated by Goncharov in \cite{Gon02}.

\smallskip

Our result on the ``cosmic motivic Galois group'' $U$ also shows that
the renormalization group appears as a canonical one parameter
subgroup $\bG_a \,\subset  U $. Thus, this realizes the hope
formulated in \cite{CoGal1} of relating concretely the
renormalization group to a Galois group.

\smallskip

As we discuss in Section \ref{Snonpert}, the group $U$ presents
similarities with the exponential torus part of the wild fundamental
 group, in the sense of Differential Galois Theory (\cf
 \cite{3Ramis}, \cite{vdp}). The latter is a modern form of the ``theory of
 ambiguity'' that Galois had in mind and takes a very concrete form in
 the work of Ramis \cite{3Ramis2}. The ``wild fundamental
 group'' is the natural object that replaces the usual fundamental
 group in extending the Riemann--Hilbert correspondence to the
 irregular case (\cf \cite{3Ramis}). At the formal level, in addition
to the monodromy representation (which is trivial in the case of the
equisingular connections), it comprises the exponential torus,
while in the non-formal case additional generators are present that
account for the Stokes phenomena in the resummation of divergent series.
 The Stokes part of the wild fundamental group (\cf
 \cite{3Ramis}) in fact appears when taking into account the
 presence of non-perturbative effects. We formulate some questions
 related to extending the CK theory of perturbative
 renormalization to the nonperturbative case.

 \smallskip

  We also bring further evidence for the
 interpretation of the renormalization group in terms of a theory
 of ambiguity. Indeed, one aspect of QFT that appears intriguing to the novice is
 the fact that many quantities called ``constants'', such as the
 fine structure constant in QED, are only nominally constant,
 while in fact they depend on a scale parameter $\mu$. Such examples
 are abundant, as most of the relevant physical quantities, including
 the coupling ``constants'', share this implicit dependence on the
 scale $\mu$. Thus, one is really dealing with functions $g( \mu)$
instead of scalars. 
This suggests the idea that a suitable ``unramified'' extension $K$ of the field
${\mathbb C}$ of complex numbers might play a role in
QFT as a natural extension of the ``field of constants" to
 a field containing functions whose basic
 behaviour is dictated by the renormalization group equations.
 The group of automorphisms of the resulting field, generated by $ \mu \partial
 / \partial \mu$, is the group of ambiguity of the physical theory and
 it should appear as the Galois group of the
 unramified extension. Here the beta function of renormalization can
be seen as logarithm of 
 the monodromy in a regular-singular local Riemann--Hilbert
 problem associated to this scaling action as in \cite{mk}.
The true constants are then the
 fixed points of this group, which form the field
 ${\mathbb C}$ of complex numbers, but a mathematically rigorous
 formulation of QFT may require extending the field of scalars first,
 instead of proving existence ``over ${\mathbb C}$''.

\smallskip

This leads naturally to a different set of questions, related to
the geometry of arithmetic varieties at the infinite primes, 
and a possible Galois interpretation of the connected component of the
identity in the id\`ele class group in class field theory (\cf
\cite{3CoZ}, \cite{3KM}). This set of questions will be dealt with
in \cite{CoMa2}. 

 \bigskip

  {\bf Acknowledgements.} We are very grateful to Jean--Pierre Ramis
 for many useful comments on an early draft of this
 paper, for the kind invitation to Toulouse, and for the many stimulating
 discussions we had there with him, Fr\'ed\'eric Fauvet, and Laurent
 Stolovitch. We thank Fr\'ed\'eric Menous and Giorgio Parisi for some
 useful correspondence. Many thanks go to Dirk Kreimer, whose
 joint work with AC on perturbative renormalization is a main topic
 of this Chapter.

 \bigskip
 \section{Renormalization in Quantum Field Theory}\label{Srenorm}

  The physical motivation behind the renormalization technique is
 quite clear and goes back to the concept of effective mass
 and to the work of Green in nineteenth century hydrodynamics \cite{3Green}.
 To appreciate it, one should \footnote{See the QFT course by
  Sidney Coleman.}
 dive under water with a ping-pong ball and start applying Newton's
 law,
 \begin{equation}
 F = m \, a \label{ball}
 \end{equation}
 to compute the initial acceleration of the ball $B$ when we let it
 loose (at zero speed relative to the still water). If one naively
 applies (\ref{ball}), one finds an unrealistic initial acceleration of
 about $11.4\, g$. \footnote{The ping-pong ball weights $m_0=2,7$
 grams and its diameter is $4$ cm so that $M=33,5$ grams.}
 In fact, if one performs the experiment, one finds
 an initial acceleration of about $1.6 \,g$.
 As explained by Green in \cite{3Green}, due to the interaction of $B$ with the
 surrounding field of water, the inertial mass $m$ involved in
 \eqref{ball} is not the bare mass $m_0$ of $B$, but it is modified to
 \begin{equation}
 m = m_0 + {\textstyle \frac{1}{2}} \, M
 \end{equation}
 where $M$ is the mass of the water occupied by $B$.
 It follows for instance that the initial acceleration $a$ of $B$ is
 given, using the Archimedean law, by
 \begin{equation}
 -(M-m_0)\, g = \left( m_0 + {\textstyle \frac{1}{2}} \, M \right) a
 \end{equation}
 and is always of magnitude less than $2g$.

 \smallskip

  The additional inertial mass $\d \, m = m - m_0$ is due to the
 interaction of $B$ with the surrounding field of water and if this
 interaction could not be turned off (which is the case if we deal
 with an electron instead of a ping-pong ball) there would be no
 way to measure the bare mass $m_0$.

 \smallskip

  The analogy between hydrodynamics and electromagnetism led,
 through the work of Thomson, Lorentz, Kramers, etc. (\cf
 \cite{3MD}), to the crucial distinction between the bare parameters,
 such as $m_0$, which enter the field theoretic equations, and the
 observed parameters, such as the inertial mass $m$.

 \smallskip

  Around 1947, motivated by the experimental findings
 of spectroscopy of the fine structure of spectra,
 physicists were able to exploit the above distinction
 between these two notions of mass (bare and observed),
 and similar distinctions for the charge and field strength,
 in order to eliminate the unwanted infinities which
 plagued the computations of QFT, due to the
 pointwise nature of the electron. We refer to \cite{3MD}
 for an excellent historical account of that period.

 \medskip

 \subsection{Basic formulas of QFT}

\bigskip

  A quantum field theory in $D=4$ dimensions is given by a
 classical action functional
 \begin{equation}
 S \, (A) =  \int \Lc \, (A) \, d^4 x ,  \label{action}
 \end{equation}
 where $A$ is a classical field and the
 Lagrangian is of the form
 \begin{equation}
  \Lc \, (A) =  \frac{1}{2} (\part A)^2  -
 \frac{m^2}{2} \, A^2 - \Lc_{\rm int} (A), \label{lag}
 \end{equation}
 with $(\part A)^2=(\part_0 A)^2-\sum_{\mu \neq 0}(\part_\mu A)^2$. The
 term $\Lc_{\rm int} (A)$ is usually a polynomial in $A$.

 \smallskip

   The basic transition from ``classical field theory"
 to  ``quantum field theory" replaces the classical notion of
 probabilities by {\em probability amplitudes} and asserts that the
 probability amplitude of a classical field configuration $A$
 is given by the formula of Dirac and Feynman
 \begin{equation}
 e^{i \, \frac{S(A)}{\hbar}}, \label{16-4}
 \end{equation}
 where $S(A)$ is the classical action (\ref{action}) and $\hbar$ is the
 unit of action, so that $iS(A)/\hbar$ is a dimensionless quantity.

 \smallskip

  Thus, one can {\em define} the quantum expectation value of a
 classical observable (\ie of a function $\Oc$ of the classical fields)
 by the expression
 \begin{equation}\label{expvalue}
 \langle \Oc \rangle =\,\Nc\, \int \Oc(A)\,e^{i \,
 \frac{S(A)}{\hbar}}\, D[A], \,
 \end{equation}
 where $\Nc$ is a normalization factor. The (Feynman) integral has only
 formal meaning, but this suffices in the case where the space of
 classical fields $A$ is a linear space in order to define without
 difficulty the terms in the perturbative expansion, which make the
 renormalization problem manifest.

 \smallskip

   One way to describe the quantum fields $\phi (x)$ is by means of
 the time ordered Green's functions
 \begin{equation}
 G_N (x_1 , \ldots , x_N) = \lgl \, 0 \, \vert T \, \phi (x_1)
 \ldots \phi (x_N) \vert \, 0 \, \rgl , \label{16-3}
 \end{equation}
 where the time ordering symbol $T$ means that the $\phi (x_j)$'s
 are written in order of increasing time from right to left.
 If one could ignore the renormalization problem, the Green's
 functions would then be computed as
 \begin{equation}
 G_N (x_1 , \ldots , x_N) = \Nc \int e^{i \, \frac{S(A)}{\hbar}} \
 A (x_1) \ldots A (x_N) \, [dA], \label{16-5}
 \end{equation}
 where the factor $\Nc$ ensures the
 normalization of the vacuum state
 \begin{equation}
 \lgl \, 0 \mid 0 \, \rgl = 1 \, . \label{16-6}
 \end{equation}

 \smallskip

   If one could ignore renormalization, the functional integral
 (\ref{16-5}) would be easy to compute in perturbation theory, \ie
 by treating the term $\Lc_{\rm int}$ in (\ref{lag}) as a
 perturbation of
 \begin{equation}
 \Lc_0 (A) = \frac{1}{2}(\part A)^2  - \frac{m^2}{2} \, A^2 \, .
 \label{16-7}
 \end{equation}
 The action functional correspondingly splits as the
 sum of two terms
 \begin{equation}
 S (A) = S_0 (A) + S_{\rm int} (A), \label{16-8}
 \end{equation}
 where the free action $S_0$ generates a Gaussian measure
 $$\exp \,
 (i \, S_0 (A)) \, [d A] = d \mu, $$
 where we have set $\hbar=1$.

 \smallskip

   The series expansion of the Green's functions is then of the form
 \begin{eqnarray}
 G_N (x_1 , \ldots , x_N) = \left( \sum_{n=0}^{\ify} \, i^n / n!
 \int A (x_1) \ldots A (x_N) \, (S_{\rm int} (A))^n \, d \mu
 \right) \nonumber \\
 \quad \left( \sum_{n=0}^{\ify} \, i^n / n! \int S_{\rm int} (A)^n
 \, d \mu \right)^{-1} \, . \nonumber
 \end{eqnarray}

 \medskip

 \subsection{Feynman diagrams}

\bigskip

  The various terms
 \begin{eqnarray}\label{feyn1}
  \int
 A (x_1) \ldots A (x_N) \, (S_{\rm int} (A))^n \, d \mu
 \end{eqnarray}
 of this expansion are integrals of polynomials
 under a Gaussian measure $d\mu$. When these are computed using
 integration by parts, the process generates a large
 number of terms $U(\G)$. The combinatorial data labelling each of
 these terms are encoded in the Feynman graph $\G$, which determines
 the terms that appear in the calculation of the corresponding
 numerical value $U(\G)$, obtained as a multiple
 integral in a finite number of space-time variables. The $U(\G)$ is
 called the unrenormalized value of the graph $\G$.

 \smallskip

  One can simplify the combinatorics of the graphs
 involved in these calculations, by introducing a suitable generating
 function. The generating function for the Green's functions
 is given by the Fourier transform
 \begin{equation}\label{ZJ}
 Z(J) =  \Nc \int \exp\left(i\frac{S(A)+ \langle J, A
 \rangle}{\hbar}\right)\,
 [dA] \end{equation}
 $$ =  \sum_{N=0}^{\ify} \frac{ i^N}{ N!}
 \int J (x_1) \ldots J (x_N)\, G_N(x_1,..x_N)\, dx_1.. dx_N , $$
 where the {\em source} $J$ is an element of the dual of the linear
 space of classical fields $A$.

 \smallskip

   The zoology of the diagrams involved in the
 perturbative expansion  is substantially
 simplified by first passing to the logarithm
 of $Z(J)$ which is the generating function for {\em connected}
 Green's functions $G_c$,
 \begin{equation}\label{ZJ1}
 i W(J) = {\rm Log} (Z(J))
 =\,\sum_{N=0}^{\ify} \frac{ i^N}{ N!}
 \int J (x_1) \ldots J (x_N) G_{N,c}(x_1,..x_N) dx_1.. dx_N .
 \end{equation}

  At the formal combinatorial level,
  while the original sum \eqref{ZJ} is on all graphs (including
 non-connected ones), taking the $\log$ in the expression \eqref{ZJ1}
 for $W(J)$ has the effect of dropping
 all disconnected graphs, while the normalization factor $\Nc$
 in (\ref{ZJ}) eliminates all the
 ``vacuum bubbles'', that is, all the graphs that do not have
 external legs. Moreover, the number $L$ of loops in a connected graph
 determines the power $\hbar^{L-1}$
 of the unit of action that multiplies the corresponding term, so that
 (\ref{ZJ1}) has the form of a semiclassical expansion.

 \smallskip

  The next step in simplifying the combinatorics of graphs consists
 of passing to  the {\em effective action} $S_{eff}(A)$.
 By definition, $S_{eff}(A)$ is the Legendre transform of $W(J)$.

 \smallskip

  The  effective action gives the quantum corrections of
 the original action. By its definition as a Legendre transform,
 one can see that the calculation obtained by
 applying the stationary phase method to $S_{eff}(A)$
 yields the same result as the full calculation of the integrals with
 respect to the original action $S(A)$. Thus the knowledge
 of the effective action, viewed as a non-linear
 functional of classical fields, is an essential step
 in the understanding of a given Quantum Field Theory.

 \smallskip

  Exactly as above, the effective action admits a formal expansion in
 terms of graphs. In terms of the combinatorics of graphs, passing from
 $S(A)$ to the effective action $S_{eff}(A)$ has the effect of dropping
 all graphs of the form
\bigskip
\bigskip

\centerline{\fig{0.38}{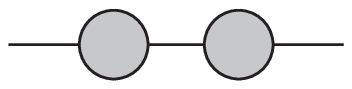}}

\bigskip

\medskip

 that can be disconnected by removal of one edge. In the figure,
the shaded areas are a shorthand notation for an arbitrary graph
with the specified external legs structure.
 The graphs that remain in this process are called {\em one particle
 irreducible} (1PI) graphs. They are by definition graphs that cannot
 be disconnected by removing a single edge.

 \smallskip

  The contribution of a 1PI graph $\Gamma$
 to the non-linear functional  $S_{eff}(A)$
 can be spelled out very concretely as follows. If
 $N$ is the number of external legs of $\Gamma$, at the formal level
 (ignoring the divergences) we have
 $$
 \Gamma(A)= \frac{1}{ N!}\int_{\sum p_j =0}\,\hat{A}(p_1)...
 \hat{A}(p_N)\,U( \Gamma(p_1,...,p_N))\,dp_1...dp_N .
 $$
 Here $\hat{A}$ is the Fourier transform of $A$
 and the {\em unrenormalized} value
 $$
 U( \Gamma(p_1,...,p_N))
 $$
 of the graph is defined by applying simple rules
 (the Feynman rules) which assign to each
 {\it internal} line in the graph a propagator
 \ie a term of the form
 \begin{equation}\label{propagator}
  \frac{1}{k^2-m^2}\,
 \end{equation}
 where $k$ is the momentum flowing through that
line. The propagators for external lines are
 eliminated for 1PI graphs.

 \smallskip

There is nothing mysterious in the appearance of the
propagator
 \eqref{propagator}, which has the role of the inverse of the quadratic
 form $S_0$ and comes from the rule of integration by parts
 \begin{equation}\label{partsrule}
 \int \,f(A)\,\langle J, A \rangle\,\exp \, (i
 \, S_0 (A)) \, [dA]= \,\int \,\partial_X f(A)\,\exp \, (i
 \, S_0 (A)) \, [dA]
 \end{equation}
 provided that
 $$
 -i\,\partial_X S_0 (A)=\,\langle J, A \rangle\,.
 $$

 \smallskip

  One then has to integrate over all momenta
 $k$
 that are left after imposing the law of conservation
 of momentum at each vertex, \ie the fact that
 the sum of ingoing momenta vanishes.
 The number of remaining integration variables is
 exactly the loop number $L$ of the graph.

 \smallskip

  As we shall see shortly, the integrals obtained this way
 are in general divergent, but by proceeding at the formal
 level we can write the effective action as
 a formal series of the form
 \begin{equation}\label{seff}
 S_{eff}(A) = S_0(A) + \sum_{\Gamma \in 1PI}
 \frac{\Gamma(A)}{S(\Gamma)},
 \end{equation}
  where the factor $S(\Gamma)$
 is the order of the symmetry group of the graph. This
 accounts for repetitions as usual in combinatorics.

 \smallskip

  Summarizing, we have the following situation. The basic unknown in
 a given Quantum Field Theory is the  effective action, which is a
 non-linear functional of classical fields and contains all quantum
 corrections to the classical action. Once known, one can obtain from it
 the Green's functions from tree level calculations (applying the
 stationary phase approximation). The formal series expansion of the
 effective action is given in terms of polynomials in the classical
 fields, but the coefficients of these polynomials are given
 by divergent integrals.

 \medskip
 \subsection{Divergences and subdivergences}

 \bigskip

  As a rule, the unrenormalized values $U(\G(p_1,\ldots,p_N))$ are
 given by divergent integrals, whose computation is governed by Feynman
 rules. The simplest of such integrals (with the corresponding graph)
 is of the form (up to powers of $2\pi$ and of the coupling constant
 $g$ and after a Wick
 rotation to Euclidean variables),
 \begin{equation}\label{Fint}
  \hbox{\psfig{figure=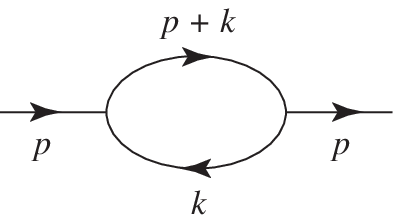}\quad =}  \int \frac{1}{k^2+m^2}\,
 \frac{1}{((p+k)^2+m^2)} \, d^D k.
 \end{equation}

 \medskip
  The integral is divergent in dimension $D=4$.
 In general, the most serious divergences in the expression of the
 unrenormalized values $U(\G)$ appear when the domain of
 integration involves arbitrarily large momenta (ultraviolet).
 Equivalently, when one
 attempts to integrate in coordinate space, one confronts
 divergences along diagonals, reflecting the fact that products of
 field operators are defined only on the configuration space of
 distinct spacetime points.

 \smallskip

   The renormalization techniques starts with the
 introduction of a regularization procedure, for instance by imposing a
 cut-off $\Lambda$ in momentum space, which restricts the corresponding
 domain of integration. This gives finite integrals, which continue to
 diverge as $\Lambda \to \infty$. One can then introduce a dependence
 on $\Lambda$ in the terms of the Lagrangian, using the unobservability
 of the bare parameters, such as the bare mass $m_0$.
 By adjusting the dependence of the bare parameters on the
 cut-off $\Lambda$, term by term in the perturbative expansion, it
 is possible, for a large class of theories called renormalizable,
 to eliminate the unwanted ultraviolet divergences.

 \smallskip

  This procedure that cancels divergences by
 correcting the {\em bare} parameters (masses, coupling constants,
 etc.) can be illustrated in the specific example of the $\phi^3$
 theory with Lagrangian
 \begin{equation}\label{phi3Lag}
  \frac{1}{2} (\partial_\mu \phi)^2 -
 \frac{m^2}{2}  \phi^2 - \frac{g}{6} \phi^3,
 \end{equation}
 which is sufficiently generic. The Lagrangian will now depend
 on the cutoff in the form
 \begin{equation}\label{cutoffterms}
 \frac{1}{2} (\partial_\mu \phi)^2 (1-\delta Z(\Lambda)) -
 \left( \frac{m^2+\delta m^2(\Lambda)}{2} \right) \phi^2 - \frac{g
 + \delta g (\Lambda)}{6} \phi^3.
 \end{equation}
 Terms such as $\delta g (\Lambda)$ are called ``counterterms''. They  do
 not have any limit as $\Lambda\to \infty$.

 \smallskip

  In the special case of asymptotically free theories, the explicit
 form of the dependence of the bare constants on the regularization
 parameter $\Lambda$ made it possible in important cases
 (\cf \cite{GK}, \cite{FMRS}) to develop successfully a constructive
 field theory, \cite{GJ}.

 \smallskip

  In the procedure of perturbative renormalization, one introduces a
 counterterm $C(\G)$ in the initial Lagrangian $\Lc$ every time one
 encounters a divergent 1PI diagram, so as to cancel the divergence. In
 the case of {\em renormalizable} theories, all the necessary
 counterterms $C(\G)$ can be obtained from the terms of the Lagrangian
 $\Lc$, just using the fact that the numerical parameters
 appearing in the expression of $\Lc$ are not observable, unlike the
 actual physical quantities which have to be finite.

 \smallskip

  The cutoff procedure is very clumsy in practice, since,
 for instance, it necessarily breaks Lorentz invariance.
 A more efficient procedure of regularization is called Dim-Reg.
 It consists in writing the integrals to be performed in dimension $D$
 and to ``integrate in dimension $D-z$ instead of $D$'', where
 now $D-z\in \C$
 (dimensional regularization).

 \smallskip

  This makes sense, since in integral dimension the Gaussian
 integrals are given by  simple functions \eqref{gaussienne}
 which continue
 to make sense at non-integral points, and provide a working definition of
 ``Gaussian integral in dimension $D-z$''.

 \smallskip

  More precisely, one first passes to the {\em Schwinger
 parameters}. In the case of the graph \eqref{Fint} this corresponds to
 writing
 \begin{equation}\label{Schwinger}
 \frac{1}{k^2+m^2}\,
 \frac{1}{(p+k)^2+m^2}=\, \int_{s>0\,t>0}\,
 e^{-s(k^2+m^2)-t((p+k)^2+m^2)}\, ds\,dt
 \end{equation}

 \smallskip

  Next, after diagonalizing the quadratic form in the
 exponential, the Gaussian integral in dimension $D$ takes the form
 \begin{equation}\label{gaussienne}
 \int\,e^{-\lambda\,k^2}\,d^D k\,=\,{\pi^{D/2}}\,{\lambda^{-D/2}}\,.
 \end{equation}
 This provides the unrenormalized value of the graph \eqref{Fint} in
 dimension $D$ as
 \begin{equation}\label{dimrg}
 \int_0^1\,\int_0^\infty\,e^{-(y(x-x^2)p^2+y
 \,m^2)}\int\,e^{-y\,k^2}\,d^D k\,\,y\,dy\,\,dx
 \end{equation}
 $$
 =\,\pi^{D/2}\,\int_0^1\,\int_0^\infty\,e^{-(y(x-x^2)p^2+y
 \,m^2)}\,y^{-D/2}\,y\,dy\,dx
 $$
 $$
 =\,\pi^{D/2}\,\Gamma(2-D/2)\,\int_0^1\,((x-x^2)p^2+\,m^2)^{D/2-2}\,dx .
 $$
 The remaining integral can be computed in terms of hypergeometric
 functions, but here the essential point is the presence of
 singularities of the $\Gamma$ function at the points $D\in 4 +
 2\N$, such that the coefficient of the pole is a polynomial
 in $p$ and the Fourier transform is a {\em local} term.

 \smallskip

  These properties are not sufficient for a theory to be
 renormalizable. For instance at $D=8$ the coefficient of pole is of
 degree 4 and the theory is not renormalizable. At $D=6$ on the other
 hand the pole coefficient has degree 2 and there are terms in the
 original Lagrangian $\Lc$ that can be used to eliminate the divergence
 by introducing suitable counterterms $\delta Z(z)$ and $\delta
 m^2 (z)$.

 \smallskip

  The procedure illustrated above works fine as long as the graph
 does not contain subdivergences. In such cases the counter terms are
 local in the sence that they appear as residues. In other
 words, one only gets simple poles in $z$.

 \smallskip

  The problem becomes far more complicated when one considers
 diagrams that possess non-trivial subdivergences.
 In this case the procedure no longer consists of a simple
 subtraction and becomes very involved, due to the following
 reasons:

 \begin{enumerate}
 \item[i)] The divergences
 of $U(\G)$
 are no longer given by local terms.
 \item[ii)] The previous
 corrections (those for the subdivergences) have to be taken into
 account in a coherent way.
 \end{enumerate}

 \smallskip

  The problem of non-local terms appears when there are poles of
 order $> 1$ in the dimensional regularization. This produces as a
 coefficient of the term in $1/z$ derivatives in $D$ of
 expressions such as
 $$ \int_0^1\,((x-x^2)p^2+\,m^2)^{D/2-2}\,dx $$
 which are no longer polynomial in $p$, even for integer values of
 $D/2-2$, but involve terms such as $\log (p^2 + 4m^2)$.

 \smallskip

  The second problem is the source of the main calculational
 complication of the subtraction procedure, namely accounting for
 subdiagrams which are already divergent.

 \smallskip

 \noindent
 The two problems in fact compensate and can be treated simultaneously,
 provided one uses the precise combinatorial recipe, due to
 Bogoliubov--Parasiuk, Hepp and Zimmermann (\cite{3BPHZ}, \cite{3BP},
 \cite{3hepp}, \cite{3zim}).

 \smallskip

  This is of an inductive nature.
 Given a graph $\G$, one first ``prepares" $\G$, by replacing the
 unrenormalized value $U(\G)$ by the formal expression
 \begin{equation}
 \overline{R}( \G) = U(\G) + \sum_{\g \sbs \G} C(\g) U( \G /
 \g) , \label{b1}
 \end{equation}
 where $\g$ varies among all divergent subgraphs.
 One then shows that the divergences of the prepared graph
 are now local terms which, for renormalisable theories, are
 already present in the original Lagrangian ${\Lc}$.
 This provides a way to define inductively the counterterm $C(\G)$ as
 \begin{equation}
 C(\G) =  -T(\overline{R}( \G)) = -T\left(U(\G) + \sum_{\g \sbs \G}
 C(\g) U( \G /
 \g)\right) , \label{b2}
 \end{equation}
 where the operation $T$ is the  projection on the
 pole part of the
 Laurent series, applied here in the parameter $z$ of DimReg.
 The renormalized value of the graph is given by
 \begin{equation}
 R(\G) =  \overline{R}( \G) +C(\G) =U(\G)  +C(\G) + \sum_{\g \sbs \G}
 C(\g) U( \G /
 \g) . \label{b3}
 \end{equation}

\section{Affine group schemes}\label{scheme}


In this section we recall some aspects of the 
general formalism of affine group schemes and
Tannakian categories, which we will need to use
later. A complete treatment of affine group schemes
and Tannakian categories can be found 
in SGA 3 \cite{SGA3} and in Deligne's \cite{De2}. 
A brief account of the formalism of affine 
group schemes in the context of differential Galois 
theory can be found in \cite{vdp}.

\smallskip

Let $\Hc$ be a commutative Hopf algebra over a field $k$ (which we assume of
characteristic zero, though the formalism of affine group schemes
extends to positive characteristic). Thus, $\Hc$ is a commutative
algebra over $k$, endowed with a (not necessarily commutative)
coproduct $\Delta: \Hc \to \Hc\otimes_k \Hc$, a 
counit $\ve: \Hc \to k$, which are
$k$-algebra morphisms  and an antipode $S: \Hc \to \Hc$
which is a
$k$-algebra antihomomorphism,
satisfying the co-rules
\begin{equation}\label{corules}
 \begin{array}{ll}
(\Delta \otimes id)\Delta = (id\otimes \Delta)\Delta & : \Hc \to
\Hc\otimes_k \Hc \otimes_k \Hc , \\[2mm]
 (id\otimes \ve)\Delta =id = (\ve \otimes id)\Delta & : \Hc \to \Hc , \\[2mm]
 m (id \otimes S)\Delta = m (S\otimes id) \Delta = 1\,\ve & : \Hc \to \Hc, 
\end{array} \end{equation}
where we use $m$ to denote the multiplication in $\Hc$.

\smallskip

Affine group schemes are the geometric counterpart of
Hopf algebras, in the following sense. One lets $G=\,{\rm Spec}\,\Hc$ be
the set of prime ideals of the commutative $k$-algebra $\Hc$, with the
Zariski topology and the structure sheaf. Here notice that the Zariski
topology by itself is too coarse to fully recover the ``algebra of
coordinates'' $\Hc$ from the topological space $\Sp(\Hc)$, while it
is recovered as global sections of the ``sheaf of functions''
on $\Sp(\Hc)$.

\smallskip

The co-rules \eqref{corules} translate on $G=\Sp(\Hc)$ to give a
product operation, a unit, and an inverse, satisfying the axioms of a
group. The scheme $G=\Sp(\Hc)$ endowed with this group structure is
called an affine group scheme.

\smallskip

One can view such $G$ as a functor that associates to any unital commutative
algebra $A$ over $k$ a group $G(A)$, whose elements 
are the $k$-algebra homomorphisms
$$
\phi \,: \Hc \to A\,,\quad \phi(X\,Y)= \phi(X) \phi(Y) \qq
X,Y\in \Hc\,, \quad\phi(1)=1\,.
$$
The product in $G(A)$ is given as the dual of the coproduct, by
 \begin{equation} \label{dualprod}
 \phi_1\,\star\,\phi_2(X)=\,\langle  \phi_1\otimes
\phi_2\,,\;\Delta(X)\rangle\,. 
 \end{equation}
This defines a group structure on $G(A)$. 
The resulting covariant functor
$$
A \,\rightarrow G(A)
$$
from commutative algebras to groups is representable (in fact by
$\Hc$). 
Conversely any covariant representable functor from the
category of commutative algebras over $k$ to groups, is defined by
an affine group scheme $G$, uniquely determined up to canonical
isomorphism.

%
%

\smallskip

We mention some basic examples of affine group schemes.

\smallskip

The additive group $G=\bG_a$: this corresponds to the Hopf algebra
$\Hc=k[t]$ with coproduct $\Delta(t)=t\otimes 1 + 1 \otimes t$.

\smallskip

The affine group scheme $G=\GL_n$: this corresponds to the Hopf
algebra $$\Hc=k[x_{i,j},t]_{i,j=1,\ldots,n} / \det(x_{i,j})t-1,$$
with coproduct $\Delta(x_{i,j})= \sum_k x_{i,k}\otimes x_{k,j}$.

\smallskip

The latter example is quite general in the following sense. If $\Hc$
is finitely generated as an algebra over $k$, then the corresponding
affine group scheme $G$ is a linear algebraic group over $k$, and can
be embedded as a Zariski closed subset in some $\GL_n$.

\smallskip

In the most general case, one can find a collection $\Hc_i\subset \Hc$
of finitely generated algebras over $k$ such that 
$\Delta(\Hc_i)\subset \Hc_i\otimes \Hc_i$, $S(\Hc_i)\subset \Hc_i$,
for all $i$, and such that, for all $i,j$ there exists a $k$ with
$\Hc_i \cup \Hc_j \subset \Hc_k$, and $\Hc=\cup_i \Hc_i$.

\smallskip

In this case, we have linear algebraic groups $G_i=\Sp(\Hc_i)$ such
that 
\begin{equation}\label{Gprojlim}
 G=\varprojlim_i G_i. 
\end{equation}
Thus, in general, an affine group scheme is a projective limit of
linear algebraic groups. 

\medskip
 \subsection{Tannakian categories}\label{tannaka}

It is natural to consider representations of an affine group scheme
$G$. A finite dimensional $k$-vector space $V$ is a $G$-module if
there is a morphism of affine group schemes $G\to \GL(V)$. This means
that we obtain, functorially, representations $G(A) \to
\Aut_A(V\otimes_k A)$, for commutative $k$-algebras $A$.
One can then consider the category ${\rm Rep}_G$ of finite dimensional
linear representations of an affine group scheme $G$.

\smallskip

We recall the notion of a Tannakian category. The main point of this
formal approach is that,
when such a category is considered over a base scheme $\cS=\Sp(k)$ (a
point), it turns out to be the category ${\rm Rep}_G$ for a uniquely
determined affine group scheme $G$. (The case of a more general scheme
$\cS$ corresponds to extending the above notions to groupoids, \cf
\cite{De2}). 

\smallskip

An abelian category is a category to which the tools of
homological algebra apply, that is, a category
where the sets of morphisms are abelian groups, there 
are products and coproducts,  kernels and cokernels always exist
and satisfy the same basic rules as in the category of modules
over a ring. 

\smallskip

A tensor category over a field $k$ of characteristic zero is a
$k$-linear abelian category $\T$ endowed with a tensor functor
$\otimes: \T\times \T \to \T$ satisfying associativity and
commutativity (given by functorial isomorphisms) and with a unit
object. Moreover, for each object $X$ there exists an object
$X^\vee$ and maps $e: X\otimes X^\vee \to 1$ and $\delta: 1 \to X
\otimes X^\vee$, such that the composites $(e\otimes 1) \circ (1\otimes
\delta)$ and $(1\otimes e) \circ (\delta \otimes 1)$ are the
identity. There is also an identification $k \simeq \End(1)$.

\smallskip

A Tannakian category $\T$ over $k$ is a tensor category endowed with a
fiber functor over a scheme $\cS$. That means a functor $\omega$ from
$\T$ to finite rank locally free sheaves over $\cS$ satisfying
$\omega(X)\otimes \omega(Y)\simeq \omega(X\otimes Y)$ compatibly with
associativity commutativity and unit. In the case where the base
scheme is a point $\cS=\Sp(k)$, the fiber functor maps to the category
$\cV_k$ of finite dimensional $k$-vector spaces.

\smallskip

The category ${\rm Rep}_G$ of finite dimensional linear
representations of an affine group scheme is a Tannakian category,
with an exact faithful fiber functor to $\cV_k$ (a neutral Tannakian
category). What is remarkable is
that the converse also holds, namely, if $\T$ is a neutral Tannakian
category, then it is equivalent to the category   
${\rm Rep}_G$ for a uniquely determined affine group scheme $G$, which
is obtained as automorphisms of the fiber functor. 

\smallskip

Thus, a neutral Tannakian category is indeed a more geometric
notion than might at first appear from the axiomatic definition,
namely it is just the category of 
finite dimensional linear representations of an affine group scheme.

\smallskip

This means, for instance, that when one considers only finite dimensional
linear representations of a group (these also form a neutral
Tannakian category), one can as well replace the given group by its
``algebraic hull", which is the affine group scheme underlying the
neutral Tannakian category.

\medskip
 \subsection{The Lie algebra and the Milnor-Moore theorem}
\bigskip

\smallskip

Let $G$ be an affine group scheme over a field $k$ of characteristic
zero. The Lie algebra $\fg(k)=\Lie\, G(k)$ is given by
the set of linear maps $ L\,:\Hc \to k$ satisfying
\begin{equation}\label{Liescheme}
L(X\,Y)=\, L(X)\,\ve(Y) +\, \ve(X)\, L(Y)\,,\quad \forall X\,,Y
\in \Hc\,,
\end{equation}
where $\ve$ is the augmentation of $\Hc$, playing the role of the
unit in the dual algebra.

\smallskip

Notice that the above formulation is equivalent to defining
the Lie algebra $\fg(k)$ in terms of left invariant derivations on $\Hc$,
namely linear maps $D:\Hc\to \Hc$ satisfying 
$D(XY)=XD(Y)+D(X)Y$ and $\Delta D=(id\otimes D)\Delta$, which
expresses the left invariance in Hopf algebra terms. The
isomorphism between the two constructions is easily obtained as
$$ D \mapsto L=\ve\, D\, , \ \ \ \ \ \ L \mapsto D=(id\otimes L)\Delta. $$
Thus, in terms of left invariant derivations, the Lie bracket is just
$[D_1,D_2]=D_1D_2-D_2D_1$. 


\smallskip

The above extends to a covariant functor $\fg=\hbox{Lie}
\ G$, 
\begin{equation}\label{LieGA}
A \,\rightarrow \fg(A)\,,
\end{equation}
from commutative $k$-algebras to Lie algebras, where $\fg(A)$ is the
Lie algebra of linear maps $ L\,:\Hc \to A$ satisfying
\eqref{Liescheme}. 

\smallskip

In general, the Lie algebra $\Lie \, G$ of an affine group scheme does
not contain enough information to recover its algebra of coordinates
$\Hc$. 
However, under suitable hypothesis, one can in fact recover the Hopf
algebra from the Lie algebra.

\smallskip

In fact, assume that $\Hc$ is a connected graded Hopf algebra, namely
$\Hc=\oplus_{n\geq 0} \Hc_n$, with $\Hc_0=k$, with commutative
multiplication. Let $\cL$ be the Lie algebra
of primitive elements of the dual $\Hc^\vee$. 
We assume that $\Hc$ is, in
each degree, a finite dimensional vector space.
Then, by (the dual of) the Milnor--Moore theorem
\cite{3MM}, we have a canonical isomorphism of Hopf algebras
\begin{equation}\label{HUL}
\Hc \simeq U(\cL)^\vee,
\end{equation}
where $U(\cL)$ is the universal enveloping algebra of $\cL$. Moreover,
$\cL= \Lie\,\, G(k)$.  

\smallskip

As above, we consider a Hopf algebra $\Hc$ endowed with an
integral positive grading. We assume that it is 
connected, so that all elements of the augmentation ideal have
strictly positive degree. We let $Y$ be the generator of the
grading so that for $X\in \Hc$ homogeneous of degree $n$ one has
$Y(X)=\,n\, X$. 

\smallskip

Let $\bG_m$ be the multiplicative group, namely the affine group
scheme with Hopf algebra $k[t,t^{-1}]$ and coproduct
$\Delta(t)=\,t\otimes t$.

\smallskip

Since the grading is integral, 
we can define, for $u\in \bG_m$, an action
$u^Y$ on $\Hc$ (or on its dual) by
\begin{equation}\label{uYaction}
u^Y(X)=\,u^n\,X \qq X\in \Hc\,,\quad {\rm degree} \,X=n\,.
\end{equation}
We can then form the semidirect product
 \begin{equation}
 G^* = G \, \rtimes \, \bG_m . \label{GGm}
 \end{equation}
This is also an affine group
scheme, and one has a natural morphism of group schemes
$$
G^*\,\to \bG_m \,.
$$
The Lie algebra of $G^*$ has an  additional generator  such that
\begin{equation}
 [Z_0 , X] = Y(X) \qquad \forall \, X \in \hbox{Lie} \ G
\,. \label{LieGGm} 
\end{equation}

 \bigskip
 \section{The Hopf algebra of Feynman graphs
and diffeographisms}\label{Sgraphs}
\bigskip

 \noindent In '97, Dirk Kreimer got the remarkable idea (see
 \cite{3dhopf}) to encode the substraction procedure by a Hopf algebra.
 His algebra of rooted trees was then refined in \cite{3CK} to a
 Hopf algebra ${{\mathcal H}}$ directly defined in terms of graphs.

 \smallskip

  The result is that one can associate to any renormalizable theory $\Tc$ a
 Hopf algebra $\Hc=\Hc(\Tc)$ over $\C$, where the coproduct reflects
 the structure of the 
 preparation formula \eqref{b1}. We discuss this explicitly for the
 case of $\Tc=\phi^3_6$, the theory $\phi^3$ in dimension $D=6$, which
is notationally simple and at the same time sufficiently generic
to illustrate all the main aspects of the general case.

 \smallskip

  In this case, the graphs have three kinds of vertices, which
 correspond to the three terms in the Lagrangian \eqref{phi3Lag}:

 \begin{itemize}

 \item Three legs vertex $ \hbox{ \psfig{figure=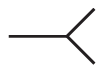}
 } $ associated to the $\phi^3$ term in the Lagrangian

 \item Two legs vertex $ \hbox{ \psfig{figure=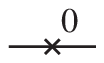} }
 $ associated to the term $\phi^2$.

 \item  Two legs vertex  $ \hbox{ \psfig{figure=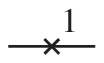} }
 $ associated to the term $(\part \, \phi)^2$.

 \end{itemize}

  The rule is that the number of edges at a vertex equals the degree of
 the corresponding monomial in the Lagrangian. Each edge either
 connects two vertices (internal line) or a single vertex (external
 line). In the case of a massless theory the term $\phi^2$ is absent
 and so is the corresponding type of vertex.

 \smallskip

  As we discussed in the previous section, the value
 $U(\Gamma(p_1,\ldots ,p_N))$ depends on the datum of the {\em
 incoming} momenta
\bigskip

\centerline{\fig{0.35}{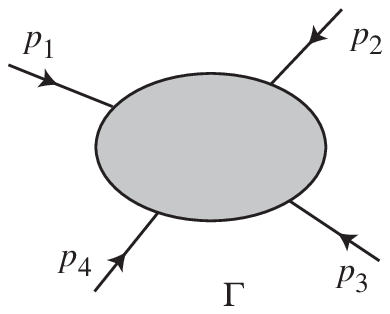}}

\bigskip

 attached to the external edges of the graph $\G$,
 subject to the conservation law
 $$ \sum p_i=0. $$

 \smallskip

  As an algebra, the Hopf algebra ${{\mathcal H}}$ is
 the free commutative algebra generated by the $\G(p_1,\ldots,p_N)$
 with $\G$ running over 1PI graphs.
 It is convenient to encode the external datum of the momenta in the
 form of a distribution $\sigma : C^\infty(E_\Gamma)\ra \Cb $ on the
 space of $C^\infty$-functions on
 \begin{equation}\label{egama}
  E_\Gamma=\,\left\{ (p_i)_{i=1 , \ldots , N} \ ; \
 \sum \, p_i = 0 \right\}  \, .  \end{equation}
 where the set $\{1, \ldots , N\}$
 of indices is the set of external legs of $\G$.
 Thus, the algebra ${{\mathcal H}}$ is identified with the symmetric
 algebra on a linear space that is the direct sum of spaces of
 distributions $C_c^{-\infty}(E_\Gamma)$, that is,
 \begin{equation}\label{algH}
 \Hc=\,S(C_c^{-\infty}(\cup E_\Gamma))\,.
 \end{equation}

 \smallskip

  In particular, we introduce the notation $\G_{(0)}$ for graphs  with
at least three external legs to mean
 $\G$ with the external structure given by the distribution $\sigma$
 that is a Dirac mass at $0 \in E_{\G}$,
 \begin{equation}\label{3Gamma0}
  \G_{(0)} = \,\left( \G (p) \right)_{p=0}
 \end{equation}

\bigskip

\centerline{\fig{0.30}{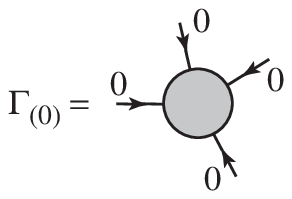}}

\bigskip

 \smallskip

 For self energy graphs, \ie graphs $\G$
 with just two external lines, we  use the two external structures
 $\sigma_j$ such that
 \begin{equation}
 \label{3Gamma1}
 \G_{(0)} = m^{-2}\,\left( \G (p) \right)_{p=0}\,,
\qquad \G_{(1)} =
 \left( \frac{\part}{\part \, p^2} \ \G (p) \right)_{p=0} .  \end{equation}
 There is a lot of freedom in the  choice
 of the external structures $\sigma_j$, the only important
 property  being
 \begin{equation}\label{sigj} \sigma_0 \, (a \, m^2+ b
 \, p^2) = a \ , \ \sigma_1 \, (a \, m^2+ b \, p^2) = b \,.  \end{equation}
 In the case of a massless theory, one does not take
 $p^2=0$  to avoid a possible pole
 at $p=0$ due to infrared divergences. It is however easy to adapt the
 above discussion to that situation.

 \smallskip

  In order to define the coproduct
 \begin{equation}
 \D : {{\mathcal H}} \ra {{\mathcal H}} \ot {{\mathcal H}} \label{h3}
 \end{equation}
 it is enough to specify it on 1PI graphs. One sets
 \begin{equation}
 \Delta \G = \G \ot 1 + 1 \ot \G + \sum_{\g \sbs \G} \g_{(i)} \ot \G /
 \g_{(i)} . \label{h4}
 \end{equation}
 Here $\g$ is a non-trivial (non empty as well as its complement)
 subset $\g \sbs \tilde\G$
 of the graph $\tilde\G$ formed by the internal edges of $\G$. The
 connected components $\g'$ of $\g$ are 1PI graphs with the property
 that the set $\epsilon(\g')$ of egdes of $\G$ that meet $\g'$ without
 being edges of $\g'$ consists of two or three elements (\cf
 \cite{3CK}). One denotes by $\gamma'_{(i)}$ the graph that has
 $\gamma'$ as set of internal edges and $\epsilon(\g')$ as external
 edges. The index $i$ can take the values 0 or 1 in the case of
 two external edges and 0 in the case of three. We assign to
 $\gamma'_{(i)}$ the external structure of momenta given by the
 distribution $\s_i$
 for two external edges and \eqref{3Gamma0} in the case of three.
 The summation in \eqref{h4} is over all multi-indices $i$
 attached to the connected components of $\g$. In \eqref{h4} $\g_{(i)}$
 denotes the product of the graphs $\g'_{(i)}$ associated to the
 connected components of $\g$. The graph $\G/\g_{(i)}$ is obtained by
 replacing each $\g'_{(i)}$ by a corresponding vertex of type
 $(i)$. One can check that $\G/\g_{(i)}$ is a 1PI graph.

 \smallskip

  Notice that,
 even if the $\g'$ are disjoint by construction, the graphs $\g'_{(i)}$
 need not be, as they may have external edges in common, as one can see
 in the example of the graph

\bigskip

\centerline{\fig{0.5}{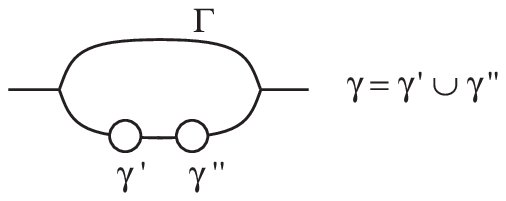}}

\bigskip

 for which the external structure of $\G/\g_{(i)}$ is identical to that
 of $\G$.

 \smallskip

  An interesting property of the coproduct $\D$ of \eqref{h4} is a
 ``linearity on the right'', which means the following (\cite{3CK}):

 \begin{prop}\label{linear}
 Let $\Hc_1$ be the linear subspace of $\Hc$ generated
 by 1 and the 1PI graphs, then
 for all $\G\in \Hc_1$ the coproduct satisfies
 $$\D (\G) \in \Hc \ot \Hc_{1}\,.$$
 \end{prop}

 \smallskip

  This properties reveals the similarity between $\D$ and
 the coproduct defined by composition of formal series. One can see
 this property illustrated in the following explicit examples taken
 from \cite{3CK}:

 \medskip

\bigskip

\centerline{\fig{0.5}{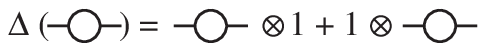}}

\bigskip

 \medskip

\bigskip

\centerline{\fig{0.5}{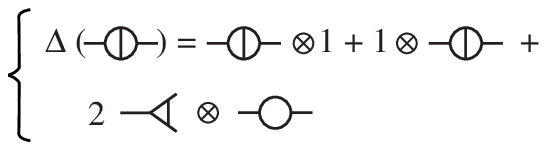}}

\bigskip

 \medskip

\bigskip

\centerline{\fig{0.5}{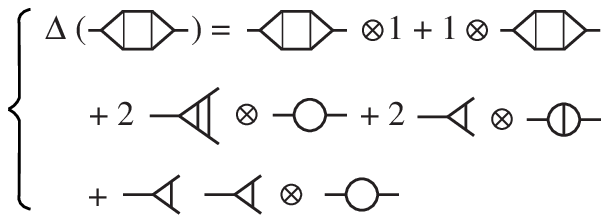}}

\bigskip

 \medskip

\bigskip

\centerline{\fig{0.5}{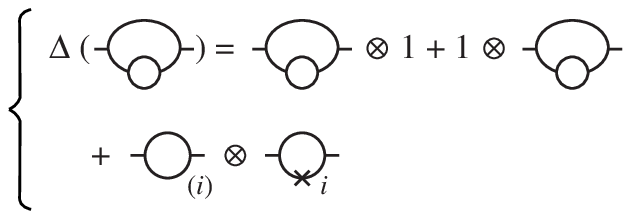}}

\bigskip

 \medskip

 \noindent The coproduct $\D$ defined by (\ref{h4}) for 1PI graphs extends
 uniquely to a homomorphism from ${{\mathcal H}}$ to ${{\mathcal H}}
 \ot {{\mathcal H}}$. The main result then is the following
 (\cite{3dhopf},\cite{3CK}):

 \begin{thm}\label{Hopfalg}
 The pair $({\mathcal H} , \D)$ is a Hopf
 algebra.
 \end{thm}

This Hopf algebra defines an affine group scheme $G$
canonically associated to the quantum field theory
according to the general formalism of section \ref{scheme}.
We refer to $G$ as the group of {\em diffeographisms}
of the theory
\begin{equation}\label{difg}
G=\,{\rm Difg}(\Tc)\,.
\end{equation}
We have illustrated the construction in the specific case
of the $\phi^3$
theory in dimension $6$, namely for $G={\rm Difg}(\phi^3_6)$.

\medskip

The presence of the external structure of graphs plays only a
minor role in the coproduct except for the explicit external
structures $\sigma_j$ used for internal graphs. We shall now see
that this corresponds to a simple decomposition at the level of
the associated Lie algebras.

 \bigskip
 \section{The Lie algebra of graphs}\label{SLie}
 \medskip

  The next main step in the CK theory of perturbative
 renormalization (\cite{3CK}) is the analysis of the Hopf algebra
 ${{\mathcal H}}$ of graphs of \cite{3CK} through the Milnor-Moore
 theorem (\cf \cite{3MM}). This
 allows one to view ${{\mathcal H}}$ as the dual
 of the enveloping algebra of a graded Lie algebra,
 with a linear basis given by 1PI graphs.
 The Lie bracket between two graphs is obtained by
 insertion of one graph in the other.
  We recall here the structure of this
 Lie algebra.

 \smallskip

  The Hopf algebra ${{\mathcal H}}$ admits several
 natural choices of grading. To define a grading it suffices
 to assign the degree of 1PI graphs together with the rule
 \begin{equation}
 \deg \, (\G_1 \ldots \G_e) = \sum \deg \, (\G_j) \ , \quad \deg \, (1)
 = 0 .
 \label{g1}
 \end{equation}
 One then has to check that, for any admissible subgraph $\g$,
 \begin{equation}
 \deg \, (\g) + \deg \, (\G / \g) = \deg \, (\G) . \label{g2}
 \end{equation}

 \smallskip

 \noindent The two simplest choices of grading are
 \begin{equation}
 I \, (\G) = \hbox{number of internal edges of} \ \G \label{g3}
 \end{equation}
 and
 \begin{equation}
 v \, (\G) = V \, (\G) - 1 = \hbox{number of vertices of} \ \G - 1 \, ,
 \label{g4}
 \end{equation}
 as well as the ``loop number" which is the difference
 \begin{equation}
 L = I - v = I - V + 1 .\label{g5}
 \end{equation}

 \smallskip

  The recipe of the Milnor-Moore theorem (\cf \cite{3MM})
applied to the bigraded Hopf algebra
 ${{\mathcal H}}$  gives a Lie algebra
 structure on the linear space
 \begin{equation}
 L =\, \bigoplus_\Gamma C^{\infty}( E_\Gamma)
 \end{equation}
 where $C^{\infty} (E_{\G})$ denotes
 the space of smooth functions on $E_{\G}$ as in \eqref{egama}, and the
 direct sum is taken over 1PI graphs $\G$.

 \smallskip

  For $X \in L$ let $Z_X$ be the linear form on ${{\mathcal H}}$
 given, on monomials $\G$, by
 \begin{equation}
  \lgl \G , Z_X
 \rgl =\lgl \sigma_{\G} , X_{\G} \rgl  , \label{g9}
 \end{equation}
 when $\G$ is connected and 1PI, and
 \begin{equation}
  \lgl \G , Z_X \rgl = 0  \label{g8}
 \end{equation}
 otherwise. Namely, for a connected 1PI graph \eqref{g9} is
 the evaluation of the external structure
 $\sigma_{\G}$ on the component $X_{\G}$ of $X$.

 \smallskip

  By construction, $Z_X$ is an infinitesimal character of ${{\mathcal
 H}}$, \ie a linear map $Z\,: \Hc \to \C$ such that
 \begin{equation}\label{infc}
Z(x y)=Z(x)\,\ve(y)+\,\ve(x)\,Z(y)\,,\quad \forall x,y\in\Hc
\end{equation}
where $\ve$ is the augmentation.

The same holds for the  commutators
 \begin{equation}
  [Z_{X_1} , Z_{X_2}] =
 Z_{X_1} \, Z_{X_2} - Z_{X_2} \, Z_{X_1} , \label{g10}
   \end{equation}
 where the product is obtained by transposing
 the coproduct of ${{\mathcal H}}$, \ie
 \begin{equation}
  \lgl Z_1 \, Z_2 , \G \rgl = \lgl Z_1 \ot Z_2 , \D \, \G
 \rgl \, . \label{g11}
  \end{equation}

 \smallskip

 \noindent  Let $\G_j$, for $j = 1,2$, be 1PI graphs, and let $\vp_j
 \in  C^{\infty} (E_{\G_j})$ be the corresponding test functions.
 For $i \in \{ 0,1 \}$, let $n_i \, (\G_1 , \G_2 ; \G)$ be the
 number of subgraphs of $\G$ isomorphic to  $\G_1$ and such that
 \begin{equation}
 \G / {\G_1}_{(i)} \sm \G_2 \, , \label{G1G2g11}
 \end{equation}
 with the notation $\G_{(i)}$, for $i \in \{ 0,1 \}$, as in
 \eqref{3Gamma0} and \eqref{3Gamma1}.

 \smallskip

   One then has the following (\cite{3CK}):

 \begin{lem}\label{Liebracket}
 Let $(\G , \vp)$ be an element of $L$, with $\vp \in C^{\infty}
 \, (E_{\G})$.
 The Lie bracket of $(\G_1 , \vp_1)$ with $(\G_2 ,
 \vp_2)$ is then given by the formula
 \begin{equation}
 \sum_{\G , i} \, \sigma_i \, (\vp_1) \ n_i \, (\G_1 , \G_2 ; \G) \,
 (\G , \vp_2)  - \sigma_i \, (\vp_2) \ n_i \, (\G_2 , \G_1 ; \G) \, (\G
 , \vp_1) \, . \label{g12}
 \end{equation}
  where  $\s_i$ is as in \eqref{3Gamma1}
 for two external edges and \eqref{3Gamma0} in the case of three.
 \end{lem}

 \medskip

  The main result on the structure of the Lie algebra is
 the following (\cite{3CK}):

 \begin{thm}\label{Liealg}
 The Lie algebra $L$ is
 the semi-direct product of an abelian Lie algebra $L_{{\rm ab}}$ with $L'$
 where $L'$ admits a canonical linear basis indexed by graphs
 with
 $$
 [\G , \G'] = \sum_v \, \G \circ_v \G' - \sum_{v'} \, \G' \circ_{v'} \G
 $$
 where $\G \circ_v \G'$ is obtained by inserting $\G'$ in $\G$ at $v$.
 \end{thm}

 \smallskip

 The corresponding Lie group $G(\C)$ is the group of characters of the Hopf
 algebra ${{\mathcal H}}$, \ie the set of complex points of the
corresponding affine group scheme $G={\rm Difg}(\Tc)$.
 
\smallskip

We see from the structure
of the Lie algebra that the group scheme ${\rm Difg}(\Tc)$ 
is a semi-direct product, $$ {\rm Difg} =
 {\rm Difg}_{{\rm ab}} \, \semi \, {\rm Difg}'  $$ of an abelian group
 ${\rm Difg}_{{\rm ab}}$ by the  
 group scheme ${\rm Difg}'$ associated to the Hopf subalgebra $\Hc'$
 constructed 
 on 1PI graphs with two or three external legs and fixed external
 structure. Passing from ${\rm Difg}'$ to ${\rm Difg}$ is a trivial
 step and we 
 shall thus restrict our attention to the group ${\rm Difg}'$
 in the sequel.

The Hopf algebra $\Hc'$ of coordinates on ${\rm Difg}'$ is now finite
dimensional in each degree for the grading given by the loop
number, so that all technical problems associated to dualities of
infinite dimensional linear spaces disappear in that context. 
In particular the Milnor-Moore theorem applies and
shows that $\Hc'$ is the dual of the enveloping algebra
of $L'$. The
conceptual structure of ${\rm Difg}'$ is that of a graded affine group
scheme (\cf Section \ref{scheme}).
 Its complex points form a pro-unipotent Lie group,
intimately related to the group of 
 formal diffeomorphisms of the dimensionless coupling constants of the
 physical theory, as we shall recall in Section \ref{difgdiff}.

 \bigskip
 \section{Birkhoff decomposition and renormalization}\label{SBirk}

  With the setting described in the previous sections, the main
 subsequent conceptual breakthrough in the CK theory of renormalization
 \cite{3CK} consisted of the discovery that formulas identical to
 equations \eqref{b1}, \eqref{b2}, \eqref{b3} occur in the Birkhoff
 decomposition of loops, for an arbitrary graded 
complex pro-unipotent Lie group
 $G$.

 \smallskip

  This unveils a neat and simple conceptual picture underlying
 the seemingly complicated combinatorics of the
 Bogoliubov--Parasiuk--Hepp--Zimmermann procedure,
 and shows that it is a special case of a general mathematical
 method of extraction of finite values given by the Birkhoff
 decomposition.

 \smallskip

We first recall some general facts about the Birkhoff
 decomposition and then describe the specific case of
interest, for the setting of renormalization.

\smallskip

The Birkhoff decomposition of loops is a factorization of the form
 \begin{equation}
 \g \, (z) = \g_- (z)^{-1} \, \g_+ (z) \qquad z \in C \,
 ,\label{birkhoffpm}
 \end{equation}
 where $C \sbs \P^1 (\Cb)$ is a smooth simple curve, $C_-$ denotes
 the component of the complement of $C$ containing $\ify \not\in C$
 and $C_+$ the other component. Both $\g$ and $\g_{\pm}$ are loops
 with values in a complex Lie group $G$
 \begin{equation}
 \g \, (z) \in G  \qquad \fl \, z \in \Cb
 \label{renorm16}
 \end{equation}
 and $\g_{\pm}$ are boundary values of holomorphic maps (which we still
 denote by the same symbol)
 \begin{equation}
 \g_{\pm} : C_{\pm} \ra G \, . \label{renorm17}
 \end{equation}
 The normalization condition $\g_- (\ify) = 1$ ensures that, if it
 exists, the decomposition \eqref{birkhoffpm} is unique (under
 suitable regularity conditions). When the loop $\g : C \ra
 G$ extends to a holomorphic loop $\g_+ : C_+ \ra G$, the Birkhoff
 decomposition is given by $\g_+ = \g$, with $\g_- = 1$.

 \smallskip

  In general, for $z_0 \in C_+$, the evaluation
 \begin{equation}
 \g \ra \g_+ (z_0) \in G \label{renorm21}
 \end{equation}
 is a natural principle to extract a finite value from the singular
 expression $\g (z_0)$. This extraction of finite values is a
 multiplicative removal of the pole part for a meromorphic loop
 $\g$ when we let $C$ be an {\em infinitesimal} circle centered at $z_0$.

 \smallskip

 This procedure is closely related to the classification of holomorphic
 vector bundles on the Riemann sphere $\P^1 (\Cb)$ (\cf
 \cite{Groth}). In fact, consider
 as above a curve $C \subset \P^1(\C)$. Let us assume for simplicity that
 $C=\{ z: |z|=1\}$, so that
 $$ C_- =\{ z: |z|> 1 \} \ \ \ \text{ and } \ \ \  C_+=\{ z: |z|<1
 \}. $$
 We consider the Lie group $G= {\rm GL}_n (\Cb)$. In this case, any loop
 $\gamma: C \to G$ can be decomposed as a product
 \begin{equation}\label{factorization}
 \gamma(z) = \gamma_-(z)^{-1}\, \lambda(z)\, \gamma_+(z),
 \end{equation}
 where $\gamma_\pm$ are boundary values of holomorphic maps
 \eqref{renorm17} and
 $\lambda$ is a homomorphism of $S^1$ into the subgroup of diagonal
 matrices in ${\rm GL}_n (\Cb)$,
 \begin{equation}\label{middleterm}
  \lambda(z)=\left( \begin{array}{cccc} z^{k_1} & & & \\
 & z^{k_2} & & \\
 & & \ddots & \\
 & & & z^{k_n} \end{array} \right),
 \end{equation}
 for integers $k_i$. There is a dense open subset $\Omega$ of the
 identity component of the loop group $\Lc G$ for which the Birkhoff
 factorization \eqref{factorization} is of the form \eqref{birkhoffpm},
 namely where $\lambda=1$. Then \eqref{birkhoffpm} gives an isomorphism
 between $\Lc^-_1 \times \Lc^+$ and $\Omega\subset LG$, where
 $$ \Lc^\pm =\{ \gamma \in \Lc G : \gamma \text{ extends to a holomorphic
 function on } C_\pm \} $$
 and $\Lc^-_1=\{ \gamma \in \Lc^- : \gamma(\infty) =1 \}$ (see \eg
 \cite{PS}).

 \smallskip

 Let $U_\pm$ be the open sets in $\P^1(\C)$
 $$ U_+=\P^1(\C)\smallsetminus \{ \infty \} \ \ \ \
 U_-=\P^1(\C)\smallsetminus \{ 0 \} . $$
 Gluing together trivial line bundles on $U_\pm$ via the transition
 function on $U_+\cap U_-$ that multiplies by $z^k$, yields a
 holomorphic line bundle $L^k$ on $\P^1(\C)$. Similarly, a holomorphic
 vector bundle $E$ is obtained by gluing trivial vector bundles on $U_\pm$
 via a transition function that is a holomorphic function
 $$ \gamma: U_+\cap U_- \to G. $$
 Equivalently,
 \begin{equation}
 E= (U_+ \ts \Cb^n)\cup_{\g} (U_- \ts \Cb^n) \, . \label{renorm20}
 \end{equation}

 \smallskip

 The Birkhoff factorization \eqref{factorization} for $\gamma$ then
 gives the Birkhoff--Grothendieck decomposition of $E$ as
 \begin{equation}
 E = L^{k_1} \op \ldots \op L^{k_n} \, .\label{renorm19}
 \end{equation}

 \smallskip

 The existence of a Birkhoff decomposition of the form
 \eqref{birkhoffpm} is then clearly equivalent to the vanishing of the
 Chern numbers
 \begin{equation}
 c_1 \, (L^{k_i}) = 0 \label{renorm18}
 \end{equation}
 of the holomorphic line bundles in the Birkhoff--Grothendieck
 decomposition \eqref{renorm19}, \ie to the condition $k_i=0$ for
 $i=1,\ldots, n$.

 \smallskip

 The above discussion for $G = {\rm GL}_n (\Cb)$ extends to
 arbitrary complex Lie groups.
 When $G$ is a simply connected nilpotent complex Lie group, the
 existence (and uniqueness) of the Birkhoff decomposition
 \eqref{birkhoffpm} is valid for any $\g$.

 \smallskip

  We now describe explicitly
 the Birkhoff decomposition with respect to an infinitesimal
 circle centered at $z_0$, and express the result
 in algebraic terms using the standard translation from the geometric
 to the algebraic language. 

\smallskip

Here we consider a graded connected  commutative Hopf algebra
${\mathcal H}$ over $\C$ and we let $G=\Sp(\Hc)$ be the associated
affine group scheme as described in Section \ref{scheme}. This
is, by definition, the set of prime ideals of
${\mathcal H}$ with the Zariski topology and a structure sheaf.
 What matters for us is the corresponding covariant
functor from commutative algebras $A$ over $\C$ to groups, given
by the set of algebra homomorphisms,
 \begin{equation}
G(A)=\, {\rm Hom}(\Hc,A)
 \end{equation}
where the group structure on $G(A)$ is dual to the coproduct \ie
is given by
$$
\phi_1\star\phi_2(h)=\,\langle \phi_1\otimes\phi_2,
\Delta(h)\rangle
$$
By construction $G$ appears in this way as a representable
covariant functor from the category of commutative $\C$-algebras
to groups.

\smallskip

 In the physics framework we are  interested in the evaluation
of loops at a specific complex number say
 $z_0=0$. We let $K=\C(\{ z \})$ (also denoted by $\C\{z\}[z^{-1}]$) be the
field of convergent Laurent series, with arbitrary radius of
convergence. We denote by $\O=\C\{ z \}$ be the ring of convergent
power series, and $\Qc=\,z^{-1}\,\C([z^{-1}])$, with
$\tilde\Qc=\C([z^{-1}])$ the corresponding unital ring.

 \smallskip

  Let us first recall the standard dictionary from the geometric to
 the algebraic language, summarized by the following diagram.

 \smallskip

 \begin{equation}
 \begin{array}{rcl}
 \underline{\mbox{{ Loops}  $\gamma\,:C\to G$}}
 & \mid &
 \underline{G(K)=\{\mbox{{ homomorphisms}
 $\phi\,:{\mathcal H}\to K$}\}} \\
 & \mid & \\
 \mbox{Loops $\gamma\,:P_1 (\Cb)\backslash\{z_0\}\to G$}
 & \mid & G(\tilde\Qc)=\{
 \phi\,,
 \phi({\mathcal
 H})\subset\;\tilde\Qc \}\\
 & \mid & \\
 \mbox{$\gamma(z_0)$
 is finite}
 & \mid &
G(\O)=\{
 \phi\,,
 \phi({\mathcal
 H})\subset\;\O \} \\
 & \mid & \\
 \gamma(z)=\gamma_1(z)\,\gamma_2(z)\,\, \forall z\in C
 & \mid &
 \phi= \phi_1\star\phi_2 \\
 & \mid & \\
 z\mapsto \gamma(z)^{-1} & \mid & \phi \circ S \\
 & \mid & \\
 \end{array}
 \end{equation}

 \medskip

For loops $\gamma\,:P_1 (\Cb)\backslash\{z_0\}\to G$ the
normalization condition $\gamma(\infty)=1$ translates
algebraically into the condition
$$
\ve_-\circ \phi=\,\ve
$$
where $\ve_-$ is the augmentation in the ring $\tilde\Qc$ and
$\ve$ the augmentation in $\Hc$.

 \noindent As a preparation
to  the main result of \cite{3CK} on renormalization and
 the Birkhoff decomposition, we reproduce in full the proof
given in \cite{3CK} of the following basic
algebraic fact, where the Hopf algebra ${\mathcal H}$ is graded in
positive degree and connected (the scalars are the only elements
of degree $0$).

 \begin{thm}\label{RenBirk}
 Let $\phi:
 {\mathcal H}\to K$ be an algebra homomorphism. The
 Birkhoff decomposition of the corresponding loop is obtained
 recursively from the equalities
 \begin{equation}\label{Hbirkhoff1}
 \phi_-(X)=-T\left(\phi(X)+\sum\phi_-(X^\prime)
 \phi(X^{\prime\prime}) \right)
 \end{equation}
 and
 \begin{equation}\label{Hbirkhoff2}
 \phi_+(X)=\phi(X)+\phi_-(X)+\sum\phi_-(X^\prime)
 \phi(X^{\prime\prime}).
 \end{equation}
 \end{thm}

  Here $T$ is, as in \eqref{b2}, the operator of projection on
 the pole part, \ie the projection on the augmentation
ideal of $\tilde\Qc$, parallel to
 $\O$. Also $X'$ and $X''$ denote the terms of lower
 degree that appear in the coproduct
 $$ \Delta(X)= X \otimes 1 + 1 \otimes X + \sum X^\prime
 \otimes X^{\prime\prime}, $$
 for $X \in{\mathcal H}$.

\smallskip

To prove that the Birkhoff decomposition corresponds to the
expressions \eqref{Hbirkhoff1} and \eqref{Hbirkhoff2}, one proceeds by
defining inductively a homomorphism $\phi_-:  \Hc \to K$ by
\eqref{Hbirkhoff1}. One then shows by induction that it is
multiplicative. 

\smallskip

Explicitly, let $\wt \Hc=\ker \ve$ be the augmentation ideal.
For $X , Y \in \wt \Hc$, one has
\begin{equation}\label{Birk1}
\begin{array}{c}
\D (XY) = XY \ot 1 + 1 \ot XY + X \ot Y + Y \ot X + XY' \ot Y'' \, + \\[2mm]
Y' \ot X Y'' + X' Y \ot X'' + X' \ot X'' Y + X' Y' \ot X'' Y'' \, . 
\end{array}
\end{equation}
We then get
\begin{equation}\label{Birk2}
\begin{array}{c}
\phi_- (XY) = -T (\phi (XY)) - T (\phi_- (X) \, \phi(Y) +  \phi_-(Y)
\, \phi(X) \, + \\[2mm] 
\phi_- (XY') \, \phi (Y'') + \phi_- (Y') \, \phi(XY'') + \phi_- (X' Y)
\, \phi(X'')\\[2mm] 
+ \, \phi_- (X') \, \phi (X'' Y) + \phi_- (X' Y') \, \phi (X'' Y'')) \, . 
\end{array}
\end{equation}
Now $\phi$ is a homomorphism and we can assume that we have shown
$\phi_-$ to be 
multiplicative, $ \phi_-(AB) = \phi_- (A) \, \phi_-(B)$, for $\deg A +
\deg B < \deg X + 
\deg Y$. This allows us to rewrite \eqref{Birk2} as
\begin{equation}\label{Birk3}
\begin{array}{c}
 \phi_-(XY) = -T (\phi (X) \, \phi(Y) + \phi_- (X) \, \phi (Y) +
\phi_- (Y) \, \phi(X) \\[2mm] 
+ \,  \phi_-(X) \,  \phi_-(Y') \, \phi (Y'') +  \phi_-(Y') \, \phi (X)
\, \phi (Y'') +  \phi_-(X') \, 
\phi_-
(Y) \,  \phi(X'')\\[2mm]
+ \, \phi_- (X') \, \phi (X'') \,  \phi(Y) + \phi_- (X') \, \phi_-
(Y') \, \phi (X'') \, \phi (Y'')) 
\, .
\end{array}
\end{equation}
Let us now compute $\phi_- (X) \, \phi_- (Y)$ using the
multiplicativity constraint 
fulfilled by $T$ in the form
\begin{equation}\label{Tmult}
T(x) \, T(y) = -T (xy) + T (T(x) \, y) + T (x \, T(y)) \, .
\end{equation}
We thus get
\begin{equation}\label{Birk4}
\begin{array}{c}
\phi_- (X) \, \phi_- (Y) = -T ((\phi(X) + \phi_- (X') \, \phi (X''))
\, (\phi (Y) \, + \\[2mm] 
 \phi_-(Y') \,  \phi(Y'')) + T (T (\phi(X) +  \phi_-(X') \, \phi
(X'')) \, (\phi(Y) \, + \\[2mm] 
\phi_- (Y') \, \phi (Y'')) + T ((\phi(X) +  \phi_-(X') \, \phi(X''))
\, T (\phi(Y) + \phi_- (Y') \, 
\phi (Y'')))\, ,\end{array}
\end{equation}
by applying \eqref{Tmult} to $x = \phi(X) + \phi_-(X') \, \phi(X'')$, $y =
\phi(Y) + \phi_- (Y') \, \phi 
(Y'')$. Since $T(x) = - \phi_-(X)$, $T(y) = -\phi_- (Y)$, we can
rewrite \eqref{Birk4} as 
\begin{equation}\label{Birk5}
\begin{array}{c}
\phi_- (X) \,  \phi_-(Y) = -T (\phi(X) \, \phi (Y) + \phi_- (X') \,
\phi (X'') \, \phi (Y)\\[2mm] 
+ \, \phi(X) \,  \phi_-(Y') \, \phi(Y'') + \phi_- (X') \, \phi (X'')
\, \phi_- (Y') \,  \phi(Y'')) 
\\[2mm]
-T (\phi_-(X) (\phi(Y) + \phi_- (Y') \, \phi(Y'')) - T ((\phi(X) +
\phi_- (X') \, \phi (X'')) \, \phi_-(Y)) 
\, .\end{array}
\end{equation}
We now compare \eqref{Birk3} with \eqref{Birk5}. Both of them contain
8 terms of the form $-T(a)$
and one checks that they correspond pairwise. This yields the
multiplicativity
of $\phi_-$ and hence the validity of \eqref{Hbirkhoff1}.

\smallskip

We then define $\phi_+$ by \eqref{Hbirkhoff2}. Since $\phi_-$ is
multiplicative, so is $\phi_+$. It remains to check that $\phi_-$ is
an element in $G(\Qc)$, while $\phi_+$ is in $G(\O)$. This is clear
for $\phi_-$ by construction, since it is a pure polar part. 
In the case of $\phi_+$ the result follows, since we have
\begin{equation}\label{phi+part}
\phi_+(X)=  \phi(X)+\sum\phi_-(X^\prime)
 \phi(X^{\prime\prime})
 -T\left(\phi(X)+\sum\phi_-(X^\prime)
 \phi(X^{\prime\prime}) \right).
\end{equation}
$\Box$

\medskip

  Then the key observation in the CK theory (\cite{3CK}) is
 that the formulae \eqref{Hbirkhoff1} \eqref{Hbirkhoff2} are in fact
 identical to the formulae \eqref{b1}, \eqref{b2}, \eqref{b3}
 that govern the combinatorics of renormalization, for $G={\rm Difg}$,
upon setting $\phi=U$, $\phi_-=C$, and $\phi_+=R$.

 \medskip

 Thus, given a renormalisable theory $\Tc$ in $D$ dimensions, the
 unrenormalised theory gives (using DimReg) a loop
 $\g(z)$ of elements of the group ${\rm Difg}(\Tc)$, associated
 to the theory (see also Section \ref{unitofmass}
 for more details).

 \smallskip

  The parameter $z$ of the loop $\g \, (z)$ is a complex variable and
 $\g \, (z)$ is meromorphic for $d=D-z$ in a neighborhood of $D$ (\ie
defines a corresponding homomorphism from $\Hc$ to germs of
meromorphic functions at $D$).

 The main result of \cite{3CK} is that the renormalised theory is given
 by the evaluation at  $d=D$ (\ie $z=0$) of the non-singular part
$\g_+$ of the Birkhoff decomposition of $\g$,
 $$
 \g \, (z) = \g_- \, (z)^{-1} \ \g_+ \, (z) .
 $$
The precise form of the loop $\gamma$ (depending on a mass parameter
$\mu$) will be discussed below in Section \ref{unitofmass}.

\smallskip

We then have the following statement (\cite{3CK}):

 \begin{thm}\label{RenBirk2}
 The following properties hold:
 \begin{enumerate}

 \item  There exists a unique meromorphic map $\g (z)
 \in {\rm Difg}(\Tc)$, for $z \in \Cb$ with  $D-z \ne D$, whose
 $\G$-coordinates are given by $U \, (\G)_{d=D-z}$.

 \item  The renormalized value of a physical observable
 $O$ is obtained by replacing $\g \, (0)$ in the
 perturbative expansion of $O$ by $\g_+ \, (0)$,
 where
 $$ \g \, (z) = \g_- \,
 (z)^{-1} \ \g_+ \, (z) $$
 is the Birkhoff decomposition of the loop
 $\g \, (z)$ around an infinitesimal circle centered at
 $d=D$ (\ie $z=0$).

 \end{enumerate}
 \end{thm}

 \medskip

  In other words, the renormalized theory is just the evaluation at
 the integer dimension $d = D$ of space-time of the holomorphic part
 $\g_+$ of the Birkhoff decomposition of $\g$. This shows that
 renormalization is a special case of the general recipe of {\em
 multiplicative} extraction of finite value given by the Birkhoff
 decomposition.

 \smallskip

  Another remarkable fact in this result is that the same infinite series
 yields simultaneously the unrenormalized effective action, the
 counterterms, and the renormalized effective action, corresponding
 to $\gamma$, $\gamma_-$, and $\gamma_+$, respectively.

 \bigskip
 \section{Unit of Mass}\label{unitofmass}

 In order to perform the extraction of
 pole part $T$ it is necessary to be a bit more
 careful than we were so far in our description of
 dimensional regularization.
 In fact, when integrating in dimension $d=D-z$, and comparing the
 values obtained for different values of  $z$, it is necessary to
 respect physical dimensions (dimensionality).
 The general principle is to only apply the operator
 $T$ of extraction of the pole part to expressions of a fixed
 dimensionality, which is independent of $z$.
 
\smallskip

 This requires the introduction of an arbitrary unit of mass (or
 momentum) $\mu$,
 to be able to  replace in the integration $d^{D-z} k$ by
 $\mu^z \,d^{D-z} k $
 which is now of a fixed dimensionality (\ie mass$^D$).

\smallskip

  Thus, the loop $\g \, (z)$ depends on the arbitrary choice of
 $\mu$. We shall now describe in more details
 the Feynman rules in $d=(D-z)$-dimensions
for $\vp^3_6$ (so that $D=6$) and exhibit this
 $\mu$-dependence.
 By definition $\g_\mu \, (z)$ is obtained by applying dimensional
 regularization (Dim-Reg) in the evaluation of the bare values of Feynman
 graphs $\G$, and  the Feynman rules
 associate an integral
 \begin{equation}\label{diag}
 U_{\G} \, (p_1 , \ldots , p_N) = \int d^{D-z} \, k_1 \ldots d^{D-z}
\, k_L \ I_{\G} 
 \, (p_1 , \ldots , p_N , k_1 , \ldots , k_L)
 \end{equation}
 to every graph $\G$, with $L$ the loop number \eqref{g5}.
 We shall formulate them in Euclidean
 space-time to eliminate irrelevant singularities on the mass shell and
 powers of $i = \sqrt{-1}$. In order to write these rules directly in
$d=D-z$ space-time dimensions, one uses the unit of mass
 $\mu$ and  replaces the coupling constant $g$ which appears in the
 Lagrangian as the coefficient of $\vp^3 / 3!$ by $\mu^{3-d/2} \, g$. The
 effect then is that $g$ is dimensionless for any value of $d$ since the
 dimension of the field $\vp$ is $\frac{d}{2} - 1$ in a $d$-dimensional
 space-time.

 \smallskip

 \noindent The integrand $I_{\G} \, (p_1 , \ldots , p_N , k_1 ,
 \ldots , k_L)$ contains $L$ internal momenta $k_j$, where $L$ is
 the loop number of the graph $\G$, and is obtained from the following
 rules,

 \begin{itemize}

 \item Assign a factor $\frac{1}{k^2+m^2}$ to each
 internal line.

 \item Assign a momentum
 conservation rule to each vertex.

 \item Assign a
 factor $\mu^{3-d/2} \, g$ to each 3-point vertex.

 \item Assign a factor $m^2$ to each 2-point vertex$_{(0)}$.

 \item  Assign a factor $p^2$ to each 2-point
 vertex$_{(1)}$.
 \end{itemize}

  The 2-point vertex$_{(0)}$
 does not appear in the case of a massless theory, and in that case
 one can in fact ignore all two point vertices.

 \noindent There is, moreover, an overall normalization factor 
  $(2\pi)^{-dL}$ where
 $L$ is the loop number of the graph, \ie the number of internal
 momenta.

\smallskip

For instance, for the one-loop graph of \eqref{Fint}, \eqref{dimrg}, the
unrenormalized value is, up to a multiplicative constant, 
$$ U_\Gamma(p)=(4\pi \mu^2)^{3-d/2} \, {g^2\, \Gamma(2-d/2)}
\, \int_0^1 (p^2 (x-x^2)+m^2)^{d/2-2} \, dx. $$

\medskip

 Let us now define precisely the character $\g_\mu(z)$ of $\Hc$
 given by the unrenormalized value of the graphs
 in Dim-Reg in dimension $d=D-z$.

 Since $\g_\mu(z)$ is a character, it is entirely specified
 by its value on 1PI graphs.
  If we let $\s$ be the external  structure of the graph
 $\G$ we would like to define
 $\g_\mu(z) \, (\G_\s)$ simply by evaluating
 $\s$ on the test function
 $U_{\G} \, (p_1 , \ldots , p_N)$, but we need to fulfill two
 requirements. First we want this evaluation
 $ \lgl \s , U_{\G} \rgl $ to be a pure number, \ie to be a
 dimensionless quantity. To achieve this we simply multiply $\lgl \s
 , U_{\G} \rgl $
  by the appropriate power of $\mu$ to make it
 dimensionless.

 \smallskip

 \noindent The second requirement is to ensure that $\g_\mu(z) \, (\G_\s)$
 is a monomial of the
 correct power  of the dimensionless coupling constant  $g$, corresponding to
the {\em order} of the graph. This is defined as $ V_3 -
 (N-2) $, where $V_3$ is the number of 3-point vertices. The order
defines a grading of $\Hc$.
To the purpose of fulfilling this requirement, for a graph with $N$
external legs, it suffices to divide by $g^{N-2}$, where $g$ is the coupling
constant. 

 \smallskip

 \noindent Thus, we let
 \begin{equation}\label{defdiag}
 \g_\mu(z) \, (\G_\s) = g^{(2-N)} \, \mu^{-B} \, \lgl \s , U_{\G} \rgl
 \end{equation}
 where $B = B \, (d)$ is the dimension of $ \lgl \s , U_{\G} \rgl $.

 \smallskip

 \noindent Using the Feynman rules this dimension
 is easy to compute and one gets \cite{3CK}
 \begin{equation}\label{dimB}
 B = \left( 1 - \frac{N}{2} \right) \, d + N +{\rm dim}\,\s \, .
 \end{equation}

 \smallskip
  Let $\g_\mu(z)$ be the character of $\Hc'$
 obtained by (\ref{defdiag}). We first need to see the exact $\mu$
 dependence of this
 loop. We consider the grading of $\Hc'$ and $G'$ given by the
 loop number of a graph,
 \begin{equation} L (\G) = I - V + 1= \hbox{loop number of} \ \G ,
 \label{Ren5} \end{equation}  where
 $I$ is the number of internal lines and $V$ the number of
 vertices and let \begin{equation} \t_t \in {\rm Aut} \, G' \ , \quad
 t \in \Rb \,
  , \label{Ren4} \end{equation} be the corresponding one parameter group
 of automorphisms.  \smallskip

  \begin{prop}\label{loopnumber}  The  loop $\g_{\mu} (z)$ fulfills
 \begin{equation} \g_{e^t \mu} (z) = \t_{t z} (\g_{\mu} (z))
 \qquad \fl \, t \in \Rb \, , \ z = D-d \label{Ren6a} \end{equation}
 \end{prop}

  The simple idea is that each of the $L$
 internal integration variables
 $d^{D-z} k$ is responsible for a factor of
 $\mu^z$ by the alteration $$d^{D-z} k \mapsto \mu^z\,\,d^{D-z}k. $$

 Let us check that this fits with the above conventions.
 Since we are on $\Hc'$ we only deal with
 1PI graphs with two or three external legs and fixed external
 structure. For $N=2$ external legs the
 dimension $B$ of $\lgl \s , U_{\G} \rgl$ is equal to 0 since the dimension
 of the external structures $\s_j$ of
 \eqref{3Gamma1} is $-2$.
  Thus, by the Feynman rules, at $D=6$, with $d = 6 - z$,
  the $\mu$ dependence is given  by $$
 \mu^{\frac{z}{2} V_3}  $$ where $V_3$ is the number of
 3-point vertices of $\G$. One checks that for such graphs
 $\frac{1}{2} \, V_3 = L$ is the loop number
 as required. Similarly if $N = 3$ the dimension $B$ of $\lgl \s ,
 U_{\G} \rgl$ is equal to $\left( 1 - \frac{3}{2} \right) \, d +
 3$, $d = 6 - z$ so that the $\mu$-dependence
 is, $$ \mu^{\frac{z}{2} V_3} \, \mu^{-z / 2} \, .
 $$ But for such graphs $V_3 = 2L+1$ and we get $ \mu^{z L}
  $ as required.

 We now reformulate a well known result, the fact that
 counterterms, once appropriately normalized, are independent of
 $m^2$ and $\mu^2$,

  \smallskip      We have (\cite{3cknew}):
 \begin{prop}\label{negnomu}
 The negative part $\g_{\mu^-}$ in the Birkhoff decomposition
 \begin{equation}
 \g_{\mu} (z) = \g_{\mu^-} (z)^{-1} \, \g_{\mu^+} (z) \label{Ren2}
 \end{equation} satisfies
 \begin{equation}
 \frac{\partial}{\partial \mu} \, \g_{\mu^-} (z) = 0 \, . \label{Ren3a}
  \end{equation} \end{prop}  \smallskip

 \underline{\em Proof.} By Theorem \ref{RenBirk} and the identification
 $\g=U$, $\g_-=C$, $\g_+=R$, this amounts to the
 fact that the counterterms do not depend on the choice of $\mu$
 (\cf \cite{Collins} 7.1.4 p. 170). Indeed
 the dependence in $m^2$ has in the minimal subtraction scheme the
 same origin as the dependence in $p^2$ and we have chosen the
 external structure of graphs  so that no $m^2$
 dependence is left. But then, since the parameter $\mu^2$ has nontrivial
 dimensionality (mass$^2$), it cannot be involved any longer. $\Box$

 \bigskip
 \section{Expansional}\label{expansional}
 \bigskip

Let $\Hc$ be a Hopf algebra  over $\C$ and
  $G=\,{\rm Spec}\,\Hc$ the corresponding
affine group scheme.

\smallskip
Given a differential field $K\supset \C$ with differentiation
$f\mapsto f'=\delta(f)$, let us describe at the Hopf algebra level
the logarithmic derivative
$$
D(g)=\,g^{-1}\,g'\in \fg(K)\qq g\in G(K)\,.
$$
Given $g\in G(K)$ one lets $g'=\delta(g)$ be the linear map from
$\Hc$ to $K$ defined by
$$
g'(X)=\, \delta(g(X)) \qq X\in \Hc\,.
$$
One then defines $D(g)$ as the linear map from $\Hc$ to $K$
\begin{equation}  \label{logder}
D(g)=\,g^{-1}\star\, g'\,.
 \end{equation}
One checks that
$$
\langle D(g),X\,Y\rangle=\,\langle D(g),X\rangle\,\ve(Y)
+\,\ve(X)\,\langle D(g),Y\rangle \qq X,Y \in \Hc\,,
$$
  so that $D(g) \in \fg(K)$.

\smallskip

In order to write down explicit solutions of $G$-valued
differential equations we shall use the ``expansional",
which is the mathematical formulation of the ``time
ordered exponential" of physicists. In the mathematical setting, 
the time ordered exponential can be formulated in terms of
the formalism of Chen's iterated integrals (\cf \cite{Chen1}
\cite{Chen2}). A mathematical formulation of the time ordered exponential
as expansional in the operator algebra setting was given by Araki in
\cite{Araki}. 

\smallskip

  Given a $\fg(\C)$-valued smooth function $\a(t)$
 where $t\in[a,b]\subset \R$ is a real parameter,
 one defines
 the time ordered exponential or expansional by the equality (\cf \cite{Araki}) 
\begin{equation}\label{expansional0}
 {\bf {\rm T}e^{\int_a^b\,\a(t)\,dt}}=\,1+\, \sum_1^\infty \int_{a\leq
s_1\leq \cdots\leq 
s_n\leq b} \,\a(s_1)\cdots\,\a(s_n) \prod ds_j \,,
 \end{equation}
where the  product is the product in $\Hc^*$ and $1\in \Hc^*$ is
the unit given by the augmentation $\ve$. One has the following
result, which in particular shows
how the expansional only depends on the one form $\a(t) dt$.

\begin{prop} \label{expprop0} The expansional satisfies the following
properties: 
\begin{enumerate}
\item When paired with any $X\in \Hc$ the sum
\eqref{expansional0} is finite and the obtained linear form
 defines an element of $G(\C)$.
\item The expansional \eqref{expansional0} is the value $g(b)$ at $b$ of
the unique solution $g(t)\in G(\C)$ which takes the value $g(a)=1$ at
$x=a$ for the differential equation
 \begin{equation}\label{diffexp0}
 dg(t)=\,g(t)\,\a(t)\,dt\,.
\end{equation}
\end{enumerate}
 \end{prop}

\underline{Proof.} The  elements $\a(t)\in\fg$ viewed as linear
forms on $\Hc$ vanish on any element of degree $0$. Thus for
$X\in \Hc$ of degree $n$, one has
$$
\langle \a(s_1)\cdots\,\a(s_m),\,X \rangle\,=\,0\qq m>n\,,
$$
 so that the sum $g(b)$ given by
\eqref{expansional0} is finite.

\smallskip
Let us show that it fulfills \eqref{diffexp0} \ie that with $X$ as
above, one has
$$
\partial_b\,\langle g(b),X\rangle=\,\langle g(b)\,\a(b),X\rangle\,.
$$
Indeed, differentiating in $b$ amounts to fix the last variable
$s_n$ to $s_n=b$.

\smallskip
One can then show that $g(b)\in G(\C)$, \ie that
$$
\langle g(b),X\,Y\rangle=\,\langle g(b),X\rangle\,\langle
g(b),Y\rangle \qq X,Y\in \Hc\,,
$$
for homogeneous elements, by induction on the sum of their
degrees. Indeed, one has, with the notation
$$
\Delta(X)= X_{(1)}\otimes X_{(2)}=X\otimes 1 + 1 \otimes X + \sum
X'\otimes X''
$$
where only terms of lower degree appear in the last sum,
$$
\partial_b\,\langle g(b),X\,Y\rangle=
\,\langle g(b)\,\a(b),X\,Y\rangle\,= \,\langle
g(b)\otimes\a(b),\Delta X\,\Delta Y\rangle\,.
$$
Using the derivation property of $\a(b)$ one gets,
$$
\partial_b\,\langle g(b),X\,Y\rangle=
\,\langle g(b),X_{(1)}\,Y\rangle\,\,\langle \a(b),X_{(2)}\rangle\,
+ \,\langle g(b),X\,Y_{(1)}\rangle\,\,\langle
\a(b),Y_{(2)}\rangle\,
$$
and the induction hypothesis applies to get
$$
\partial_b\,(\langle g(b),X\,Y\rangle-\,\langle g(b),X\rangle\,
\langle g(b),Y\rangle)=\,0\,.
$$
Since $g(a)=1$ is a character one thus gets $g(b)\in G(\C)$.

We already proved 2) so that the proof is complete. $\Box$

\smallskip
The main properties of the expansional in our context are
summarized in the following result. 

\begin{prop} \label{expprop}
1) One has
 \begin{equation}\label{expan1}
 {\bf {\rm T}e^{\int_a^c\,\a(t)\,dt}}=\,{\bf {\rm T}e^{\int_a^b\,\a(t)\,dt}}\,
 {\bf {\rm T}e^{\int_b^c\,\a(t)\,dt}}
 \end{equation}

2) Let $\Omega \subset \R^2$ be an open set and $\omega= \a(s,t)
ds + \b(s,t) dt$, $(s,t)\in \Omega$ be a {\em flat}
$\fg(\C)$-valued connection \ie such that
$$
\partial_s\,\b-\,\partial_t\,\a+\,[\a,\,\b]=\,0
$$
then $\displaystyle {\bf {\rm T}e^{\int_0^1\,\g^*\omega}}\, $ only
depends on the homotopy class of the path $\g$, $\g(0)=a$,
$\g(1)=b$.
 \end{prop}

\underline{Proof.} 1) Consider both sides as $G(\C)$-valued
functions of $c$. They both fulfill equation \eqref{diffexp0} and
agree for $c=b$ and are therefore equal.

2) One can for instance use the existence of enough finite
dimensional representations of $G$ to separate the elements of
$G(\C)$, but it is also an exercise to give a direct argument.
$\Box$

\smallskip
Let $ K$ be the field $ \C(\{z\})$ of convergent Laurent series in
$z$. Let us define the monodromy of an element $\omega\in \fg(K)$.
As explained above we can write $G$ as the projective limit of
linear algebraic groups $G_i$ with finitely generated Hopf
algebras  $\Hc_i \subset \Hc$ and can assume in fact that each
$\Hc_i$ is globally invariant under the grading $Y$.

Let us first work with $G_i$ \ie assume that $\Hc$ is finitely
generated. Then the element $\omega\in \fg(K)$ is specified by
finitely many elements of $K$ and thus there exists $\rho >0$ such
that all elements of $K$ which are involved  converge in the
punctured disk $\Delta^*$ with radius $\rho$. Let then $z_0\in
\Delta^*$ be a base point, and define the monodromy by
 \begin{equation}\label{mono}
M =\;{\bf {\rm T}e^{\int_0^1\,\g^*\omega}}\,,
 \end{equation}
where $\g$ is a path in the class of the generator of
$\pi_1(\Delta^*,z_0)$. By proposition \ref{expprop} and the
flatness of the connection $\omega$, viewed as a connection in two
real variables, it only depends on the homotopy class of $\g$.

By construction the conjugacy class of $M$ does not depend on the
choice of the base point. When passing to the projective limit one
has to take care of the change of base point, but the condition of
{\em trivial monodromy},
$$
M=1\,,
$$
is  well defined at the level of the projective limit $G$ of the
groups $G_i$.

One then has,

\begin{prop} \label{trimono} Let $\omega\in \fg(K)$ have trivial
monodromy. Then there exists a solution $g\in G(K)$ of the
equation
\begin{equation}\label{diffsys}
 D(g)=\, \omega\,.
\end{equation}
 \end{prop}

\underline{Proof.}  We view as above $G$ as the projective limit
of the $G_i$ and treat the case of $G_i$ first. With the above
notations we let
\begin{equation}\label{sol00}
g(z) =\;{\bf {\rm T}e^{\int_{z_0}^z\,\omega}}\,,
 \end{equation}
independently of the path in $\Delta^*$ from $z_0$ to $z$. One
needs to show that for any $X\in \Hc$ the evaluation
$$
h(z)=\,\langle g(z),\,X\rangle
$$

is a  convergent Laurent series in $\Delta^*$, \ie that $h\in K$.
It follows, from the same property for $\omega(z)$ and the
finiteness (proposition \ref{expprop0}) of the number of non-zero
terms in the pairing with $X$ of the infinite sum
\eqref{expansional0} defining $g(z)$, that $z^N\,h(z)$ is bounded
for $N$ large enough. Moreover, by proposition \ref{expprop0}, one
has $\overline{\partial} h=\,0$, which gives $h\in K$.

Finally, the second part of Proposition \ref{expprop0} shows that
one gets a solution of \eqref{diffsys}. To pass to the projective
limit one constructs by induction a projective system of solutions
$g_i \in G_i(K)$ modifying the solution in $G_{i+1}(K)$ by left
multiplication by an element of $G_{i+1}(\C)$ so that it projects
on $g_i$. $\Box$

\medskip

The simplest example shows that the condition of triviality of the
monodromy is not superfluous. For instance, let $\bG_a$ be the additive
group, \ie the group scheme with
  Hopf
algebra the algebra $\C[X]$ of  polynomials in one variable $X$
and coproduct given by,
$$
\Delta X=\,X\otimes 1+ \,1\otimes X \,.
$$
Then, with $ K$ the field $ \C(\{z\})$ of convergent Laurent
series in $z$, one has $\bG_a(K)=K$ and the logarithmic derivative
$D$ \eqref{logder}
 is just given by $D(f)=f'$ for $f\in K$.
The residue of $\omega \in K$ is then a non-trivial obstruction to
the existence of solutions of $D(f)=\omega$.

 \bigskip
 \section{Renormalization group}\label{Rensect}

    Another result of the CK theory of renormalization in
 \cite{3cknew} shows that the renormalization group
 appears in a  conceptual manner from the geometric
  point of view described in Section \ref{SBirk}.
 It is shown in \cite{3cknew} that the mathematical
 formalism recalled here in the previous section 
 provides a way to lift the usual notions of
 $\b$-function and renormalization group from the
 space of coupling constants of the theory $\Tc$ to the group ${\rm
Difg}'(\Tc)$.

\medskip

The principle at work can be summarized as
 \begin{equation} \label{divamb}
\text{Divergence} \Longrightarrow \text{Ambiguity}.
\end{equation}
Let us explain in what sense it is the {\em divergence} of the
theory that generates the renormalization group as a group of
ambiguity. 
As we saw in the previous section, the  regularization process
requires  the introduction of an arbitrary unit of mass $\mu$. The
way the theory (when viewed as an element of the group ${\rm
Difg}'(\Tc)$ by
evaluation of the positive part of the Birkhoff decomposition at
$z =0$) depends on the choice of $\mu$ is through the grading rescaled
by $z= D-d$ (\cf Proposition \ref{loopnumber}). If the
resulting expressions in $z$ were regular at $z=0$, this dependence
would disappear at $z=0$. As we shall see below, this dependence
will in fact still be present
and generate a one parameter subgroup $F_t=e^{t\beta}$ of ${\rm
Difg}'(\Tc)$ as a group of ambiguity of the physical theory.

\smallskip

After recalling the results of \cite{3cknew}
we shall go further and improve on the 
scattering formula (Theorem \ref{scatteringthm})
and give an explicit formula (Theorem \ref{genmu}) for the
 families
 $\g_\mu(z)$ of ${\rm
Difg}'(\Tc)$-valued loops which fulfill the
 properties proved in Propositions \ref{loopnumber}
 and \ref{negnomu}, in the context of
 quantum field theory, namely
 \begin{equation}
  \g_{e^t \mu} (z) = \t_{t z} (\g_{\mu} (z)) \qquad \fl \, t \in
 \Rb \, ,  \label{Ren6}
 \end{equation}
and
 \begin{equation}
 \frac{\partial}{\partial \mu} \, \g_{\mu^-} (z) = 0 \,
 . \label{Ren3} \end{equation}
  where $\g_{\mu^-}$ is the negative piece of the Birkhoff
 decomposition of $\g_{\mu}$.  

\medskip

 The discussion which follows will be quite general, the framework
 is given by a complex graded pro-unipotent Lie group $G(\C)$, which
we can think of as the complex points of an affine group scheme $G$
and is identified with ${\rm Difg}'(\Tc)$ in the context above.  We let
$\Lie\, G(\C)$ be its Lie algebra and we let $\t_t$ be
the one parameter group of automorphisms implementing the grading $Y$.

\smallskip

We then consider the Lie group given by the semidirect product
 \begin{equation}
 G(\C) \, \rtimes_{\t} \, \R  \label{Ren15}
 \end{equation}
 of $G(\C)$ by the action of the grading $\t_t$.
 The Lie algebra of \eqref{Ren15} has an  additional generator satisfying
 \begin{equation}
 [Z_0 , X] = Y(X) \qquad \forall \, X \in \hbox{Lie} \, G(\C) \,. \label{Ren13}
 \end{equation}

 \smallskip

  Let then  $\g_\mu(z)$ be a family of $G(\C)$-valued 
loops which fulfill \eqref{Ren6}
and \eqref{Ren3}. Since $\g_{\mu^-}$ is independent of $\mu$ we denote it
 simply by $\g_-$. One has the following which we recall from \cite{3cknew}:

 \begin{lem}\label{negpart}
 \begin{equation} \label{limren}
 \g_-
 (z) \, \t_{t z} (\g_- (z)^{-1}) \ \hbox{is regular at} \
 z = 0 \, .
 \end{equation}
Moreover, the limit
\begin{equation}
 F_t= {\rm lim}_{z \ra 0} \:\g_- (z) \, \t_{t z} (\g_-
 (z)^{-1})  \  \,  \label{Ren9}
 \end{equation}
defines a 1-parameter group, which depends polynomially on $t$ when
evaluated on an element $x\in \Hc$.
  \end{lem}

 \underline{\em Proof.}  
Notice first that
 both $\g_{-} (z)\,\g_\mu (z)$ and $y(z)=\g_{-} (z)\,\t_{-t z}
 (\g_{\mu} (z))$
 are regular at $z=0$, as well as
 $\t_{tz}(y(z))=\,\t_{tz}(\g_{-} (z))\,\g_\mu (z)$, so that the ratio
 $\g_-
 (z) \, \t_{t z} (\g_- (z)^{-1}) $ is regular at $z = 0$. 

\smallskip

We know thus that, for any $t \in \Rb$, the
limit 
\begin{equation}\label{limitx}
 \lim_{z \ra 0} \ \lgl
\g_-(z) \, \t_{tz} (\g_-(z)^{-1}) , x \rgl 
\end{equation} 
exists, for any $x \in \Hc$. 
We let the grading $\t_t$ act by automorphisms of both $\Hc$ and
the dual algebra $\Hc^*$ so that $$\langle \theta_t(u),x\rangle=
\langle u,\theta_t(x)\rangle \,,\quad
 \forall x\in \Hc\,, u\in \Hc^*\,.$$
 We then have
\begin{equation}\label{limitform}
\lgl \g_-(z) \, \t_{tz} (\g_-(z)^{-1}) , x \rgl = 
\lgl \g_-(z)^{-1} \ot \g_-(z)^{-1} , (S
\ot \t_{tz}) \, \D x \rgl \, ,  
\end{equation} 
 
so that, writing the coproduct $\D x
= \sum x_{(1)} \, \ot \, x_{(2)}$ as a sum of
homogeneous elements, we express \eqref{limitform}
as a sum of terms 
\begin{equation}\label{limitform1}
\lgl
\g_-(z)^{-1} , S \, x_{(1)} \rgl \, \lgl \g_-(z)^{-1} , \t_{t z} \,
x_{(2)} \rgl = P_1 \left( \frac{1}{z} \right) \, e^{ktz} \,
P_2 \left( \frac{1}{z} \right),
\end{equation} 
for polynomials $P_1$, $P_2$.

\smallskip

The existence of the limit \eqref{limitx} means that the
sum \eqref{limitform} 
of these terms is holomorphic at $z = 0$. 
Replacing the exponentials $ e^{ktz}$ by their Taylor expansion 
at $z=0$ shows that
the value of \eqref{limitform} at $z = 0$, 
 $$ \lgl F_t ,
x \rgl = \lim_{z \ra 0} \ \lgl \g_-(z) \, \t_{tz}
(\g_-(z)^{-1}) , x \rgl \, , $$ 
is a polynomial in $t$.

\smallskip
 
 Let us check that $F_t$ is a one parameter subgroup
 \begin{equation}
  F_t \in G(\C) \,  \ \ \ \forall  t \in \Rb ,
  \ \ \ \text{ with } \ \ \  F_{s+t} =F_s \,
  F_t \ \ \ \forall  s,t \in \Rb. \label{Ren10}
 \end{equation}

In fact, first notice that 
the group
$G(\C)$ is a topological group for the topology of simple convergence,
\ie that
\begin{equation}\label{convergG}
 \g_n \ra \g \quad \hbox{iff} \quad \lgl \g_n , x \rgl
\ra \lgl \g , x \rgl \qquad \fl \, x \in \Hc \, . 
\end{equation}
Moreover, using the first part of Lemma \ref{negpart}, one gets 
\begin{equation}\label{thetaeps}
\t_{t_1 z}
(\g_-(z) \, \t_{t_2 z} (\g_-(z)^{-1})) \ra F_{t_2} \qquad
\hbox{when} \ z \ra 0 \, . 
\end{equation} 
We then have 
$$ F_{t_1 +t_2} = \lim_{z \to 0} \ \g_-(z) \,
\t_{(t_1 + t_2) z} \, (\g_-(z)^{-1}) $$
$$ = \lim_{z
\ra 0} \   \g_-(z)  \, \t_{t_1 z} ( \g_-(z)^{-1} ) \, \t_{t_1
z} (\g_-(z) \, \t_{t_2 z} (\g_-(z)^{-1})) = F_{t_1} \,
F_{t_2}. $$
$\Box$

\smallskip

 As shown in \cite{3cknew}
 and recalled below (\cf Lemma \ref{lemYd})
the generator $\b = \left( \frac{d}{d t} \, F_t \right)_{t=0}$
  of this one parameter group is related to
 the {\it residue} of $\g$,
 \begin{equation}
 {\rm Res}_{z = 0}^{} \g = - \left( \frac{\partial}{\partial u} \,
  \g_- \left( \frac{1}{u} \right) \right)_{u=0}\, , \label{Ren11}
  \end{equation}
 by the simple equation
 \begin{equation} \b = Y \, {\rm Res} \, \g \, , \label{Ren12}
  \end{equation}
 where $Y = \left( \frac{d}{d t} \, \t_t \right)_{t=0}$ is the
grading.

\smallskip

    When applied to the finite renormalized theory,
  the one parameter group \eqref{Ren10}
  acts as the {\em renormalization group},
 rescaling the unit of mass $\mu$.
 One has (see \cite{3cknew}):

 \begin{prop}\label{scale}
  The finite value $\g_{\mu}^+(0)$ of the Birkhoff
 decomposition satisfies  \begin{equation}\label{scalevalue}
  \g^+_{e^t\,\mu}(0)= F_t\,\g^+_{\mu}(0)\,
 , \ \ \ \forall t\in \Rb.  \end{equation}
 \end{prop}

 Indeed $\g^+_{\mu}(0)$ is the regular value of
  $\g_{-} (z)\,\g_\mu (z)$ at $z =0$ and
 $\g^+_{e^t\,\mu}(0)$ that of  $\g_{-} (z)\,\t_{tz}(\g_\mu (z))$
 or equivalently of
 $\t_{-tz}(\g_{-} (z))\,\g_\mu (z)$ at $z =0$.
 But the ratio $$\t_{-tz}(\g_{-} (z))\,\g_{-} (z)^{-1} \rightarrow F_t$$
 when $z \rightarrow 0$, whence the result. $\Box$

 \smallskip
  In terms of the infinitesimal generator
 $\beta$, equation \eqref{scalevalue}
 can be rephrased as the equation
 \begin{equation}\label{renaction}
  \mu \frac{\partial}{\partial \mu}
 \,\g^+_{\mu}(0)=\, \b\,\g^+_{\mu}(0)\,.
  \end{equation}

 \smallskip  Notice that, for a loop $\g_\mu(z)$ regular at $z=0$
 and fulfilling \eqref{Ren6}, the value $\g_\mu(0)$
 is independent of $\mu$, hence the presence of
 the divergence is the real source of the
 ambiguity manifest in the renormalization group equation
 \eqref{renaction}, as claimed in \eqref{divamb}.

\medskip

We now take the key step
in the characterization of loops fulfilling 
\eqref{Ren6}
and \eqref{Ren3} and
reproduce in full the following argument from \cite{3cknew}. Let
$\Hc^*$ denote the linear dual of $\Hc$.

\begin{lem}\label{lemYd} Let $z \ra \gamma_-(z) \in G(\C)$
satisfy \eqref{limren}
 with 
\begin{equation}\label{gammaminussum}
\gamma_-(z)^{-1} = 1 + {\displaystyle
\sum_{n=1}^{\ify}} \ \frac{d_n}{z^n},
\end{equation} 
where we have $d_n \in \Hc^*$. One then
has $$ Y \, d_{n+1} = d_n \, \b \qquad \fl \, n \geq 1 \, , \ Y
(d_1) = \b \, . $$ 
\end{lem}

\medskip

\underline{\it Proof.} Let $x \in \Hc$ and let us show that $$
\lgl \b , x \rgl = \lim_{z \ra 0} \, z \lgl \gamma_-(z)^{-1} \ot
\gamma_-(z)^{-1} , (S \ot Y) \, \D (x) \rgl \, .  $$

We know by \eqref{Ren9} and \eqref{limitform} 
that when $z \ra 0$, 
\begin{equation}\label{conver} \lgl \gamma_-(z)^{-1} \ot
\gamma_-(z)^{-1} , (S \ot \t_{tz}) \, \D (x) \rgl \ra \lgl F_t , x \rgl ,
\end{equation} 
 where the left hand side is, by \eqref{limitform1}, 
a finite sum 
$S=\sum P_k(z^{-1})\,e^{ktz}$
for polynomials $P_k$. Let $N$ be the maximal degree 
of the $P_k$, the regularity of  $S$ at $z=0$
is unaltered if one replaces the $e^{ktz}$ by their
Taylor expansion to order $N$ in $z$. The obtained 
expression is a polynomial in $t$ with coefficients
which are Laurent polynomials in $z$. Since 
the regularity at $z=0$ 
holds for all values of $t$ these
coefficients are all regular at $z=0$ \ie they 
are polynomials in $z$. Thus the left hand side of \eqref{conver}
is a uniform family of
holomorphic functions of $t$ in, say, $\vert t \vert \leq 1$, and
its derivative $\partial_t S$ at $t = 0$ 
converges to $\partial_t \lgl F_t , x \rgl$ when $z\ra 0$,
$$
z \,\lgl \gamma_-(z)^{-1} \ot
\gamma_-(z)^{-1} , (S \ot Y) \, \D (x) \rgl \ra \lgl \b , x \rgl .
$$

Now the function $z \ra z \, \lgl \gamma_-(z)^{-1} \ot \gamma_-(z)^{-1} , (S
\ot Y) \, \D x \rgl$ is holomorphic for $z \in \Cb \backslash \{
0 \}$ and also at $z = \ify \in P_1 (\Cb)$, since $\gamma_-(\ify) =
1$ so that $Y(\gamma_-(\ify)) =
0$. Moreover, by the above it is also holomorphic at $z = 0$ and is
therefore a constant, which gives $$ \lgl \gamma_-(z)^{-1} \ot
\gamma_-(z)^{-1} , (S 
\ot Y) \, \D (x) \rgl = \frac{1}{z} \ \lgl \b , x \rgl \, .
 $$ Using the product in $\Hc^*$, this means that $$
\gamma_-(z) \, Y (\gamma_-(z)^{-1}) = \frac{1}{z} \, \b \, . $$ 
Multiplying by $\gamma_-(z)^{-1}$ on the left, we get $$ Y
(\gamma_-(z)^{-1}) = \frac{1}{z} \ \gamma_-(z)^{-1} \, \b \, .  $$
One has $Y (\gamma_-(z)^{-1}) = {\displaystyle \sum_{n=1}^{\ify}} \
\frac{Y (d_n)}{z^n}$ and $\frac{1}{z} \, \gamma_-(z)^{-1} \, \b =
\frac{1}{z} \, \b + {\displaystyle \sum_{n=1}^{\ify}} \
\frac{1}{z^{n+1}} \, d_n \, \b$ which gives the desired equality.
$\Box$

\medskip

In particular we get $Y (d_1) = \b$ and, since $d_1$ is the
residue ${\rm Res} \, \vp$, this shows that $\b$ is uniquely
determined by the residue of $\gamma_-(z)^{-1}$.

\medskip

The following result (\cf \cite{3cknew}) shows
  that the higher pole structure of the divergences is
 uniquely determined by their residue and
 can be seen as a strong form of the t'Hooft relations \cite{tf}.

 \smallskip
 \begin{thm}\label{scatteringthm}
 The negative part $\g_- (z)$ of the Birkhoff
 decomposition is completely determined by the residue,
 through the scattering formula
  \begin{equation} \g_- (z) = \lim_{t \ra \ify}
 e^{-t \left( \frac{\b}{z} + Z_0 \right)} \, e^{t Z_0} \, . \label{Ren14}
  \end{equation}
 \end{thm}

Both factors in the right hand side belong to
the semi-direct product \eqref{Ren15}, while the ratio (\ref{Ren14}) belongs
to $G(\C)$. 

\smallskip

We reproduce here the proof of Theorem \ref{scatteringthm} 
given in \cite{3cknew}.

\smallskip

\underline{\it Proof.} We endow  $\Hc^*$ with the topology of
simple convergence on $\Hc$.  Let us first show, using Lemma
\ref{lemYd},
 that the coefficients $d_n$ of \eqref{gammaminussum} 
are given by
iterated integrals of the form
\begin{equation}\label{dncond}
 d_n = \int_{s_1 \geq s_2 \geq \cdots \geq s_n \geq 0} \t_{-s_1}
(\b) \, \t_{-s_2} (\b) \ldots \t_{-s_n} (\b) \, \Pi  \, ds_i \, .
\end{equation} 
For $n = 1$, this just means that $$ d_1 =
\int_0^{\ify} \t_{-s} (\b) \, ds \, , $$ which follows
from $\beta=Y (d_1)$ and the equality 
\begin{equation}\label{Yeq}
Y^{-1} (x) = \int_0^{\ify} \t_{-s} 
(x) \, ds \qquad \fl \, x \in \Hc \, , \ \epsilon (x) = 0 \, . 
\end{equation} 
We see from \eqref{Yeq} that, for $\a , \a' \in \Hc^*$ such that
$$ 
Y (\a) = \a' \, , \ \lgl \a , 1 \rgl = \lgl \a' , 1 \rgl = 0,
$$ 
one has 
$$ \a = \int_0^{\ify} \t_{-s} (\a') \, ds
\, . $$ 
Combining this equality with Lemma \ref{lemYd} and the
fact that $\t_s \in {\rm Aut} \, \Hc^*$ is an automorphism, gives
an inductive proof of \eqref{dncond}. The meaning of this formula should be
clear: we pair both sides with $x \in \Hc$, and let $$ \D^{(n-1)}
\, x = \sum x_{(1)} \ot x_{(2)} \ot \cdots \ot x_{(n)} \, . $$ 
Then the right hand side of \eqref{dncond} is just 
\begin{equation}\label{dnrhs} \int_{s_1
\geq \cdots \geq s_n \geq 0}  \, \lgl \b \ot \cdots \ot \b \ , \
\t_{-s_1} (x_{(1)}) \ot \t_{-s_2} (x_{(2)}) \cdots \ot \t_{-s_n}
(x_{(n)}) \rgl  \Pi  \, ds_i 
\end{equation}
and the convergence of
the multiple integral is exponential, since $$ \lgl \b , \t_{-s}
(x_{(i)}) \rgl = O \, (e^{-s}) \qquad \hbox{for} \quad s \ra +
\ify \, . $$ We see, moreover, that, if $x$ is
homogeneous of degree $\deg (x)$ and if $n
> \deg (x)$, then at least one of the $x_{(i)}$ has degree 0, so that
$\lgl \b , 
\t_{-s} (x_{(i)}) \rgl = 0$ and \eqref{dnrhs} gives 0. This shows that the
pairing of $\gamma_-(z)^{-1}$ with $x \in \Hc$ only involves finitely
many non zero terms in the formula $$ \lgl \gamma_-(z)^{-1} , x \rgl =
\ve (x) + \sum_{n=1}^{\ify} \frac{1}{z^n} \ \lgl d_n , x \rgl
\, .  $$ 
Thus to get formula \eqref{Ren14}, we dont need 
to worry about possible convergence problems
of the series in $n$.
The proof of \eqref{Ren14} 
involves the expansional formula (\cf \cite{Araki})
$$
e^{(A+B)} = \sum_{n=0}^{\ify} \, \int_{\sum u_j = 1 , \, u_j \geq
0} \, e^{u_0 A} \, B e^{u_1 A} \ldots B e^{u_n A} \, \Pi  \, du_j .
$$ 

We apply this with $A = t Z_0$, $B = t \b$, $t > 0$ and get 
$$
e^{t(\b + Z_0)} = \sum_{n=0}^{\ify} \ \int_{\sum v_j = t , \, v_j
\geq 0} \ e^{v_0 Z_0} \, \b e^{v_1 Z_0} \, \b \ldots \b e^{v_n
Z_0} \, \Pi  \, dv_j \, . 
$$ 
Thus, with $s_1 = t -
v_0$, $s_1 - s_2 = v_1 , \ldots ,s_{n-1} - s_n = v_{n-1}$, $s_n =
v_n$ and replacing $\b$ by $\frac{1}{z} \, \b$, we obtain
$$
e^{t (\b / z + Z_0)} = \sum_{n=0}^{\ify} \ \frac{1}{z^n} \
\int_{t \geq s_1 \geq s_2 \geq \cdots \geq s_n \geq 0} \, e^{t
Z_0} \, \t_{-s_1} (\b) \ldots \t_{-s_n} (\b) \,  \Pi  \,ds_i .
$$ 
Multiplying by $e^{-t Z_0}$ on the left and using
\eqref{dnrhs}, we obtain 
$$ \gamma_-(z)^{-1} = \lim_{t \ra \ify} \ e^{-t Z_0} \,
e^{t (\b / z + Z_0)} \, . $$ 
$\Box$

\bigskip

 One inconvenient of formula \eqref{Ren14} is that it hides the
 geometric reason for the convergence of the right hand side when
 $t\ra \infty$. This convergence is in fact related to the
 role of the horocycle foliation as the stable foliation
 of the geodesic flow. The simplest non-trivial case, which illustrates an
 interesting analogy between the renormalization group and the
 horocycle flow, was analyzed in \cite{mk}.

 \smallskip

 This suggests to use the formalism developed in section 
\ref{expansional} and express directly the 
negative part $\g_- (z)$ of the Birkhoff decomposition
as an expansional using \eqref{gammaminussum}
 combined with the 
iterated integral expression \eqref{dncond}.
This also amounts in fact to
analyze the convergence of
 $$
 X(t)=e^{-t \left( \frac{\b}{z} + Z_0 \right)} \, e^{t Z_0} \in
G(\C)\rtimes_\theta \R 
 $$
 in the following manner.

\smallskip

 By construction, $X(t)$ fulfills a simple differential equation as follows.

 \begin{lem} Let  $X(t)=e^{-t \left( \frac{\b}{z} + Z_0 \right)}
  \, e^{t Z_0}$. Then, for all $t$, 
 $$ X(t)^{-1}dX(t)=\,-\frac{1}{z}\t_{-t}(\beta)\,dt $$
  \label{eqdiff}
 \end{lem}

  \underline{Proof.}
 One has $X(t)=e^{tA}\,e^{tB}$ so that
 $$dX(t)=\,(e^{tA}\,A\, e^{tB}+\,e^{tA}\,B\,
  e^{tB})dt$$
 One has $A+B=-(\frac{\b}{z} + Z_0) + Z_0=\,-\frac{\b}{z}$ and  $$ e^{tA}\,
 (-\frac{\b}{z})\, e^{tB}= e^{tA}\, e^{tB}(-\frac{1}{z}\t_{-t}(\beta)) $$
 which gives the result. $\Box$

   \medskip

 With the notations of section \ref{expansional}
 we can thus rewrite Theorem \ref{scatteringthm} in the following form.

 \begin{cor}\label{prep}
 The negative part $\g_- (z)$ of the Birkhoff decomposition is given by
 \begin{equation}\label{expan}
 \g_- (z) =\,{\bf {\rm T}e^{-\frac{1}{z}\,\int_0^\infty\,\t_{-t}(\beta)\,dt}}
 \end{equation}

 \end{cor}  \smallskip

  This formulation is very suggestive of:

 \begin{itemize}

\item The convergence of the ordered product.

\item The value of the residue.

\item The special case when $\b$  is an eigenvector for
 the grading.

\item The regularity in $\frac{1}{z}$.

\end{itemize}

 \medskip
  We now show that we obtain the general solution to
 equations \eqref{Ren6} and \eqref{Ren3}.
 For any loop $\g_{\rm reg}(z)$ which is regular at $z=0$
 one obtains an easy solution by setting $\g_\mu(z)=
 \; \t_{z\log\mu}(\g_{\rm reg}(z))$. The following result
 shows that the most general solution depends in fact of
 an additional parameter $\beta$ in the Lie algebra of $G(\C)$.  \medskip

 \begin{thm}\label{genmu}  Let $\g_\mu(z)$ be a family of $G(\C)$-valued loops
 fulfilling \eqref{Ren6} and \eqref{Ren3}.
 Then there exists a unique $\beta \in {\rm Lie}\,G(\C)$
 and a loop $\g_{\rm reg}(z)$ regular  at $z=0$
 such that
\begin{equation}\label{solexp}
  \g_\mu(z) =\,{\bf {\rm T}e^{-\frac{1}{z}\,
 \int^{-z \log\mu}_\infty\,\t_{-t}(\beta)\,dt}}\;
  \t_{z\log\mu}(\g_{\rm reg}(z))\,. 
\end{equation}
Conversely, 
 for any $\beta$ and regular loop  $\g_{\rm reg}(z)$ the expression
\eqref{solexp} gives a solution to equations \eqref{Ren6} and \eqref{Ren3}.

The Birkhoff
 decomposition of the  loop $\g_\mu(z)$ is given by
\begin{equation}\label{gammaplusminus}
\begin{array}{ll} \g_\mu^+(z)= & {\bf {\rm T}e^{-\frac{1}{z}
  \,\int_0^{-z \log\mu}\,\t_{-t}(\beta)\,dt}}\;
 \t_{z\log\mu}(\g_{\rm reg}(z))\,, \\[3mm]
 \g_\mu^-(z) = & \,{\bf {\rm
 T}e^{-\frac{1}{z}\,\int_0^\infty\,\t_{-t}(\beta)\,dt}}\,.
 \end{array} \end{equation}
  \end{thm}
   \medskip

 \underline{Proof.} Let $\g_\mu(z)$ be a family of $G(\C)$-valued loops
 fulfilling \eqref{Ren6} and \eqref{Ren3}.
 Consider the loops  $\a_\mu(z)$ given by $$ \a_\mu(z)=\, \t_{sz}
 (\g_- (z)^{-1})\,,\quad s={\rm log}\,\mu $$ which fulfill
 \eqref{Ren6} by construction so that
 $ \a_{e^s\,\mu}(z)=\,\t_{sz}(\a_\mu(z)).
 $ The ratio $\a_\mu(z)^{-1}\,\g_\mu(z)$
 still fulfills \eqref{Ren6}
 and is moreover regular at $z =0$. Thus there is a unique loop
 $\g_{\rm reg}(z)$ regular  at $z=0$
 such that
 $$
 \a_\mu(z)^{-1}\,\g_\mu(z)=\,\t_{z\log\mu}(\g_{\rm reg}(z))\,.
 $$
 We can thus assume that $\g_\mu(z)=\a_\mu(z)$.
    By corollary \ref{prep}, applying $\t_{sz}$ to both sides
and using Proposition \ref{expprop}
to change variables in the integral,  one gets
 \begin{equation}\label{expan1a}
  \g_\mu(z)^{-1} =\,{\bf{\rm
 T}e^{-\frac{1}{z}\,\int_{-sz}^\infty\,\t_{-t}(\beta)\,dt}}
   \end{equation}
 and this proves the first statement of the theorem
using the appropriate notation for the inverse.

 \medskip  For the second part we can
again assume $\g_{\rm reg}(z)=1$  and let $\g_\mu(z)$ be given by
\eqref{expan1a}. Note that the basic properties
 of the time ordered exponential, Proposition \eqref{expprop},
 show that
 \begin{equation}\label{expan3}
  \g_\mu(z)^{-1} =\,{\bf {\rm T}e^{-\frac{1}{z}\,
 \int_{-sz}^0\,\t_{-t}(\beta)\,dt}}\,
 {\bf {\rm T}e^{-\frac{1}{z}\,\int_{0}^\infty\,\t_{-t}(\beta)\,dt}}
 \end{equation}
 so that
 \begin{equation}\label{expan4}
  \g_\mu(z)^{-1} =\,{\bf {\rm T}e^{-\frac{1}{z}\,
 \int_{-sz}^0\,\t_{-t}(\beta)\,dt}}\,  \g_- (z)
 \end{equation}
 where $\g_- (z) $ is a regular function of $1/z$.

  \smallskip
 By Proposition \eqref{expprop} one then obtains
 \begin{equation}\label{expan5}
 {\bf {\rm T}e^{-\frac{1}{z}\,\int_{-sz}^0\,\t_{-t}(\beta)\,dt}}\,
  {\bf {\rm T}e^{-\frac{1}{z}\,\int_0^{-sz}\,\t_{-t}(\beta)\,dt}}\, =\,1
 \end{equation}

  \smallskip
 We thus get
 $$ \g_\mu^+(z)=\,{\bf {\rm T}e^{-\frac{1}{z}
 \,\int_0^{-sz}\,\t_{-t}(\beta)\,dt}}\, $$
  Indeed taking the inverse of both sides in  (\ref{expan4}),
 it is enough to check the  regularity of the given expression for
 $\g_\mu^+(z)$ at $z=0$.
 One has in fact
 \begin{equation}\label{lim}
  \lim_{z\ra 0} {\bf {\rm T}e^{-\frac{1}{z}
 \,\int_0^{-sz}\,\t_{-t}(\beta)\,dt}}=\,e^{s\b} .
  \end{equation} $\Box$

 \bigskip

In the physics context, in order to 
preserve the homogeneity of the dimensionful variable $\mu$, 
it is better to replace everywhere $\mu$ by
$\mu/\lambda$ in the right hand side of the formulae of Theorem
\ref{genmu}, where $\lambda$ is an arbitrarily chosen unit.

 \bigskip
 \section{Diffeographisms and diffeomorphisms}\label{difgdiff}

 Up to what we described in Section \ref{Rensect},
 perturbative renormalization
 is formulated in terms of the group $G={\rm Difg}(\Tc)$, whose
 construction is still 
 based on the Feynman graphs of the theory $\Tc$. This does not
completely clarify the nature of the renormalization process. 

\smallskip

Two successive steps provide a solution to this problem. The first,
which we discuss in this section, is
part of the CK theory and consists of the relation
 established in \cite{3cknew} between the group ${\rm Difg}(\Tc)$ and
the group of formal diffeomorphisms. The other will be the main
result of the following sections, namely the construction of a
universal affine group scheme $U$, independent of the physical theory,
that maps to each particular $G={\rm Difg}(\Tc)$
and suffices to achieve the renormalization of the theory
in the minimal subtraction scheme.

 \smallskip

   The extreme complexity of the computations
 required for the tranverse index formula
 for foliations led to the
 introduction (Connes--Moscovici \cite{3cm1}) of the Hopf algebra of transverse
 geometry. This is neither commutative nor cocommutative, but is
 intimately related to the group of formal diffeomorphisms, whose Lie
 algebra appears from the Milnor-Moore theorem (\cf \cite{3MM}) applied
 to a large commutative subalgebra. A motivation for the CK work on
 renormalization was in fact, since
 the beginning, the appearance of intriguing similarities between
 the Kreimer Hopf algebra of rooted trees in \cite{3dhopf} and the
 Hopf algebra of transverse geometry introduced in \cite{3cm1}.

 \smallskip

  Consider the group of formal diffeomorphisms $\varphi$ of $\C$
 tangent to the identity, \ie satisfying
 \begin{equation}\label{tgid}
 \vp (0) = 0 \, , \ \ \ \vp' (0) = {\rm id}\, .
 \end{equation}
 Let ${{\mathcal H}}_{\diff}$ denote its Hopf algebra of coordinates.
 This has generators $a_n$ satisfying
 \begin{equation}\label{cm-an}
 \varphi(x) = x + \sum_{n\geq 2} a_n(\varphi)\, x^n .
 \end{equation}
 The coproduct in ${{\mathcal H}}_{\diff}$ is defined by
 \begin{equation}\label{coprodHdiff}
 \lgl \D a_n  \, , \, \vp_1 \ot \vp_2   \rgl = a_n(\vp_2 \circ \vp_1).
 \end{equation}

 \smallskip

 \noindent We describe then the result of \cite{3cknew}, specializing to the
 massless case and again taking $\Tc=\vp^3_6$, the $\vp^3$ theory with
$D=6$, as a sufficiently general 
 illustrative example. When, by rescaling the field,
 one rewrites the term of \eqref{cutoffterms} with the
 change of variable
 $$ \frac{1}{2}(\partial_\mu \phi)^2 (1-\delta Z) \ \ \ \leadsto \ \ \
 \frac{1}{2}(\partial_\mu \tilde\phi)^2, $$
 one obtains a corresponding formula for the effective
 coupling constant, of the form
 \begin{equation}
 g_{{\rm eff}} = \left( g + \quad \sum_{\hbox{\psfig{figure=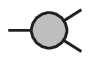}}}
 \quad g^{2\ell + 1} \,
 \frac{\G}{S(\G)} \right) \left( 1 - \quad
 \sum_{\hbox{\psfig{figure=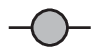}}} \quad g^{2\ell} \,
 \frac{\G}{S(\G)} \right)^{-3/2}\, , \label{Ren16}
 \end{equation}
 thought of as  a power series (in
 $g$) of elements of the Hopf algebra ${{\mathcal H}}=\Hc(\vp^3_6)$.
 Here  both $g \, Z_1 =g + \d g$ and the field strength
 renormalization $Z_3$ are thought of as power series (in
 $g$) of elements of the Hopf algebra ${{\mathcal H}}$.

 \smallskip

  Then one has the following result (\cite{3cknew}):
 \begin{thm}\label{hom}
 The expression \eqref{Ren16} defines a Hopf algebra homomorphism
 \begin{equation}\label{HcmH}
 \Phi: {{\mathcal H}}_{\diff} \stackrel{g_{{\rm eff}}}{\longrightarrow}
 {{\mathcal H}} \, ,
 \end{equation}
 from the Hopf algebra ${{\mathcal H}}_{\diff}$ of coordinates on the
 group of formal diffeomorphisms of $\Cb$ satisfying \eqref{tgid}
 to the CK Hopf algebra ${{\mathcal H}}$ of the massless theory.
 \end{thm}

 \smallskip

  The Hopf algebra homomorphism \eqref{HcmH} is obtained by
 considering the formal series \eqref{Ren16}
 expressing the effective coupling constant
 \begin{equation}\label{powergeff}
 g_{{\rm eff}}(g) = g + \sum_{n\geq 2} \alpha_n \, g^n \ \ \ \ \
 \alpha_n \in \cH,
 \end{equation}
 where all the coefficients $\alpha_{2n}=0$ and the $\alpha_{2n+1}$ are
 finite linear combinations of products of graphs, so that
 $$ \alpha_{2n+1} \in \Hc \ \ \  \forall n\geq 1. $$

 \smallskip

  The homomorphism
 \eqref{HcmH} at the level of Hopf algebras, and the corresponding
 group homomorphism \eqref{Ren19} from $G$ to the group of formal
 diffeomorphisms ${\rm Diff}(\C)$, are obtained then by assigning
 \begin{equation}\label{mapHcmH}
  \Phi(a_n)=\alpha_n.
 \end{equation}

 \smallskip

  The transposed group homomorphism
 \begin{equation}
 {\rm Difg}(\vp^3_6) \stackrel{\rho}{\longrightarrow} {\rm Diff} (\Cb)
\label{Ren19} 
 \end{equation}
 lands in the subgroup of {\it odd} formal diffeomorphisms,
 \begin{equation}
 \vp (-x) = -\vp (x) \qquad \fl \, x \, . \label{Ren20}
 \end{equation}

 \smallskip

  The physical significance of \eqref{HcmH} is transparent:
 it defines a natural action of ${\rm Difg}(\vp^3_6)$ by (formal)
 diffeomorphisms on the 
 coupling constant. The image under $\rho$ of $\b = Y \, {\rm Res} \,
 \g$ is the usual $\b$-function of the coupling constant $g$.

 \smallskip

  The Birkhoff decomposition can then be formulated {\it
 directly} in the group of formal diffeomorphisms of the space of
 coupling constants.

 \smallskip

  The result can be stated as follows (\cite{3cknew}):
 \begin{thm}\label{Theorem10}
 Let the unrenormalized effective coupling constant $g_{\rm eff}
 (z)$ be viewed as a formal power series in $g$ and let
 \begin{equation}\label{Birkgeff}
 g_{\rm eff}(z) =  g_{{\rm eff}_+} (z)\, (g_{{\rm
 eff}_-}(z))^{-1}
 \end{equation}
 be its (opposite) Birkhoff decomposition in the group of formal
 diffeomorphisms. Then the loop $g_{{\rm eff}_-} (z)$ is the bare
 coupling constant and $g_{{\rm eff}_+} (0)$ is the renormalized
 effective coupling.
 \end{thm}

 \smallskip

  This result is now, in its statement, no longer depending upon our
 group ${\rm Difg}$ or the Hopf algebra ${{\mathcal H}}$, though of
 course the proof 
 makes heavy use of the above ingredients. It is a challenge to
 physicists to find a direct proof of this result.

 \bigskip
 \section{Riemann--Hilbert problem}\label{RHsect}

Before we present our main result formulating perturbative
renormalization as a Riemann--Hilbert correspondence, we 
recall in this section several standard facts about the
Riemann--Hilbert problem, both in the regular singular case
and in the irregular singular case. This will prepare the ground
for our understanding of renormalization and of the
renormalization group in these terms.

\smallskip

  In its original formulation, Hilbert's 21st problem is a
 reconstruction problem for differential equations from the data of
 their monodromy representation. Namely, the problem asks whether
 there always exists a linear differential equation of Fuchsian
 type on $\bP^1(\C)$ with specified singular points and specified
 monodromy.

 \smallskip

  More precisely, consider an algebraic linear ordinary
 differential equation, in the form of a system of rank $n$
 \begin{equation}\label{ODE}
 \frac{d}{dz} f(z) + A(z) f(z) =0
 \end{equation}
 on some open set $U=
 \bP^1(\C)\smallsetminus \{ a_1,\ldots a_r \}$,  
 where $A(z)$ is an $n\times n$ matrix of rational
 functions on $U$. In particular, this includes the case of a
 linear scalar $n$th order differential equation.

 \smallskip

  The system \eqref{ODE} is Fuchsian if $A(z)$ has a pole at $a_i$ of
 order at most one, for all the points $\{ a_1,\ldots a_r \}$. 
 Assuming that all $a_i\neq \infty$,
 this means a system \eqref{ODE} with
 \begin{equation}\label{Fuchsyst}
  A(z) = \sum_{i=1}^r \frac{A_i}{z-a_i},
 \end{equation}
 where the complex matrices $A_i$ satisfy the additional condition
 $$ \sum_{i=1}^r A_i=0 $$
 to avoid singularities at infinity.

 \smallskip

  The space ${\mathcal S}$ of germs of holomorphic solutions of
 \eqref{ODE} at a point $z_0\in U$ is an $n$-dimensional complex
 vector space. Moreover, given any element $\ell \in
 \pi_1(U,z_0)$, analytic continuation along a loop representing the
 homotopy class $\ell$ defines a linear automorphism of
 ${\mathcal S}$, which only depends on the homotopy class $\ell$.
 This defines the {\em monodromy representation} $\rho:
 \pi_1(U,z_0) \to {\rm Aut}({\mathcal S})$ of the differential
 system \eqref{ODE}.

 \smallskip

  The Hilbert 21st problem then asks whether any finite
 dimensional complex linear representation of $\pi_1(U,z_0)$ is the
 monodromy representation of a differential system \eqref{ODE} with
 Fuchsian singularities at the points of $\bP^1(\C)\smallsetminus
 U$.

 \medskip
\subsection{Regular-singular case}
\bigskip

  Although the problem in this form was solved {\em negatively}
 by Bolibruch in 1989 (cf. \cite{3AnBol}), the original formulation of
 the Riemann--Hilbert problem was also given in terms of a different
 but sufficiently close condition on the differential equation
 \eqref{ODE}, with
 which the problem does admit a positive answer, not just in the
 case of the projective line, but in much greater generality. It is
 sufficient to relax the Fuchsian condition on \eqref{ODE} to the
 notion of {\em regular singular points}. The regularity
 condition at a singular point $a_i \in \bP^1(\C)$ is a growth
 condition on the solutions, namely all solutions
in any strict angular sector centered at
 $a_i$ have at most
 polynomial growth in $1/|z-a_i|$. The system \eqref{ODE} is 
 regular singular if every $a_i
 \in \bP^1(\C)\smallsetminus U$ is a regular singular point. 
 
\smallskip

An order $n$ differential equation written in the form
$$
\delta^n\,f + \, \sum_{k<n} a_k\,\delta^k\,f=\,0
$$
where $\delta=\,z \,\frac{d}{dz}$, is regular singular at $0$ iff
all the functions $a_k(z)$ are regular at $z=0$ ({\em Fuchs
criterion}).

\smallskip

For example, the two singular
 points $x=\pm \,\Lambda$ of the prolate spheroidal wave equation
$$
(\frac{d}{dx}\,(x^2-\Lambda^2)\,\frac{d}{dx}+
\,\Lambda^2\,x^2)\,f=\,0
$$
are regular singular since one can write the equation in the
variable $z=x-\Lambda$ in the form
$$
\delta^2\,f +\,\frac{z}{z+2\,\Lambda}\,\delta\,f+
\,\Lambda^2\,\frac{z\,(\Lambda +z)^2}{z+2\,\Lambda}\,\,f=\,0\,.
$$

 \smallskip

Though
 for scalar equations the Fuchsian and regular singular conditions
 are equivalent, the Fuchsian condition is in general a stronger
 requirement than the regular singular.

 \medskip

  In connection with the theory of renormalization,
 we look more closely at the regular singular Riemann--Hilbert problem
 on $\bP^1(\C)$. In this particular case, the solution to the problem
 is given by Plemelj's theorem (\cf \cite{3AnBol} \S 3). The argument
 first produces a system with the assigned monodromy on
 $U$, where in principle an analytic solution has no constraint on the
 behavior at the singularities.
 Then, one restricts to a {\em local problem} in small
 punctured disks $\Delta^*$ around the singularities, for which a system
 exists with the prescribed behavior of solutions at the origin. The
 global trivialization of the holomorphic bundle on $U$ determined by
 the monodromy datum yields the
 patching of these local problems that produces a global
 solution with the correct growth condition at the singularities.

 \smallskip

  More precisely (\cf \eg \cite{3AnBol} \S 3), we denote by $\tilde U$ the
 universal cover of $U$, with projection $p(\tilde z)=z$ and
 group of deck transformation $\Gamma
 \simeq \pi_1(U,x_0)$. For $G=\GL_n(\C)$, and a given monodromy
 representation $\rho: \Gamma \to G$, one considers the principal
 $G$-bundle $P$ over $U$,
 \begin{equation}\label{Pbundle}
 P= \tilde U \times G / \sim \, \ \ \ \ (\tilde z,g)\sim (\ell \tilde
 z, \rho(\ell) g), \ \  \forall \ell\in \Gamma.
 \end{equation}
 Consider the global section  
 \begin{equation}\label{branching}
 \xi: \tilde U \to P, \ \ \ \  \xi(\tilde z)= (\tilde z, 1)/\sim
 \end{equation}
of the pullback of $P$ to $\tilde U$. This satisfies
 $$ \xi(\tilde z)= \xi(\ell\,\tilde z)\, \rho(\ell), \ \ \  \forall
 \ell \in \Gamma. $$
 As a holomorphic bundle $P$ admits a global trivialization on $U$, which is given
 by a global holomorphic section $\gamma_U$. Thus, we can
 write $\xi(\tilde z) = \gamma_U(z) \sigma(\tilde z)$, for some
 holomorphic map $\sigma: \tilde U \to G$, so that we have
 \begin{equation}\label{sigmasect}
 \sigma(\tilde z) = \gamma_U(z)^{-1} \, \xi(\tilde z).
 \end{equation}
 This is the matrix of solutions to a differential system \eqref{ODE}
 with specified monodromy, where
 \begin{equation}\label{Asigma}
 A(\tilde z) = -\frac{d\sigma(\tilde z)}{dz} \, \sigma(\tilde z)^{-1}
 \end{equation}
 satisfies $A(\tilde z)=A(\ell \tilde z)$ for all $\ell\in
 \Gamma$, hence it defines the $A(z)$ on $U$ as in \eqref{ODE}. The
 prescription \eqref{Asigma} gives the flat connection on $P$ expressed
 in the trivialization given by $\gamma_U$. Due to the arbitrariness
in the choice of 
 the section $\gamma_U$, the
 differential system defined this way does not have any
 restriction on the behavior at the singularities. One can correct
 for that by looking at the local Riemann--Hilbert problem near
 the singular points and using the Birkhoff decomposition of
 loops.

 \medskip
\subsection{Local Riemann--Hilbert problem and Birkhoff decomposition}
\bigskip

  Consider a small disk $\Delta$ around a singular point, say $z=0$,
 and let $\Delta^*=\Delta\smallsetminus \{ 0 \}$. Let $V$ be a connected
component of the preimage $p^{-1}(\Delta^*)$ in $\tilde U$. Let $\ell$ be
 the element of $\Gamma$ obtained by lifting to $V$ the canonical 
 generator of the fundamental group $\Z$ of $\Delta^*$. 
One has $\ell\, V =V $ and one can identify the restriction of 
$p$ to $V$ with the 
universal cover $( \log r, \theta)\to r e^{i\theta}$ 
of $\Delta^*$. Let then
 $\rho(\ell)\in G=\GL_n(\C)$ be the
 monodromy. Let $\eta$ be such that
\begin{equation}\label{logG}
 \exp(2\pi i\,\eta) = \rho(\ell).
 \end{equation}

\smallskip

 Consider
 \begin{equation}\label{xilog}
 \gamma_{\Delta}(\tilde z)=  \exp(\eta\, \log r)\,\, \exp(\eta\, i\theta)\,,
 \end{equation}
 as a map from $V$ to $G=\GL_n(\C)$. Then with the above notations
the ratio $\sigma(\tilde z)\gamma_{\Delta}(\tilde z)^{-1}$ drops down to
a holomorphic map from $\Delta^*$ to $G=\GL_n(\C)$.
This gives a $G$-valued loop $\gamma(z)$
defined on $\Delta^*$. This loop will have a factorization of the
form \eqref{factorization}, with a possibly nontrivial diagonal
term \eqref{middleterm}. We can use the negative part $\gamma^-$,
which is holomorphic away from $0$, to correct the local frame
$\gamma_U$ so that the singularity of \eqref{Asigma} at $0$
is now a regular singularity, while the behaviour at the other
singularities has been unaltered.

 \smallskip

 When there are several singular points, we consider a small disk
 $\Delta_i$ around 
 each $a_i$, for $\bP^1(\C)\smallsetminus U=\{ a_1, \ldots, a_n
 \}$. The process described above can be applied repeatedly to
 each singular point, as the negative parts $\gamma_i^-$ are
 regular away from $a_i$.  Thus, the solution of the Riemann--Hilbert 
 problem is given by
 \eqref{sigmasect} with a new frame which is $\gamma_U$ corrected by the
 product of the $\gamma_i^-$. Then \eqref{Asigma} has the right
 behavior at the singularity.

\smallskip

 The trivial principal $G$ bundle on each $\Delta_i$ can be
 patched to the bundle $P$ on $U$ to give a holomorphic principal
 $G$-bundle ${\mathcal P}$ on $\bP^1(\C)$, with transition functions
 given by the loops $\gamma_i$ with values in $G$. The bundle ${\mathcal
 P}$ admits a global meromorphic section. If it is holomorphically
 trivial (this case corresponds to the Fuchsian Riemann--Hilbert
 problem), then it admits a global holomorphic section, while 
 when ${\mathcal P}$ is not holomorphically trivial, the
 Birkhoff decompositions only determine a meromorphic section and this
 yields a regular singular equation \eqref{Asigma}.

\smallskip

This procedure explains the relation between the Birkhoff
decomposition and the classical (regular-singular) Riemann--Hilbert
problem, namely, the negative part of the Birkhoff decomposition
can be used to correct the behavior of solutions near the
singularities, without introducing further singularities
elsewhere. We'll see, however, that in the case of renormalization,
one has to consider a more general case of the Riemann--Hilbert
problem, which is no longer regular-singular.

 \medskip
 \subsection{Geometric formulation}
 \bigskip

  In the regular singular version, the Riemann-Hilbert problem can be
 formulated in a more intrinsic form, for $U$ a punctured Riemann
 surface or
 more generally a smooth quasi-projective variety over $\C$. The
 data of the differential system \eqref{ODE} are expressed as a
 pair $(M,\nabla)$ of a locally free coherent sheaf on $U$ with a
 connection
 \begin{equation}\label{connex}
 \nabla : M \to M\otimes \Omega^1_{U/\C}.
 \end{equation}
 In the case of $U\subset \bP^1(\C)$, this is equivalent to the
 previous formulation with $M\cong {\mathcal O}_U^n$ and
 \begin{equation}\label{connA}
  \nabla f = df + A(z) f dz.
 \end{equation}
 The condition of regular singularities becomes
 the request that there exists an algebraic extension $(\bar M,
 \bar\nabla)$ of the data $(M,\nabla)$ to a smooth projective
 variety $X$, where $U$ embeds as a Zariski open set, with
 $X\smallsetminus U$ a union of divisors $D$ with normal crossing, so
 that the extended connection $\bar\nabla$ has log singularities,
 \begin{equation}\label{logsing}
 \bar\nabla : \bar M \to \bar M \otimes \Omega^1_{X/\C}(\log D).
 \end{equation}

 \smallskip

 In Deligne's work \cite{3Deligne} in 1970,
 the geometric point of view in terms of the data $(M,\nabla)$,
 was used to extend to higher dimensions the type of argument above
 based on solving the local Riemann--Hilbert problem around the divisor
 of the prescribed singularities and patching it to the analytic
 solution on the complement (\cf the survey given in
 \cite{3Katz}). From a finite dimensional complex linear
 representation of the fundamental group one obtains a local system
 $L$ on $U$. This determines a unique analytic solution
 $(M,\nabla)$ on $U$, which in principle has no constraint on the
 behavior at the singularities. However, by restricting to a {\em
 local problem} in small polydisks around the singularities divisor,
 one can show that $(M,\nabla)$ does extend to a $(\bar M, \bar\nabla)$
 with the desired property. The patching problem becomes more
 complicated in higher dimension because one can move along components
 of the divisor. The {\em Riemann--Hilbert
 correspondence}, that is, the correspondence constructed this way
 between finite dimensional complex linear representations of the
 fundamental group and algebraic linear differential systems with
 regular singularities, is in fact an equivalence of categories.
 This categorical viewpoint leads to far reaching
 generalizations of the Riemann--Hilbert correspondence (\cf  \cite{mek} and \eg
 the surveys \cite{lemek} and \cite{3GeMa} \S 8), formulated as an equivalence of
 derived categories between regular holonomic ${\mathcal
 D}$-modules and perverse sheaves. 

 \smallskip

 In any case, the basic philosophy underlying Riemann--Hilbert
 can be summarized as follows. Just like the index theorem
 describes a correspondence between certain topological and analytic
 data, the Riemann--Hilbert correspondence consists of an explicit
 equivalence between suitable classes of analytic data (differential
 systems, ${\mathcal D}$-modules) and
 representation theoretic or algebro-geometric data (monodromy,
 perverse sheaves), and it appears naturally in a variety of
 contexts\footnote{Grothendieck refers to
 Riemann--Hilbert as {\em le th\'eor\`eme du bon Dieu.}}.

\medskip
\subsection{Irregular case}\label{Sirreg}
\bigskip

The next aspect of the Riemann--Hilbert problem,
 which is relevant to the theory of renormalization is
what happens to
 the Riemann--Hilbert correspondence
 when one drops the regular singular condition. In this case, it is
 immediately clear by looking at very simple examples that finite 
 dimensional complex linear
 representations of the fundamental group no
 longer suffice to distinguish equations whose solutions can have very
 different analytic behavior at the singularities but equal monodromy.

\smallskip

 For example, consider the differential equation
\begin{equation}\label{nonregsing}
  \delta f + \frac{1}{z}\, f =0, 
\end{equation}
with the usual notation $\delta =z \frac{d}{dz}$.
 The Fuchs criterion immediately shows that it is not
 regular-singular. It is also not hard to see that the
 equation has trivial monodromy, which shows that the
 monodromy is no longer sufficient to determine the system in
 the irregular case. The equation \eqref{nonregsing} has 
 differential Galois group $\C^*$ \footnote{See below in this section
for a discussion of the  differential Galois group.}.

  \smallskip

  Differential equations of the form \eqref{ODE} satisfying the
 regular singular conditions are extremely special. For instance, in
 terms of the Newton polygon of the equation, the singular point is
 regular if the polygon has only one side with zero slope and is
 irregular otherwise (\cf Figure \ref{newton}).

 \begin{figure}
 \begin{center}
 \epsfig{file=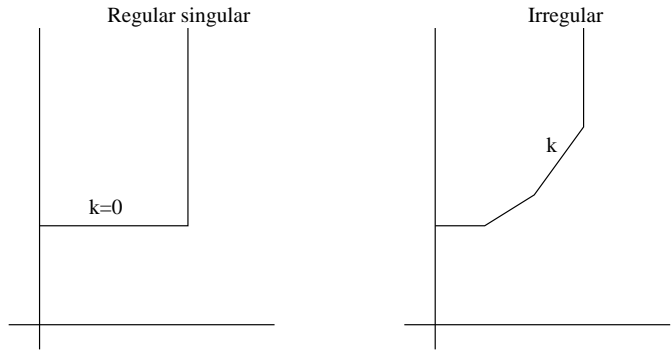}
 \end{center}
 \caption{Newton polygons \label{newton}}
 \end{figure}

 \smallskip

  There are different possible approaches to the irregular
 Riemann--Hilbert correspondence. The
 setting that is closest to what is needed in the theory of
 renormalization was developed by Martinet and Ramis \cite{3Ramis},
 by replacing the fundamental group with a
 {\em wild fundamental group}, which arises from
 the asymptotic theory of divergent series and differential
 Galois theory. In the representation datum of the Riemann--Hilbert
correspondence, in addition to the monodromy, this group contains at
the formal level (perturbative) an exponential torus related to
differential Galois theory (\cf \cite{PuSi} \S 3 and
\cite{vdp}). Moreover, at the non-formal level, which we discuss
in Section \ref{Snonpert}, it also incorporates the Stokes' phenomena
related to resummation of divergent series (\cf \cite{3Ramis}). 

 \smallskip

  As in the
 case of the usual Riemann--Hilbert correspondence of \cite{3Deligne}, the
 problem can be first reduced to a local problem on a punctured disk
 and then patched to yield the global case.
 In particular, for the purpose of renormalization, we are only
 interested in the {\em local} version of
 the irregular Riemann--Hilbert correspondence, in a small punctured
 disk $\Delta^*$ in the complex plane around a singularity $z=0$.

 \smallskip

At the formal level, one is working over the differential field of
formal complex Laurent series $\C((z))=\C[[z]][z^{-1}]$, with 
differentiation $\delta=z\frac{d}{dz}$, while at the non-formal level
one considers the subfield $\C(\{z\})$ of convergent Laurent series
and implements methods of resummation of divergent solutions of
\eqref{ODE}, with 
\begin{equation}\label{germA}
  A\in {\rm End}(n, \C(\{ z \})).
\end{equation}

\smallskip

For the purpose of the application to the theory of renormalization
that we present in the following sections, the structure of the wild
fundamental group of \cite{3Ramis} is best understood in terms of
differential Galois theory (\cf \cite{vdp}). In this setting, one
works over a differential field $(K,\delta)$, such that the field of
constants $k=\Ker(\delta)$ is an algebraically closed field of
characteristic zero. 
Given a differential system $\delta f = Af$, its Picard--Vessiot
ring is a $K$-algebra with a differentiation extending $\delta$. As a
differential algebra it is simple and is generated over $K$ by the
entries and the inverse determinant of a fundamental matrix for the
equation $\delta f = Af$. The differential Galois group of the
differential system is given by the automorphisms of the
Picard--Vessiot ring commuting with $\delta$. 

\smallskip

The set of all possible such differential systems
(differential modules over $K$) has the structure of a neutral
Tannakian category (\cf Section \ref{tannaka}), hence it can be
identified with the category of finite dimensional linear
representations of a unique affine group scheme over the field $k$. 
Any subcategory $\T$ that inherits the structure of a neutral
Tannakian category in turn corresponds to an affine group scheme $G_\T$,
that is the corresponding universal differential Galois group and
can be realized as automorphisms of the universal Picard--Vessiot ring
$R_\T$. This is generated over $K$ by the entries and inverse
determinants of the fundamental matrices of all the differential
systems considered in the category $\T$.

\smallskip

In these terms, one can recast the original regular--singular case
described above. The subcategory of differential modules over
$\C((z))$ given by regular--singular differential systems is a
neutral Tannakian category and the corresponding affine group scheme
is the algebraic hull of $\Z$, generated by the formal monodromy
$\gamma$. The latter can be seen as an automorphism of the universal
Picard--Vessiot ring acting by
$$ \gamma \, Z^a = \exp(2\pi i a)\, Z^a\, , \ \ \ \ \ \gamma\, L = L +
2\pi i, $$
where the universal Picard--Vessiot ring of the regular-singular case
is generated by $\{ Z^a \}_{a\in \C}$ and $L$, with relations dictated
by the fact that, in angular sectors, these formal generators can be
thought of, respectively, as the powers $z^a$ and the function
$\log(z)$ (\cf \cite{vdp}, \cite{PuSi}). 

\smallskip

In the irregular case, when one considers any differential system 
$\delta f =Af$ with arbitrary degree of irregularity, the universal
Picard--Vessiot ring is generated by elements $\{ Z^a \}_{a\in \C}$
and $L$ as before and by additional elements $\{ E(q) \}_{q\in \Ec}$,
where 
\begin{equation}\label{Elat}
 \Ec= \cup_{\nu\in \N^\times} \Ec_\nu, \ \ \text{ for } \ \  \Ec_\nu=
z^{-1/\nu} \C [ z^{-1/\nu} ].
 \end{equation}
These additional generators correspond, in local sectors, to functions of
the form $\exp(\int q\, \frac{dz}{z})$ and satisfy corresponding
relations $E(q_1+q_2)=E(q_1)E(q_2)$ and $\delta E(q)=q E(q)$.

\smallskip

When looking at a specific differential system \eqref{ODE}, instead of
the universal case, the decsription above of the Picard--Vessiot ring
corresponds to the fact that such system always admits a formal
fundamental solution of the form
\begin{equation}\label{formalSol}
 \hat F(x)= \hat H(u) u^{\nu \ell} e^{Q(1/u)},
 \end{equation}
with $u^\nu =z$, for some $\nu\in\N^\times$, with
$$ \ell\in {\rm End}(n,\C), \ \ \ \ \hat H\in \GL(n,\C((u))\,),$$
and with $Q$ a diagonal matrix with entries $\{ q_1, \ldots, q_n \}$ in
$u^{-1}\C[u^{-1}]$, satisfying $[e^{2\pi i \nu L}, Q]=0$ (\cf
\cite{3Ramis}). 

\smallskip

In the universal case described above, with arbitrary degrees of
irregularity in the differential systems, the corresponding universal
differential Galois group $\cG$ is described by a split exact sequence
(\cf \cite{vdp}),
\begin{equation}\label{Gspiltseq}
1 \to {\mathcal T} \to \cG \to \bar\Z \to 1,
\end{equation}
where $\bar\Z$ denotes the algebraic hull of $\Z$ generated by the
formal monodromy $\gamma$ and ${\mathcal T}=\Hom(\Ec, \C^*)$ is the Ramis
exponential torus. 

\smallskip

Now the action of the formal monodromy as an automorphism of the
universal Picard--Vessiot ring is the same as before on the $Z^a$ and
$L$, and is given by
\begin{equation}\label{actgammaEq}
\gamma\, E(q) = E(\gamma q),
\end{equation}
where the action on $\Ec$ is determined by the 
 action of $\Z/\nu\Z$ on $\Ec_\nu$ by
 \begin{equation}\label{Galnu}
 \gamma : q\left( z^{-1/\nu} \right) \mapsto q \left(
 \exp \left( \frac{-2\pi i}{\nu}\right) \, z^{-1/\nu} \right).
 \end{equation}
The exponential torus, on the other hand, acts by automorphisms of the
universal Picard--Vessiot ring by $\tau\,Z^a=Z^a$, $\tau\, L=L$ and
$\tau\, E(q)=\tau(q) E(q)$. The formal monodromy acts on the 
exponential torus by $(\gamma\tau)(q)=\tau(\gamma q)$.

\smallskip

Thus, at the formal level, the local Riemann--Hilbert correspondence
is extended beyond the regular-singular case, as a classification of 
differential systems with aribirary degree of irregularity at $z=0$ in
terms of finite dimensional linear representations of the group $\cG$.
The wild fundamental group of Ramis \cite{3Ramis} further extends this
irregular Riemann--Hilbert correspondence to the non-formal setting by
incorporating in the group further generators corresponding to the
Stokes' phenomena. We shall discuss this case in Section
\ref{Snonpert}, in relation to nonperturbative effects in
renormalization.

\bigskip
\section{Local equivalence  of meromorphic connections}\label{Sconnections}
\bigskip

We have seen in Section \ref{Rensect} that loops $\gamma_\mu(z)$
satisfying the conditions 
$$ \g_{e^t \mu} (z) = \t_{t z} (\g_{\mu} (z)) \qquad \fl \, t \in
 \Rb \ \ \ \text{ and } \ \ \ \frac{\partial}{\partial \mu} \,
\g_{\mu^-} (z) = 0 $$
can be characterized (Theorem \ref{genmu}) in expansional form 
$$ \g_\mu(z) =\,{\bf {\rm T}e^{-\frac{1}{z}\,
 \int^{-z \log\mu}_\infty\,\t_{-t}(\beta)\,dt}}\;
  \t_{z\log\mu}(\g_{\rm reg}(z)), $$
hence as solutions of certain differential equations (Proposition
\ref{expprop0}). 
In this section and the following, we examine more closely the
resulting class
of differential equations. Rephrased in geometric terms, loops
$\gamma_\mu(z)$ satisfying the conditions above correspond to
equivalence classes of flat equisingular $G$-valued connections on a
principal $\C^*$ bundle $B^*$ over a punctured disk $\Delta^*$. The
equisingularity condition (defined below in Section \ref{classif}) 
expresses geometrically the condition that $\partial_\mu\, \g_{\mu^-}
(z) = 0$. We will then provide, in Section \ref{SGalois}, the
representation theoretic datum of the Riemann--Hilbert correspondence 
for this class of differential systems. Similarly to what we recalled
in the previous section, this will be obtained in the
form of an affine group scheme of a Tannakian category of flat
equisingular bundles. Since we show in Theorem \ref{rh} below that 
flat equisingular connections on $B^*$ have trivial monodromy, it is
not surprising that the affine group scheme we will obtain in Section
\ref{SGalois} will resemble most the Ramis exponential torus described
in the previous section.

\medskip
 
We take the same notations as in Section \ref{expansional} and let
$G$ be a graded affine group scheme with positive integral grading
$Y$. We consider the local behavior of solutions of
$G$-differential systems near $z=0$ and work locally, \ie over an
infinitesimal punctured disk $\Delta^*$ centered at $z=0$ and with
convergent Laurent series.

\smallskip

As above, we let $ K$ be the field $ \C(\{z\})$ of convergent
Laurent series in $z$ and $O \subset   K$ the subring of  series
without a pole at $0$. The field $ K$ is a differential field and
we let $\Omega^1$ be the $1$-forms on $ K$ with
$$
d\,: K \to \Omega^1
$$
the differential. One has $df = \frac{df}{dz}\,dz $.

\smallskip

A connection on the trivial $G$-principal bundle $P=
\Delta^*\times G$ is specified by the restriction of the
connection form to $\Delta^*\times 1$ \ie by a $\fg$-valued
$1$-form $\omega$ on $\Delta^*$

We let $\Omega^1(\fg)$ denote $\fg$-valued $1$-forms on $\Delta^*$.
Every element of $\Omega^1(\fg)$ is of the form $A \,dz$
with $A \in \fg(K)$.

\smallskip

As in section  \ref{expansional} we consider the operator
$$ D: G( {K}) \to \Omega^1(\fg) \ \ \ \ Df=
f^{-1}\, df\,. $$ It satisfies
\begin{equation}\label{prodrule}
 D(fh)=D h + h^{-1}\, Df\, h.
\end{equation}

The differential equations we are looking at are then of the form
\begin{equation}\label{Gdiff}
 Df=\,\omega
\end{equation}
where $\omega\in \Omega^1(\fg)$ specifies the connection on the
trivial $G$-principal bundle.

The local singular behavior of solutions is the same in the
classes of connections under the following equivalence relation:

\begin{defn}\label{equivconnection}
We say that two connections $\omega$ and $\omega'$ are equivalent
iff
\begin{equation}\label{gaugetransf}
 \omega' = Dh + h^{-1} \omega\, h, 
\end{equation}
for $h \in  G(O)$ a $G$-valued map regular in $\Delta$.
\end{defn}

By proposition \ref{trimono} the triviality of the monodromy:
$M=1$, is a well defined condition which ensures the existence of
solutions $f\in G(K)$ for the equation
\begin{equation}\label{Gdiff11}
  Df=\,\omega
\end{equation}

A solution $f$ of \eqref{Gdiff11}  defines a $G$-valued loop. By
our assumptions on $G$, any $f\in G(K)$ has a unique Birkhoff
decomposition of the form
\begin{equation}\label{fBirkh}
f = f_-^{-1} f_+,
\end{equation}
where
$$
f_+ \in  G(O)\,,\quad f_-\in G(\Qc)
$$
where $O \subset  {K}$ is the subalgebra of regular functions and
$\Qc=\,z^{-1}\,\C([z^{-1}])$. Since $\Qc$ is not unital one needs
to be more precise in defining $G(\Qc)$. Let
$\tilde{\Qc}=\,\C([z^{-1}])$ and $\ve_1$ its augmentation. Then
$G(\Qc)$ is the subgroup of $G(\tilde{\Qc})$ of homomorphisms
$\phi\,:\,\Hc \to \tilde{\Qc}$ such that $\ve_1\circ
\phi=\,\ve$ where $\ve$ is the augmentation of $\Hc$.

\begin{prop}  \label{class} Two connections $\omega$ and $\omega'$
with trivial monodromy are equivalent iff the negative pieces of
the Birkhoff decompositions of any solutions $f^\omega$ of $
Df=\,\omega $ and $f^{\omega'}$ of $ Df=\,\omega' $ are the same,
$$f^\omega_-=f^{\omega'}_-\,.$$
\end{prop}

\underline{Proof.} By proposition \ref{trimono} there exists
solutions $f^\omega\in G(K)$ of $ Df=\,\omega $ and $f^{\omega'}
\in G(K)$ of $ Df=\,\omega' $. Let us show that $\omega$ is
equivalent to $D((f^\omega_-)^{-1})$. One has
$f^\omega=\,(f^\omega_-)^{-1}\,f^\omega_+$, hence the product rule
\eqref{prodrule} gives the required equivalence since $f^\omega_+
\in G(O)$. This shows that if $f^\omega_-=f^{\omega'}_-$ then
$\omega$ and $\omega'$ are equivalent. Conversely equivalence of
$\omega$ and $\omega'$ implies equivalence of
$D((f^\omega_-)^{-1})$ with $D((f^{\omega'}_-)^{-1})$ and hence an
equality of the form
$$
(f^\omega_-)^{-1}=\,(f^{\omega'}_-)^{-1}\,h\,,
$$
with $h\in G(O)$. The uniqueness of the Birkhoff decomposition
then implies $h=1$ and $f^\omega_-=f^{\omega'}_-\,.$ $\Box$

\medskip

Our notion of equivalence in Definition \ref{equivconnection} is
simply a change of local holomorphic frame, \ie by an element $h \in
G(O)$ (rather than by $h \in  G(K)$). This is quite natural in
our context, in view of the result of Proposition \ref{class} above,
that relates it to the negative part of the Birkhoff decomposition.

\bigskip
\section{Classification  of equisingular flat
connections}\label{classif}
\bigskip

 At the geometric level  we  consider a
$\bG_m$-principal bundle
\begin{equation}\label{base}
\bG_m \to B\,\to \Delta\,,
\end{equation}
over an infinitesimal  disk $\Delta$. We let
$$
 b\mapsto w(b) \quad \forall w\in \C^*\,,
$$
be the action of $\bG_m$ and  $\pi\,:B\to \Delta$ be the
projection, $$V=\pi^{-1}(\{0\})\subset B$$ be the fiber over $0\in
\Delta$ where we fix a base point $y_0 \in V$. Finally we let
$B^*\subset B$  be the complement of $V$.

With $G$ as above and $Y$ its grading we view the trivial
$G$-principal bundle $P=B\times G$ as equivariant with respect to
$\bG_m$
 using the action
\begin{equation}\label{Gdiff1}
 u(b,g)=\,(u(b),u^Y(g)) \quad \forall u\in \C^*\,,
\end{equation}
\medskip
where $u^Y$ makes sense since the grading $Y$ is integer valued.

We let $P^*=B^*\times G$ be the restriction to $B^*$ of the bundle
$P$.

\begin{defn}
We say that the  connection $\omega$ on $P^*$ is equisingular iff
it is $\bG_m$-invariant and if its restrictions to sections of the
principal bundle $B$ which agree at $0\in \Delta$
 are mutually equivalent.
\end{defn}

\medskip

Also as above we consider the operator
$$  Df=
f^{-1}\, df\,. $$ The operator $D$ satisfies
\begin{equation}\label{prodrule1}
 D(fh)=D h + h^{-1}\, Df\, h.
\end{equation}

\begin{defn}\label{GequivP}
We say that two connections $\omega$ and $\omega'$ on $P^*$ are
equivalent iff
$$ \omega' = Dh + h^{-1} \omega h, $$
for a $G$-valued $\bG_m$-invariant map $h$ regular in $B $.
\end{defn}

We are now ready to prove the main step which will allow us to
formulate renormalization as a Riemann-Hilbert correspondence. For
the statement we choose a non-canonical regular section
$$
\sigma\,: \Delta \to B\,,\quad \sigma(0)= y_0 \,,
$$
and we shall show later that the following correspondence between
flat
 equisingular $G$-connections and the Lie algebra $\fg$
is in fact independent of the choice of $\sigma$. To lighten
notations we use $\sigma$ to trivialize the bundle $B$ which we
identify with $\Delta \times \C^*$.

\begin{figure}
\begin{center}
\includegraphics[scale=0.45]{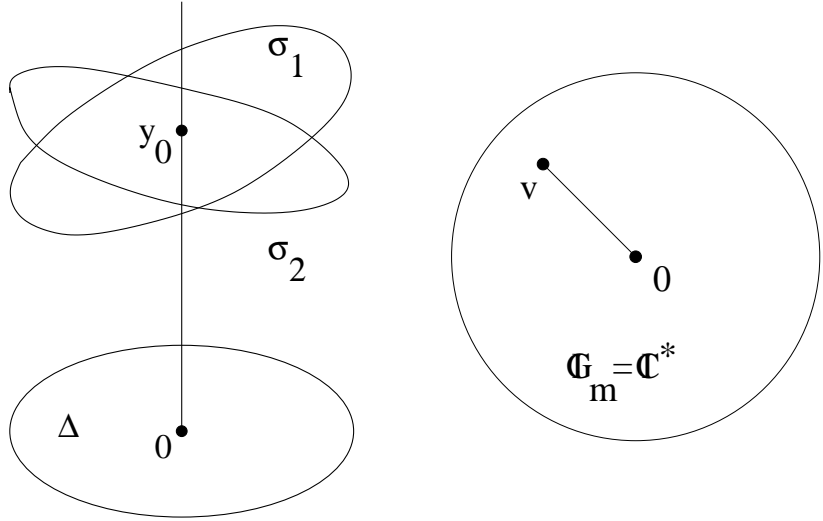}
\caption{\label{Figequising} Equisingular condition and the path of
integration}
\end{center}
\end{figure}

\medskip

\begin{thm} \label{rh} Let $\omega$ be a flat
 equisingular $G$-connection.
There exists a unique element
 $\beta \in \fg$ of
the Lie algebra of $G$ such that $\omega$ is equivalent to the
flat
 equisingular connection $D\g$ associated to the following section
\begin{equation}\label{solexpm}
 \g(z,v) =\,{\bf {\rm T}e^{-\frac{1}{z}\,
 \int^{v}_0\,u^Y(\beta)\,\frac{du}{u}}}\;
\in G\;
  \,,
\end{equation}
where the integral is performed on the straight path $u=t v$,
$t\in[0,1]$.

\end{thm}

\underline{Proof.} The proof consists of two
main steps. We first prove  the vanishing of the two monodromies
of the connection corresponding to the
two generators of the fundamental group of $B^*$.
This implies the existence of a solution of the 
equation $D \g= \omega$. We then show that the equisingularity
condition allows us to apply Theorem \ref{genmu}
to the restriction of $\g$ to a section of the bundle $B$
over $\Delta$.

We encode as above a connection on $P^*$ in
terms of $\fg$-valued $1$-forms on $B^*$ and we use the
trivialization $\sigma$ to write it as
$$
\omega =\,A\,dz\,+\,B\,\frac{dv}{v}
$$
in which both $A$ and $B$ are $\fg$-valued functions $A(z,v)$ and
$B(z,v)$ and $\frac{dv}{v}$ is the fundamental $1$-form of the
principal bundle $B$.

Let $\omega =\,A\,dz\,+\,B\,\frac{dv}{v}$ be an invariant
connection. One has
$$
\omega(z,u\,v)=\,u^Y(\omega(z,v))\,,
$$
which shows that the coefficients of $\omega$ 
are determined by their restriction to  the
section  $v=1$.
 Thus one has
$$\omega(z,u)=\,u^Y(a)\,dz\,+\,u^Y(b)\,\frac{du}{u} $$
for suitable elements $a,b\in \fg(K)$.

The flatness of the connection means that
\begin{equation}\label{flat}
\frac{db}{dz}-Y(a)+\,[a,b]=0
\end{equation}

The positivity of the integral grading $Y$ shows that the
connection $\omega$ extends to a flat connection on the product
$\Delta^*\times \C$. Moreover its restriction to $\Delta^*\times
\{0\}$ is equal to $0$ since $u^Y(a)=0$ for $u=0$. This suffices to show that the two
generators of $\pi_1(B^*)=\Z^2$ give a trivial monodromy. 
Indeed the generator corresponding to a fixed value 
of $z_0\in \Delta^*$ has trivial monodromy since the
connection $\omega$ extends to  $z_0\times \C$ which
is simply connected. The other generator corresponds to 
a fixed value of $u$ which one can choose as $u=0$, and since the 
restriction of the connection to 
$\Delta^*\times
\{0\}$ is equal to $0$ the monodromy vanishes also.
One can
then explicitely write down a solution of the differential system
\begin{equation}\label{partial}
D\g =\, \omega
\end{equation}
as in Proposition \ref{trimono},
  with a base point
of the form $(z_0,0)\in \Delta^*\times
\{0\}$. Taking a path in $\Delta^*\times
\{0\}$ from $(z_0,0)$ to $(z,0)$ and then the straight path $(z,t v)$,
$t\in[0,1]$ gives the solution (using Proposition \ref{expprop})
in the form
\begin{equation}\label{sol}
 \g(z,v) =\,{\bf {\rm T}e^{
 \int^{v}_0\,u^Y(b(z))\,\frac{du}{u}}}\;
  \,,
\end{equation}
where the integral is performed on the straight path $u=t v$,
$t\in[0,1]$.

This gives a translation invariant loop $\g$,
\begin{equation}\label{sol2}
\g(z,u)=\,u^Y\g(z)
\end{equation}
fulfilling
\begin{equation}\label{partial2}
\g(z)^{-1}d\g(z)=\,a\,dz\,,\qquad \g(z)^{-1}Y\g(z)=\,b\,.
\end{equation}

\medskip
By hypothesis $\omega$ is equisingular and thus the restrictions
$\omega_s$ to the lines $\Delta_s=\,\{(z,e^{sz});\,z\in \Delta^*
\}$ are mutually equivalent. By proposition \ref{class} and the
fact that the restriction of $\g(z,u)=u^Y\g(z)$ to $\Delta_s$ is
given by $\g_s(z)=\,\t_{sz}\g(z)$ we get that the negative parts
of the Birkhoff decomposition of the loops $\g_s(z)$ are
independent of the parameter $s$.

Thus by the results of section \ref{expan},
 there exists
an element $\beta\in \fg$ and a regular loop $\g_{\rm reg}(z)$
such that
\begin{equation}\label{solexp0}
 \g(z,1) =\,{\bf {\rm T}e^{-\frac{1}{z}\,
 \int^{0}_\infty\,\t_{-t}(\beta)\,dt}}\;
 \g_{\rm reg}(z)\,.
\end{equation}

Thus up to equivalence, (using the regular loop $u^Y(\g_{\rm
reg}(z))$ to perform the equivalence) we see that the solution is
given by
\begin{equation}\label{solexp1}
 \g(z,u) =\,u^Y({\bf {\rm T}e^{-\frac{1}{z}\,
 \int^{0}_\infty\,\t_{-t}(\beta)\,dt}})\;
  \,,
\end{equation}
which only depends upon $\beta \in \fg$. Since $u^Y$ is an
automorphism one can in fact rewrite \eqref{solexp1}
 as
\begin{equation}\label{solexp2}
  \g(z,v) =\,{\bf {\rm T}e^{-\frac{1}{z}\,
 \int^{v}_0\,u^Y(\beta)\,\frac{du}{u}}}\;
  \,,
\end{equation}
where the integral is performed on the straight path $u=t v$,
$t\in[0,1]$.

\medskip

We next need to understand in what way the class of the solution
\eqref{solexp1} depends upon $\beta \in \fg$. An equivalence
between two equisingular flat connections generates a relation
between solutions of the form

$$
\g_2(z,u) =\,\g_1(z,u) \,h(z,u)
$$
with $h$ regular. Thus the negative pieces of the Birkhoff
decomposition of both
$$
\g_j(z,1) =\,{\bf {\rm T}e^{-\frac{1}{z}\,
 \int^{0}_\infty\,\t_{-t}(\beta_j)\,dt}}\;
  \,,
$$
have to be the same which gives $\beta_1=\beta_2$
using the equality of residues at $z=0$.

\medskip

Finally we need to show that for any $\beta \in \fg$ the
connection $\omega=\,D\g$ with $\g$ given by \eqref{solexpm} is
equisingular. The equivariance follows from the invariance of the
section $\g$. Let then $v(z)\in\C^*$ be a regular function with
$v(0)=1$ and consider the section $v(z)\sigma(z)$ instead of
$\sigma(z)$. The restriction of $\omega=\,D\g$ to this new section
is now given by
\begin{equation}\label{solexpm1}
 \g_v(z) =\,{\bf {\rm T}e^{-\frac{1}{z}\,
 \int^{v(z)}_0\,u^Y(\beta)\,\frac{du}{u}}}\;
\in G\;
  \,.
\end{equation}
We claim that the Birkhoff decomposition of $\g_v$ is given (up to
taking the inverse of the first term) by,
\begin{equation}\label{solexpm2}
 \g_v(z) =\,{\bf {\rm T}e^{-\frac{1}{z}\,
 \int^{1}_0\,u^Y(\beta)\,\frac{du}{u}}}\;
\,{\bf {\rm T}e^{-\frac{1}{z}\,
 \int^{v(z)}_1\,u^Y(\beta)\,\frac{du}{u}}}\;
  \,.
\end{equation}
Indeed the first term in the product is a regular function of
$z^{-1}$ and gives a polynomial in $z^{-1}$ when paired with any
element of $\Hc$. The second term is a regular function of $z$
using the Taylor expansion of $v(z)$ at $z=0$,
($v(0)=1$). $\Box$

In fact the above formula \eqref{solexpm2} for changing the
choice of the section shows that the following holds.

\medskip

\begin{thm} \label{ind} The above correspondence
between  flat equisingular $G$-connections and elements
 $\beta \in \fg$ of
the Lie algebra of $G$ is independent of the choice of the local
regular section $ \sigma\,: \Delta \to B\,,\;\sigma(0)= y_0
\,. $

Given two choices  $\sigma_2=\,\alpha \,\sigma_1$ of local
sections $ \sigma_j(0)= y_0$, the regular values $\g_{{\rm reg}}(y_0)_j$ of solutions
of  the above differential system in the corresponding singular
frames are related by
$$
\g_{{\rm reg}}(y_0)_2=\,e^{-s\,\beta}\g_{{\rm reg}}(y_0)_1
$$
where $$s= \,(\frac{d\alpha(z)}{dz})_{z=0}\,.$$

\end{thm}

\medskip
It is this second statement that controls the ambiguity inherent
to the renormalization group in the physics context, where there is
no prefered choice of local regular section $\sigma$. In that
context, the group is $G={\rm Difg}(\Tc)$, and
the principal bundle $B$ over an infinitesimal disk centered
at the critical dimension $D$
admits as fiber the set of all possible normalizations for
the integration in dimension $D-z$.

Moreover, in the physics context, the choice of the base point in
the fiber $V$ over the critical dimension $D$ corresponds to a
choice of the Planck constant. The choice of the section $\sigma$ (up
to order one) corresponds to the choice of a ``unit of mass".

\bigskip
\section{The universal singular frame}\label{Sunivframe}
\bigskip

In order to reformulate the results of section \ref{classif} as a
Riemann-Hilbert correspondence, we begin to analyze the representation
theoretic datum associated to the equivalence classes
of equisingular flat connections. In Theorem \ref{rh1} below, we
classify them in terms of homomorphisms from a group
$U^*$ to $G^*$. In Section \eqref{SGalois} we will then show how to
replace homomorphisms $U^* \to G^*$ by finite dimensional linear
representations of $U^*$, which will give us the final form of the 
Riemann-Hilbert correspondence.

\smallskip

Since we need to get
both $Z_0$ and $\beta$ in the range at the Lie algebra level, 
it is natural to first think about the free Lie
algebra generated by $Z_0$ and $\beta$. It is important, though,
to keep track of the positivity and integrality of the grading
so that the formulae of the previous sections make sense.
These properties of integrality and positivity allow one to write
$\beta$ as an infinite formal sum
\begin{equation}\label{betasum}
\beta=\; \sum_1^\infty\;\beta_n\,,
\end{equation}
where, for each $n$, $\beta_n$ is homogeneous of degree $n$ for the
grading, \ie $Y(\beta_n)=n \beta_n$.

\smallskip

Assigning $\beta$ and the action of the grading on it is the same as
giving a collection of homogeneous elements $\beta_n$ that
fulfill no restriction besides $Y(\beta_n)=n \beta_n$. In
particular, there is no condition on their Lie brackets. Thus, these
data are the same as giving a homomorphism from the following
affine group scheme $U$ to $G$.

At the Lie algebra level $U$ comes  from the free graded Lie
algebra $$\cF(1,2,3,\cdots)_{\bullet}$$ generated by elements
$e_{-n}$ of degree $n$, $n>0$. At the Hopf algebra level we thus
take the graded dual of the enveloping algebra $\cU(\cF)$ so that

\begin{equation}\label{hopfu}
\Hc_u=\; \cU(\cF(1,2,3,\cdots)_{\bullet})^\vee
\end{equation}

\smallskip

As is well known, as an algebra $\Hc_u$ is isomorphic to the
linear space of noncommutative polynomials in variables $f_n$,
$n\in \N_{>0}$ with the product given by the shuffle.

\smallskip

It defines by construction a pro-unipotent affine group scheme $U$
which is graded in positive degree. This allows one to construct the
semi-direct product $U^*$ of $U$ by the grading as an affine group
scheme with a natural morphism $\,: U^*\,\to \bG_m$, where $\bG_m$
is the multiplicative group.

\smallskip

Thus, we can reformulate the main theorem of section \ref{classif}
as follows

\begin{thm} \label{rh1} Let $G$ be a positively graded
pro-unipotent Lie group.

There exists a canonical bijection between equivalence classes of
flat equisingular $G$-connections and graded representations
$$\rho \, : U\,\to G$$
 of $U$ in $G$.
\end{thm}

\smallskip

The compatibility with the grading means that $\rho$ extends to an
homomorphism
$$\rho^* \, : U^*\,\to G^*$$
which is the identity on $\bG_m$.

\smallskip

The group $ U^*$ plays in the formal theory of equisingular
connections the same role as the  Ramis exponential torus  in the
context of differential Galois theory.

\medskip

The equality
\begin{equation}\label{esum}
e=\; \sum_1^\infty\;e_{-n}\,,
\end{equation}
defines an element of the Lie algebra of $U$. As $U$ is by
construction a pro-unipotent affine group scheme we  can lift $e$
to a morphism 
\begin{equation}\label{rgU}
{\bf{rg}}\; :\,\bG_a \,\to \,U\,,
\end{equation}
of affine group schemes from the additive group $\bG_a$ to $U$.

\smallskip

It is this morphism ${\bf{rg}}$ that represents the
renormalization group in our context. The corresponding ambiguity
is generated as explained above in Theorem \ref{ind} by the
absence of a canonical trivialization for the $\bG_m$-bundle
corresponding to integration in dimension $D-z$.

\smallskip

The formulae above make sense in the universal case where
$G^*=U^*$ and allow one to define the universal singular frame by the
equality
\begin{equation}\label{univ}
\g_U(z,v) =\,{\bf {\rm T}e^{-\frac{1}{z}\,
 \int^{v}_0\,u^Y(e)\,\frac{du}{u}}}\;
\in U\;
  \,.
\end{equation}

\medskip

The frame \eqref{univ} is easily computed in terms of iterated
integrals and one obtains the following result.

\begin{prop}\label{univsin} 
The universal singular frame is given by
\begin{equation}\label{framecoeff}
\g_U(z,v) =\,\sum_{n \geq
0}\,\sum_{k_j>0}\,\frac{e(-k_1)e(-k_2)\cdots e(-k_n)}
{k_1\,(k_1+k_2)\cdots (k_1+k_2+\cdots +k_n)}\,v^{\sum k_j}\,z^{-n}.
\end{equation}
\end{prop}

 \underline{\em Proof.}
Using \eqref{esum}
and \eqref{expansional0} we get, for the coefficient of
$$e(-k_1)e(-k_2)\cdots e(-k_n)$$ the expression
$$
v^{\sum k_j}\,z^{-n} \,\int_{0\leq s_1\leq \cdots\leq s_n\leq 1}
\,s_1^{k_1-1}\cdots\,s_n^{k_n-1} \prod ds_j \,,
$$
which gives the desired result. $\Box$

\medskip
 The same expression appears in the local index formula
of \cite{cmindex}, where the renormalization group idea is used in
the case of higher poles in the dimension spectrum.

\medskip

Once one
uses this universal singular frame in the dimensional
regularization technique, all divergences disappear and one
obtains a finite theory which depends only upon the choice of
local trivialization of the $\bG_m$-principal bundle $B$, whose
base $\Delta$ is the space of complexified dimensions around 
the critical dimension $D$, and whose 
fibers correspond to normalization of the integral in
complex dimensions.

\smallskip

Namely, one can apply the Birkhoff decomposition to $\gamma_U$ in the
pro-unipotent Lie group $U(\C)$. For a given physical theory
$\sT$, the resulting $\gamma_U^+$ and $\gamma_U^-$ respectively map, via the
representation $\rho: U \to G={\rm Difg}(\sT)$, to the renormalized
values and the countertems in the minimal subtraction scheme.

\bigskip
\section{Mixed Tate motives}\label{Smotives}
\bigskip

In this section we recall some aspects and ideas from the theory of
motives that will be useful in interpreting the results of the
following Section \ref{SGalois} in terms of motivic Galois theory. 
The brief exposition given here of some aspects of the theory of mixed
Tate motives, is derived mostly from Deligne--Goncharov \cite{dg}. The
relation to the setting of renormalization described above will be the
subject of the next section. 

\smallskip

The purpose of the theory of motives, initiated by Grothendieck, is to 
develop a unified setting underlying different cohomological 
theories (Betti, de Rham, \'etale, $\ell$-adic, crystalline), by constructing an 
abelian tensor category that provides a ``linearization'' of the 
category of algebraic varieties. For smooth projective varieties
a category of motives (pure motives) is constructed, with morphisms
defined using algebraic correspondences between smooth projective 
varieties, considered modulo equivalence (\eg numerical equivalence).
The fact that this category has the desired properties depends
upon the still unproven standard conjectures of Grothendieck.

\smallskip

 For
more general (non-closed) varieties,
the construction of a category of motives (mixed motives) remains a difficult task.
Such category of mixed motives over a field (or more generally
over a scheme $\cS$) should be an abelian tensor category, with the
following properties (\cf \eg \cite{Le}). There will be a 
functor (natural in $\cS$) that assigns to each
smooth $\cS$-scheme $X$ its motive $M(X)$, with K\"unneth 
isomorphisms $M(X)\otimes M(Y) \to M(X\times_\cS Y)$.
The category will contain Tate objects $\Z(n)$, 
for $n\in \Z$, where $\Z(0)$ is the unit for $\otimes$ and
$\Z(n)\otimes \Z(m)\cong \Z(n+m)$.
The Ext functors in the category of mixed motives define a 
``motivic cohomology''
$$ H^m_{mot}(X,\Z(n)) := \Ext^m (\Z(0),M(X)\otimes \Z(n)). $$
This will come endowed with Chern classes $c^{n,m}: K_{2n-m}(X) \to
H^m_{mot}(X,\Z(n))$ from algebraic $K$-theory
that determine natural isomorphisms
$\Gr^\gamma_n K_{2n-m}(X)\otimes \Q \cong H^m_{mot}(X,\Z(n))\otimes \Q$,
where on the left hand side there is the weight $n$ eigenspace of
the Adams operations. The motivic cohomology 
will be universal with respect to all cohomology theories
satisfying certain natural properties (Bloch--Ogus axioms).
Namely, for any such cohomology $H^*(\cdot ,\Gamma(*))$ there 
will be a natural transformation 
$H^*_{mot}(\cdot, \Z(*)) \to H^*(\cdot,\Gamma(*))$.
Moreover, to a map of schemes $f: \cS_1 \to \cS_2$ 
there will correspond functors $f^*$, $f_*$, $f^{!}$, $f_{!}$, 
behaving like the corresponding functors of sheaves.

\smallskip

Though, at present, there is not yet a general construction of 
such a category of mixed motives, there are constructions
of a triangulated tensor category $DM(\cS)$, which has the right 
properties to be the bounded derived category of the category
of mixed motives. 
The constructions of $DM(\cS)$ due
to Levine \cite{Le} and Voevodsky \cite{Vo} are known to be
equivalent (\cite{Le}, VI 2.5.5).
The triangulated category of {\em mixed Tate motives} $DMT(\cS)$ 
is then defined as the full triangulated subcategory of $DM(\cS)$
generated by the Tate objects.
 One can then hope
to find a method that will reconstruct the category
knowing only the derived category. We mention briefly 
what can be achieved along these lines.

\smallskip

Recall that a triangulated category ${\mathcal D}$ is an additive category
with an automorphism $T$ and a family of distinguished
triangles $X \to Y \to Z \to T(X)$, satisfying suitable axioms (which 
we do not recall here). A $t$-structure consists of two full subcategories
${\mathcal D}^{\leq 0}$ and ${\mathcal D}^{\geq 0}$ with the properties:
${\mathcal D}^{\leq -1} \subset {\mathcal D}^{\leq 0}$ and 
${\mathcal D}^{\geq 1} \subset {\mathcal D}^{\geq 0}$; for all
$X\in {\mathcal D}^{\leq 0}$ and all $Y\in {\mathcal D}^{\geq 1}$ one has
$\Hom_{\mathcal D} (X,Y)=0$; for all $Y\in {\mathcal D}$ there exists a
distinguished triangle as above with $X\in {\mathcal D}^{\leq 0}$
and $Z\in {\mathcal D}^{\geq 1}$. Here we used the notation 
${\mathcal D}^{\geq n} = {\mathcal D}^{\geq 0}[-n]$ and
${\mathcal D}^{\leq n} = {\mathcal D}^{\leq 0}[-n]$, with
$X[n]=T^n(X)$ and $f[n]=T^n(f)$. The heart of the t-structure is the
full subcategory ${\mathcal D}^0= {\mathcal D}^{\leq 0}\cap 
{\mathcal D}^{\geq 0}$. It is an abelian category.

Thus, given a construction of the triangulated category $DMT(\cS)$ of 
mixed Tate motives, one can try to obtain from it, at least
rationally, a category $MT(\cS)$ of mixed Tate motives, as the heart
of a $t$-structure on $DMT(\cS)_\Q= DMT(\cS)\otimes \Q$. It is
possible to define such a $t$-structure when the Beilinson--Soul\'e
vanishing conjecture holds, namely when
\begin{equation}\label{BSconj}
\Hom^j(\Q(0),\Q(n))=0, \ \ \ \text{ for } n>0, j\leq 0.
\end{equation}
where $\Hom^j(M,N)=\Hom(M,N[j])$ and $\Q(n)$ is the image in
$DMT(\cS)_\Q$ of the Tate object $\Z(n)$ of $DMT(\cS)$.

\smallskip

The conjecture \eqref{BSconj} holds in the case of a number field
$\K$, by results of Borel \cite{Bo} and Beilinson \cite{Bei}.
Thus, in this case it is possible to extract from $DMT(\K)_\Q$ a
tannakian category $MT(\K)$ of mixed Tate motives over $\K$.
For a number field $\K$ one has
\begin{equation}\label{ExtK}
\Ext^1(\Q(0),\Q(n))= K_{2n-1}(\K)\otimes \Q
\end{equation}
and $\Ext^2(\Q(0),\Q(n))=0$.

The category $MT(\K)$ has a fiber functor $\omega: MT(\K)\to {\rm
Vect}$, with $M \mapsto \omega(M)=\oplus_n \omega_n(M)$ where
\begin{equation}\label{omegaGr}
\omega_n(M)=\Hom(\Q(n),\Gr_{-2n}^w(M)),
\end{equation}
with $\Gr_{-2n}^w(M)=W_{-2n}(M)/W_{-2(n+1)}(M)$ the graded
structure associated to the finite increasing weight filtration
$W_\bullet$.

If $S$ is a set of finite places of $\K$, it is possible to define
the category of mixed Tate motives $MT(\O_S)$ over the set of
$S$-integers $\O_S$ of $\K$ as mixed Tate motives over $\K$ that
are unramified at each finite place $v\notin S$. The condition of
being unramified can be checked in the $\ell$-adic realization
(\cf \cite{dg} Prop.1.7). For $MT(\O_S)$ we have
\begin{equation}\label{ExtOS}
\Ext^1(\Q(0),\Q(n))=\left\{ \begin{array}{ll} K_{2n-1}(\K)\otimes
\Q & n\geq 2 \\[2mm] \O_S^*\otimes \Q & n=1 \\[2mm] 0 & n\leq 0.
\end{array} \right.
\end{equation}
and $\Ext^2(\Q(0),\Q(n))=0$.
In fact, the difference between \eqref{ExtOS} in $MT(\O_S)$ and
\eqref{ExtK} in $MT(\K)$ is the $\Ext^1(\Q(0),\Q(1))$ which is
finite dimensional in the case \eqref{ExtOS} of $S$-integers and
infinite dimensional in the case \eqref{ExtK} of $\K$.

The category $MT(\O_S)$ is a tannakian category, hence there
exists a corresponding group scheme $G_\omega=G_\omega\langle
MT(\O_S)\rangle$, given by the automorphisms of the fiber functor
$\omega$. This functor determines an equivalence of categories
between $MT(\O_S)$ and finite dimensional linear representations
of $G_\omega$. The action of $G_\omega$ on $\omega(M)$ is
functorial in $M$ and is compatible with the weight filtration.
The action on $\omega(\Q(1))=\Q$ defines a morphism $G_\omega \to
\bG_m$ and a decomposition
\begin{equation}\label{semidirGU}
G_\omega= U_\omega \rtimes \bG_m,
\end{equation}
as a semidirect product, for a unipotent affine group scheme $U_\omega$. 
The $\bG_m$
action compatible with the weight filtration determines a positive
integer grading on the Lie algebra $\Lie(U_\omega)$. The functor
$\omega$ gives an equivalence of categories between $MT(\O_S)$ and
the category of finite dimensional graded vector spaces with an
action of $\Lie(U_\omega)$ compatible with the grading.

The fact that $\Ext^2(\Q(0),\Q(n)=0$ shows that $\Lie(U_\omega)$
is freely generated by a set of homogeneous generators in degree
$n$ identified with a basis of the dual of $\Ext^1(\Q(0),\Q(n))$
(\cf Prop. 2.3 of \cite{dg}). There is however no canonical
identification between $\Lie(U_\omega)$ and the free Lie algebra
generated by the graded vector space $\oplus
\Ext^1(\Q(0),\Q(n))^\vee$.

A tannakian category $\T$ has a canonical affine $\T$-group scheme
(\cf \cite{De2} and also Section \ref{Sfundgrp} below), which one
calls the fundamental group $\pi(\T)$. The 
morphism $G_\omega \to \bG_m$ that gives the decomposition
\eqref{semidirGU} is the $\omega$-realization of
a homomorphism
\begin{equation}\label{pitoGm}
 \pi(MT(\O_S)) \to \bG_m
\end{equation}
given by the action of $\pi(MT(\O_S))$ on $\Q(1)$, and the group
$U_\omega$ is the $\omega$-realization of the kernel $U$ of
\eqref{pitoGm}. 

\medskip

We mention, in particular, the following case (\cite{dg}, \cite{3sasha4}), 
which will be relevant in our context of renormalization. 

\begin{prop}\label{SNmotives}
Consider the case of the scheme $\cS_N=\O[1/N]$ for $\K=\Q(\zeta_N)$ the 
cyclotomic field of level $N$.
For $N=3$ or $4$, the Lie algebra $\Lie(U_\omega)$ is (noncanonically)
isomorphic to the free Lie algebra with one generator in each degree $n$.  
\end{prop}

\medskip
\subsection{Motives and noncommutative geometry: analogies}\label{MotNCG}
\medskip

There is an intriguing analogy between these motivic constructions
and those of KK-theory and cyclic cohomology in noncommutative
geometry.

\smallskip

Indeed the basic steps in the
construction of the category $DM(\cS)$
parallel the basic steps in the
construction of the Kasparov bivariant
theory KK. The basic ingredients are the same,
namely   the correspondences
which, in both cases, have a finiteness property ``on
one side". One then passes in both
cases to complexes. In the case
of KK this is achieved by simply taking formal finite
differences of ``infinite" correspondences.
Moreover, the basic equivalence relations between these
``cycles" includes homotopy in very much the same
way as in the theory of motives (\cf \eg p.7 of \cite{dg}).
Also as in the theory of motives one obtains an
additive category which can be
  viewed as a ``linearization" of the category
of algebras. Finally, one should note, in the case of KK,
that a slight improvement (concerning exactness) and
a great technical simplification are obtained if
one considers ``deformations" rather than correspondences
as the basic ``cycles" of the theory, as is achieved in 
E-theory.

\smallskip

Next, when
instead of working over $\Z$ one considers the category $DM(k)_\Q$
obtained by tensorization by $\Q$, one can pursue the
analogy much further and make contact with cyclic cohomology,
where also one works rationally, with a similar role of
filtrations. There also the obtained
``linearization" of the category
of algebras is fairly explicit and simple
in noncommutative geometry. The obtained category
is just the category of $\Lambda$-modules,
based on the cyclic category $\Lambda$.
One obtains a  functor $A \to A^\natural$,
which allows one to treat algebras as
objects in an abelian category, where many tools
such as the bifunctors $\Ext^n(X,\,Y)$ 
are readily available. The key ingredient is the {\em cyclic category}.
It is a small category which has the same classifying space
as the compact group $U(1)$ (\cf \cite{Co-cycl}).

\smallskip

Finally, it is noteworthy that algebraic K-theory and
regulators already appeared in the context of
quantum field theory and noncommutative geometry in \cite{coka}.

\medskip
\subsection{Motivic fundamental groupoid}\label{Sfundgrp}
\bigskip

Grothendieck initiated the field of ``anabelian algebraic
geometry'' meant primarily as the study of the action of
absolute Galois groups like $\Gal(\bar \Q/\Q)$ on the 
profinite fundamental group of algebraic varieties (\cf
\cite{Gro-esq}). The most celebrated example is  
the projective line
minus three points. In this case, a finite cover of
$\P^1\smallsetminus \{ 0,1,\infty \}$ defines an algebraic curve.
If the projective line is considered over $\Q$, and so are the
ramification points, the curve obtained this way is defined over
$\bar\Q$, hence the absolute Galois group $\Gal(\bar \Q/\Q)$ acts. 
Bielyi's theorem shows that, in fact, all algebraic curves defined 
over $\bar \Q$ arise as coverings of the projective line ramified 
only over the points $\{ 0,1,\infty \}$.
This has the effect of realizing the absolute Galois group
as a subgroup of outer automorphisms of the profinite fundamental
group of the projective line minus three points. Motivated by Grothendieck's
``esquisse d'un programme'' \cite{Gro-esq}, Drinfel'd introduced in
the context of transformations of structures of quasi-triangular
quasi-Hopf algebras \cite{3Drin} a Grothendieck--Teichm\"uller group
$GT$, which is a pro-unipotent version of the group of
automorphisms of the fundamental group of
$\P^1\smallsetminus \{ 0,1,\infty \}$, with an
injective homomorphism $\Gal(\bar \Q/\Q) \to GT$.

\smallskip

Deligne introduced in \cite{De} a notion of ``motivic fundamental
group'' in the context of mixed motives. 
Like Grothendieck's theory of motives provides a cohomology
theory that lies behind all the known realizations, the notion of
motivic fundamental group lies behind all notions of fundamental
group developed in the algebro-geometric context. For instance, the
motivic fundamental group has as Betti realization
a pro-unipotent algebraic envelope of the nilpotent quotient of the
classical fundamental group, and as de Rham realization a unipotent
affine group scheme whose finite-dimensional representations  
classify vector bundles with nilpotent integrable connections.  
In the case of $\P^1\smallsetminus \{ 0,1,\infty \}$, the 
motivic fundamental group is an iterated extension of Tate motives.

\smallskip

For $\K$ a number field, $X$ the complement of a finite set of
rational points on a projective line over $\K$, and $x,y\in
X(\K)$, Deligne constructed in \S 13 of \cite{De} motivic path
spaces $P_{y,x}$ and motivic fundamental groups
$\pi_1^{mot}(X,x)=P_{x,x}$. One has $\pi_1^{mot}(\bG_m,x)=\Q(1)$ as
well as local monodromies $\Q(1) \to \pi_1^{mot}(X,x)$. More
generally, the motivic path spaces can be defined for a class of
unirational arithmetic varieties over a number field, \cite{De},
\S 13.

Given an embedding $\sigma: \K \hookrightarrow \C$, the
corresponding realization of $\pi_1^{mot}(X,x)$ is the algebraic
pro-unipotent envelope of the fundamental group $\pi_1(X(\C),x)$,
namely the spectrum of the commutative Hopf algebra
\begin{equation}\label{colim}
{\rm colim}\,\, \left( \Q \left[\pi_1(X(\C),x)\right]/J^N
\right)^\vee,
\end{equation}
where $J$ is the augmentation ideal of $\Q[\pi_1(X(\C),x)]$.

We recall the notion of Ind-objects, which allows one to enrich
an abelian category by adding  
inductive limits. If $\sC$ is an abelian category, 
and $\sC^\vee$ denotes the category of contravariant functors of
$\sC$ to Sets, then $\Ind(\sC)$ is defined as the full subcategory of
$\sC^\vee$ whose objects are functors of the form
$X \mapsto \varinjlim \Hom_\sC(X,X_\alpha)$,
for $\{ X_\alpha \}$ a directed system in $\sC$. 

One can use the notion above to define ``commutative algebras''
in the context of Tannakian categories. In fact, given a   
Tannakian category $\T$, one defines a commutative algebra with unit 
as an object $A$ of $\Ind(\T)$ with a product $A\otimes A \to A$ and a
unit $1\to A$ satisfying the usual axioms. The category of affine
$\T$-schemes is dual to that of commutative algebras with unit,
with $\Sp(A)$ denoting the affine $\T$-scheme associated to $A$
(\cf \cite{De}, \cite{De2}). The motivic path spaces constructed
in \cite{De} are affine $MT(\K)$-schemes, $P_{y,x}=\Sp(A_{y,x})$.
The $P_{y,x}$ form a groupoid with respect to composition of paths
\begin{equation}\label{paths}
P_{z,y}\times P_{y,x} \to P_{z,x}.
\end{equation}

In the following we consider the case of
\begin{equation}\label{minusroots}
X=\P^1 \smallsetminus V, \ \ \text{ where }  \ \ V= \{ 0, \infty\}
\cup \mu_N,
\end{equation}
with $\mu_N$ the set of $N$th roots of unity. The $P_{y,x}$ are
unramified outside of the set of places of $\K$ over a prime
dividing $N$ (\cf Proposition 4.17 of \cite{dg}). Thus, they can
be regarded as $MT(\O[1/N])$-schemes.

For such $X=\P^1 \smallsetminus V$, one first extends the 
fundamental groupoid to base points in $V$ using ``tangent directions".
One then restricts the resulting groupoid to
points in $V$. One obtains this way
 the system of $MT(\O[1/N])$-schemes $P_{y,x}$, for
$x,y\in V$, with the composition law \eqref{paths}, the local
monodromies $\Q(1) \to P_{x,x}$ and equivariance under the action
of the dihedral group $\mu_N \rtimes \Z/2$ (or of a larger
symmetry group for $N=1,2,4$).

One then considers the $\omega$-realization $\omega(P_{y,x})$. 
There are canonical paths $\gamma_{xy}\in \omega(P_{x,y})$ associated to  pairs
of points $x,y\in V$ such that $\gamma_{xy}\circ \gamma_{yz} =\gamma_{xz}$.
This gives an explicit equivalence (analogous to a Morita equivalence) between the groupoid
$\omega(P)$ and a pro-unipotent affine group scheme $\Pi$. This is
described as
\begin{equation}\label{Pigroup}
\Pi = \varprojlim \exp (\cL/\deg \geq n),
\end{equation}
where $\cL$ is the graded Lie algebra freely generated by degree
one elements $e_0, e_\zeta$ for $\zeta\in \mu_N$.

Thus, after applying the fiber functor $\omega$, the properties of the
system of the $P_{y,x}$ translate to the data of the vector space
$\Q=\omega(\Q(1))$, a copy of the group $\Pi$ for each pair
$x,y\in V$, the group law of $\Pi$ determined by the groupoid law
\eqref{paths}, the local monodromies given by Lie algebra
morphisms
$$ \Q \to \Lie(\Pi), \ \ \ \ 1 \mapsto e_x, \ \ x\in V, $$
and group homomorphisms $\alpha: \Pi \to \Pi$ for $\alpha \in
\mu_N\rtimes \Z/2$, given at the Lie algebra level by
$$ \alpha : \Lie(\Pi) \to \Lie(\Pi) \ \ \ \ \alpha: e_x \mapsto
e_{\alpha x}. $$

One restricts the above data to $V\smallsetminus\{\infty\}$.
The structure obtained this way has a group scheme of
automorphisms $H_\omega$. Its action on $\Q=\omega(\Q(1))$
determines a semidirect product decomposition
\begin{equation}\label{semidirHV}
H_\omega = V_\omega \rtimes \bG_m,
\end{equation}
as in \eqref{semidirGU}. 
Using the image of the straight path $\g_{01}$
under the action of the automorphisms, 
one can
identify $\Lie(V_\omega)$ and $\Lie(\Pi)$ 
at the level of vector spaces (Proposition 5.11, \cite{dg}), 
while the Lie bracket
on $\Lie(V_\omega)$ defines a new bracket on $\Lie(\Pi)$ described
explicitly in Prop.5.13 of \cite{dg}.

We can then consider the $G_\omega$ action on the
$\omega(P_{y,x})$. This action does depend on $x,y$. 
In particular, for the pair $0,1$, one obtains this way a homomorphism
\begin{equation}\label{GtoH}
G_\omega = U_\omega \rtimes \bG_m \longrightarrow H_\omega=
V_\omega \rtimes \bG_m,
\end{equation}
compatible with the semidirect product decomposition given by the
$\bG_m$-actions.

Little is known explicitly about the image of $\Lie(U_\omega)$ in
$\Lie(V_\omega)$. Only in the case of $N=2,3,4$ the map $U_\omega
\to V_\omega$ is known to be injective and the dimension of the
graded pieces of the image of $\Lie(U_\omega)$ in $\Lie(V_\omega)$
is then known (Theorem 5.23 and Corollary 5.25 of \cite{dg}, \cf also
Proposition \ref{SNmotives} in Section \ref{Smotives} above).

The groups $H_\omega$ and $V_\omega$ are $\omega$-realizations of
$MT(\O[1/N])$-group schemes $H$ and $V$, as in the case of
$U_\omega$ and $U$, where $V$ is the kernel of the morphism $H \to
\bG_m$ determined by the action of $H$ on $\Q(1)$.

\subsection{Expansional and multiple polylogarithms}

Passing to complex coefficients (\ie using the
Lie algebra $\C\langle\langle e_0,e_\zeta\rangle\rangle$), 
the multiple polylogarithms at roots of unity appear as
coefficients of an expansional taken with respect to the
path $\g_{01}$ in $X =\P^1\backslash\{0, \mu_N , \infty\}$
and the universal flat connection on $X$ given below in \eqref{alphampolylogs}. 
We briefly recall here this well known fact
(\cf \S 5.16 and Prop. 5.17 of
\cite{dg} and \S 2.2 of \cite{Rac}).

\smallskip 

The multiple polylogarithms are defined 
for $k_i\in \Z_{>0}$,  $0< |z_i|\leq 1$, by 
the expression
\begin{equation}\label{mpolylogs}
{\rm Li}_{\,k_1,\ldots,k_m}(z_1,z_2,\ldots,z_m) = \sum_{0<n_1<n_2<\cdots <n_m}
\frac{z_1^{n_1} z_2^{n_2} \cdots z_m^{n_m}}{n_1^{k_1} n_2^{k_2} \cdots
n_m^{k_m}}  
\end{equation}
which converges for $(k_m,|z_m|)\neq (1,1)$. 

\smallskip 

Kontsevich's formula for multiple 
zeta values as iterated integrals was generalized by Goncharov
to multiple polylogarithms using the connection 
\begin{equation}\label{alphampolylogs}
\alpha(z)dz = \sum_{a\in  \mu_N\cup \{0\}} \, \frac{dz}{z-a} \,\, e_a. 
\end{equation}
It is possible to give meaning to the expansional 
\begin{equation}\label{expansLmu2}
\gamma={\bf {\rm T}e^{\int_0^1\,\a(z)\,dz}},
\end{equation}
using a simple regularization at $0$ and $1$ (\cf \cite{dg}) 
by dropping the logarithmic terms $(\log\epsilon)^k$,
$(\log\eta)^k$ in the expansion of 
$$
\gamma={\bf {\rm T}e^{\int_\epsilon^{1-\eta}\,\a(z)\,dz}}.
$$
when $\epsilon \ra 0$ and $\eta \ra 0$.

\smallskip 

\begin{prop}\label{expmpoly} 
For $k_i>0$, the coefficient of 
$e_{\zeta_1}\,e_0^{k_1-1}\,e_{\zeta_2}  
\,e_0^{k_2-1}\ldots \,e_{\zeta_m}\,e_0^{k_m-1}$ in the 
expansional \eqref{expansLmu2} is given by 
$$
(-1)^m\,{\rm Li}_{\,k_1,\ldots,k_m}(z_1,z_2,\ldots,z_m) 
$$
where the roots of unity $z_j$ are given by $z_j=\,\zeta_j^{-1}\, 
\zeta_{j+1}$, for $j<m$ and $z_m=\zeta_m^{-1}$.
\end{prop}

Racinet used this iterated integral description to study the shuffle
relations for values of multiple polylogarithms at roots of unity
\cite{Rac}.

\bigskip
\section{The ``cosmic Galois group" of renormalization
as a motivic Galois group}\label{SGalois}
\bigskip

In this section we construct a category of equivalence classes
of equisingular flat vector bundles. This allows us to reformulate
the Riemann--Hilbert correspondence underlying perturbative renormalization 
in terms of finite dimensional
linear representations of the ``cosmic Galois group", that is, the
 group scheme $U^*$ introduced in
Section \ref{Sunivframe} above. The relation to the formulation given
in the Section \ref{Sunivframe} consists of passing to finite dimensional
representations of the group $G^*$. In fact, since $G^*$ is an
affine group scheme, there are enough such representations, and
they are specified (\cf \cite{dg}) by assigning the data of
\begin{itemize}
\item A graded vector space $E=\oplus_{n\in \Z} E_n\,,$,
\item A graded representation $\pi$ of $G$ in $E$.
\end{itemize}

\smallskip

Notice that a graded representation of $G$ in $E$ can equivalently
be described as a graded representation of $\fg$ in $E$.
Moreover, since the Lie algebra $\fg$ is positively graded, both
representations are compatible with the {\em weight} filtration
given by
\begin{equation} \label{weight}
W^{-n}(E)=\,\oplus_{m\geq n} E_m\,.
\end{equation}

\smallskip

At the group level, the corresponding representation in the
associated graded
$$
Gr^W_n=\, W^{-n}(E)/W^{-n-1}(E)\,.
$$
is the identity.

\smallskip

We now consider equisingular flat bundles, defined as follows.

\begin{defn}\label{Wconn}
Let $(E,W)$ be  a filtered vector bundle with a given
trivialization of the associated graded $Gr^W(E)$.
\begin{enumerate}
\item A $W$-connection on $E$ is a
connection $\nabla$ on $E$, which is compatible with the
filtration (\ie restricts to all $W^k(E)$) and induces the trivial
connection on the associated graded $Gr^W(E)$.
\item Two $W$-connections on $E$ are $W$-equivalent iff there exists
an automorphism of $E$ preserving the filtration, inducing the
identity on $Gr^W(E)$, and conjugating the connections.
\end{enumerate}
\end{defn}

\smallskip

Let $B$ be the principal $\bG_m$-bundle considered in Section
\ref{classif}. The above definition \ref{Wconn}
is extended to the relative case of the pair $(B,B^*)$. Namely, 
$(E,W)$ makes sense on $B$,
the connection
$\nabla$ is defined on $B^*$ and the automorphism
implementing the equivalence extends to $B$.

\smallskip

We define a category $\Ec$ of equisingular flat bundles. The {\em
objects} of $\Ec$ are the equivalence classes of pairs
$$\Theta=(E,\nabla) ,$$
where
\begin{itemize}
\item $E$ is a $\Z$-graded finite dimensional vector space.
\item $\nabla$ is an equisingular flat $W$-connection on $B^*$,
defined on the $\bG_m$-equivariant filtered vector bundle
$(\tilde{E},W)$ induced by $E$ with its weight filtration
\eqref{weight}.
\end{itemize}

\smallskip

By construction $\tilde{E}$ is the trivial bundle $B\times E$
endowed with the action of $\bG_m$ given by the grading. The
trivialization of the associated graded $Gr^W(\tilde{E})$ is
simply given by the identification with the trivial bundle with
fiber $E$. The equisingularity of $\nabla$ here means that it is
$\bG_m$-invariant and that all restrictions to sections $\sigma$
of $B$ with $\sigma(0)=y_0$ are $W$-equivalent on $B$.

\smallskip

We refer to such pairs $\Theta=(E,\nabla)$ as {\em flat
equisingular bundles}. We only retain the datum of the
$W$-equivalence class of the connection $\nabla$ on $B$
as explained above.

\smallskip

Given two flat equisingular bundles $\Theta$, $\Theta'$ we define
the {\em morphisms}
$$
T\in{\rm Hom}(\Theta, \Theta')
$$
in the category $\Ec$ as linear maps $ T\,: \,E\to E'\,, $
compatible with the grading, fulfilling the condition that the
following $W$-connections $\nabla_j$, $j=1,2$, on
$\tilde{E'}\oplus \tilde{E}$ are $W$-equivalent (on $B$),
\begin{equation}\label{morphisms}
\nabla_1 = \left[ \begin{matrix}\nabla' &T\,\nabla-\,\nabla'\,T
\cr 0 &\nabla \cr \end{matrix} \right] \, \sim \nabla_2 = \left[
\begin{matrix}\nabla' &0 \cr 0 &\nabla \cr \end{matrix} \right]
\,.
\end{equation}

\smallskip

Notice that this is well defined, since condition
\eqref{morphisms} is independent of the choice of representatives
for the connections $\nabla$ and $\nabla'$. The condition
\eqref{morphisms} is obtained by conjugating $\nabla_2$ by the
unipotent matrix
$$
\left[ \begin{matrix}1 &T \cr 0 &1 \cr \end{matrix}\right]\,.
$$

\smallskip

In all the above we worked over $\C$, with convergent Laurent
series. However, much of it can be rephrased with formal Laurent
series. Since the universal singular frame is given in rational
terms by proposition \ref{univsin}, the results of this section 
hold over any field $k$ of characteristic zero and in 
particular over $\Q$. 

\smallskip

For  $\Theta=(E,\nabla)$, we set $\omega(\Theta)=E$ and we view
$\omega$ as a functor from the category of equisingular flat
bundles to the category of vector spaces. We then have the
following result.

\smallskip

\begin{thm} \label{tann}
Let $\Ec$ be the category of equisingular flat bundles defined
above, over a field $k$ of characteristic zero.
\begin{enumerate}
\item $\Ec$ is a Tannakian category.
\item The functor $\omega$ is a fiber functor.
\item $\Ec$ is equivalent to the category of finite dimensional
representations of $U^*$.
\end{enumerate}
\end{thm}

\underline{\em Proof.}
Let $E$ be a finite dimensional graded vector space over $k$. We
consider the unipotent algebraic group $G$ such that $G(k)$
consists of endomorphisms $S \in {\rm End}(E)$ satisfying the conditions 
\begin{equation}\label{SW1}
S \,W_{-n}(E)\subset W_{-n}(E),
\end{equation}
where $W_\cdot(E)$ is the weight filtration, and
\begin{equation}\label{SW2}
S|_{Gr_n} =1 ,
\end{equation} 
where $Gr_n$ denote the associated graded.

\smallskip

The group $G$ can be identified with the unipotent group of upper
triangular matrices. Its Lie algebra is then identified with strictly
upper triangular matrices. 

\smallskip

The following is a direct translation between $W$-connections and
$G$-valued connections.

\begin{prop}\label{WGvalued}
Let $(E,\nabla)$ be an object in $\Ec$.
\begin{enumerate}
\item $\nabla$ defines an equisingular $G$-valued connection, for $G$
as above.
\item All equisingular $G$-valued connections are obtained this way.
\item This bijection preserves equivalence.
\end{enumerate}
\end{prop}

In fact, since $W$-connections are compatible with the filtration and
trivial on the associated graded, they are obtained by adding a
$Lie(G)$-valued 1-form to the trivial connection. Similarly,
$W$-equivalence is given by the equivalence as in Definition \ref{GequivP}.

\smallskip

\begin{lem}\label{Objequiv}
Let $\Theta=(E,\nabla)$ be an object in $\Ec$. 
Then there exists a unique
representation $\rho=\rho_\Theta$ of $U^*$ in $E$, such that
\begin{equation}\label{Drhouniv}
D \rho(\gamma_U) \simeq \nabla,
\end{equation}
where $\gamma_U$ is the universal singular frame.
Given a representation $\rho$ of $U^*$ in $E$, there exists a
$\nabla$, unique up to equivalence, such that $(E,\nabla)$ is an
object in $\Ec$ and $\nabla$ satisfies \eqref{Drhouniv}.
\end{lem}

\underline{\em Proof of Lemma.} Let $G$ be as above. By Proposition
\ref{WGvalued} we view $\nabla$ as a $G$-valued connection.
By applying Theorem \ref{rh1} we get a unique element $\beta \in {\rm
Lie}(G)$ such that equation \eqref{Drhouniv} holds. For the second
statement, notice that \eqref{logder} gives a rational expression for
the operator $D$. This, together with the fact that the coefficients
of the universal singular frame are rational, implies that we obtain a
rational $\nabla$.
$\Box$

\begin{lem}\label{Morequiv}
Let $(E,\nabla)$ be an object in $\Ec$.
\begin{enumerate}
\item For any $S\in {\rm Aut}(E)$ compatible with the grading,
$S\,\nabla\, S^{-1}$ is an equisingular connection.
\item $\rho_{(E,S\,\nabla\, S^{-1})} = S\, \rho_{(E,\nabla)}\, S^{-1}$.
\item $S\, \nabla\, S^{-1} \sim \nabla \Leftrightarrow [\rho_{(E,\nabla)}, S]=0$.
\end{enumerate}
\end{lem}

\underline{\em Proof of Lemma.} The equisingular condition is
satisfied, since the $\bG_m$-invariance follows by compatibility with
the grading and restriction to sections satisfies
$$ \sigma^*(S \, \nabla\, S^{-1})= S \,  \sigma^*(\nabla) \, S^{-1}. $$
The second statement follows by compatibility of $S$ with the
grading. In fact, we have
$$  S \,\,{\bf {\rm T}e^{-\frac{1}{z}\,
 \int^{v}_0\,u^Y(\beta)\,\frac{du}{u}}}\,\, S^{-1} = {\bf {\rm T}e^{-\frac{1}{z}\,
 \int^{v}_0\,u^Y(S \, \beta\, S^{-1})\,\frac{du}{u}}}. $$
The third statement follows immediately from the second, since
equivalence corresponds to having the same $\beta$, by Theorem
\ref{rh}. $\Box$

\smallskip

\begin{prop}\label{AutS}
Let $\Theta=(E,\nabla)$ and $\Theta'=(E',\nabla')$ be object of $\Ec$. Let $T: E \to
E'$ be a linear map compatible with the grading. Then the following
two conditions are equivalent.
\begin{enumerate}
\item $T\in {\rm Hom} (\Theta,\Theta')$;
\item $T\, \rho_\Theta = \rho_{\Theta'}\, T$.
\end{enumerate}
\end{prop}

\underline{\em Proof of Proposition.} Let
$$ S = \left( \begin{array}{cc} 1 & T \\ 0 & 1 \end{array}\right). $$
By construction, $S$ is an automorphism of $E'\oplus E$, compatible
with the grading. By (3) of the previous Lemma, we have
$$ S\,\,  \left( \begin{array}{cc} \nabla' & 0 \\ 0 & \nabla
\end{array}\right) \,\, S^{-1} \sim  \left( \begin{array}{cc} \nabla' & 0 \\ 0 & \nabla
\end{array}\right) $$
if and only if 
$$ \left( \begin{array}{cc} \beta' & 0 \\ 0 & \beta \end{array}\right)
\,\, S = S \,\, \left( \begin{array}{cc} \beta' & 0 \\ 0 & \beta
\end{array}\right). $$
This holds if and only if $\beta'\,T= T\, \beta$. $\Box$

\smallskip

Finally, we check that the tensor product structures are compatible. 
We have 
$$ (E,\nabla)\otimes (E',\nabla') = (E\otimes E', \nabla \otimes 1 + 1
\otimes \nabla'). $$
The equisingularity of the resulting connection comes from the
functoriality of the construction.

\smallskip

We check that the functor $\rho \mapsto D\rho(\gamma_U)$
constructed above, from the category of representations of $U^*$ to
$\Ec$, is compatible with tensor products. This follows by the
explicit formula
$$ {\bf {\rm T}e^{-\frac{1}{z}\,
 \int^{v}_0\,u^Y(\beta\otimes 1 + 1 \otimes \beta')\,\frac{du}{u}}}
= {\bf {\rm T}e^{-\frac{1}{z}\,
 \int^{v}_0\,u^Y(\beta)\,\frac{du}{u}}} \,\otimes \, {\bf {\rm T}e^{-\frac{1}{z}\,
 \int^{v}_0\,u^Y(\beta')\,\frac{du}{u}}}. $$
On morphisms, it is sufficient to check the compatibility on $1\otimes
T$ and $T\otimes 1$.

\medskip

We have shown that the tensor category $\Ec$ is equivalent to the
category of finite dimensional representations of $U^*$.
The first two statements of the Theorem then follow from the third
(\cf \cite{De2}).

$\Box$

\medskip

\smallskip

For each integer $n \in \Z$, we then define an object $\Q(n)$ in
the category $\Ec$ of equisingular flat bundles as the trivial
bundle given by a one-dimensional $\Q$-vector space placed in
degree $n$, endowed with the trivial connection on the associated
bundle over $B$.

\smallskip

For any flat equisingular bundle $\Theta$ let
$$
\omega_n(\Theta)=\,{\rm Hom}(\Q(n), \Gr_{-n}^W(\Theta))\,,
$$
and notice that $\omega = \oplus \,\omega_n$.

\bigskip

The group $U^*$ can be regarded as a motivic Galois group. One
has, for instance, the following identification  (\cite{3sasha4},
\cite{dg}, \cf also Proposition \ref{SNmotives} in Section \ref{Smotives} above).

\begin{prop} \label{motgal} There is a
(non-canonical)  isomorphism
\begin{equation}\label{MotU}
U^* \sim G_{\cM_T}(\fO) \,.
\end{equation}
of the affine group scheme $U^*$ with the motivic Galois group
$G_{\cM_T}(\fO) $
 of  the scheme $S_4$ of $4$-cyclotomic integers.
\end{prop}

\smallskip

It is important here to stress the fact (\cf the ``mise en garde"
of \cite{dg}) that there is so far no ``canonical" choice of a
free basis in the Lie algebra of the above motivic Galois group so
that the above isomorphism still requires making a large number of
non-canonical choices. In particular it is premature to assert
that the above category of equisingular flat bundles is directly
related to the category of $4$-cyclotomic Tate motives. The
isomorphism \eqref{MotU} does not determine the scheme $S_4$
uniquely. In fact, a similar isomorphism holds with $S_3$ the
scheme of 3-cyclotomic integers.

\smallskip
On the other hand, when considering the category $\cM_T$ in
relation to physics, inverting the prime $2$ is relevant  to the
definition of geometry in terms of $K$-homology, which is at the
center stage in noncommutative geometry. We recall, in that
respect, that it is only after inverting the prime $2$ that (in
sufficiently high dimension) a manifold structure on a simply
connected homotopy type is determined by the $K$-homology
fundamental class.

\smallskip

Moreover, passing from $\Q$ to a field with a complex place, such
as the above cyclotomic fields $k$, allows for the existence of
non-trivial regulators for all algebraic $K$-theory groups
$K_{2n-1}(k)$. It is noteworthy also that algebraic K-theory and
regulators already appeared in the context of quantum field theory
and NCG in \cite{coka}. The appearance of multiple polylogarithms
in the coefficients of divergences in QFT, discovered by
Broadhurst and Kreimer (\cite{B}, \cite{BK}), as well as recent
considerations of Kreimer on analogies between residues of quantum
fields and variations of mixed Hodge--Tate structures associated
to polylogarithms (\cf \cite{Kr2}), suggest the existence for the
above category of equisingular flat bundles of suitable Hodge-Tate
realizations given by a specific choice of Quantum Field Theory.

 \bigskip
 \section{The wild fundamental group}\label{Snonpert} 
\bigskip

We return here to the general discussion of the Riemann--Hilbert
correspondence in the irregular case, which we began in Section
\ref{Sirreg}. 

\smallskip

The universal  differential Galois group $\cG$ of
\eqref{Gspiltseq} governs the irregular Riemann--Hilbert
correspondence at the formal level, namely over the differential
field $\C((z))$ of formal Laurent series. In general, when passing to
the non-formal level, over convergent Laurent series $\C(\{z\})$, the
corresponding universal differential Galois group acquires additional
generators, which depend upon resummation of divergent series and are
related to the Stokes phenomenon (see \eg the last section of
\cite{vdp} for a brief overview). 

\smallskip

At first, it may then seem surprising that, in the
Riemann--Hilbert correspondence underlying perturbative
renormalization that we derived in Sections \ref{Sunivframe} and
\ref{SGalois}, we found the same affine group scheme $U^*$, regardless
of whether we work over $\C((z))$ or over $\C(\{z\})$. This is, in
fact, not quite so strange. There are known classes of
equations (\cf \eg Proposition 3.40 of \cite{PuSi}) for which the 
differential Galois group is the same over $\C((z))$ and over
$\C(\{z\})$. Moreover, in our particular case, it is not hard to
understand the conceptual reason why this should be the case. It can
be traced to the result of Proposition \ref{expprop0}, which shows
that, due to the pro-unipotent nature of the group $G$, the
expansional formula is in fact algebraic. Thus, when considering
differential systems with $G$-valued connections, one can pass from
the formal to the non-formal case (\cf also Proposition
\ref{trimono}). 

\smallskip

This means that the Stokes part of Ramis' wild fundamental group will
only appear, in the context of renormalization, when one incorporates
non-perturbative effects. In fact, in the non-perturbative setting, 
the group $G={\rm Difg}(\sT)$ of diffeographisms, or rather its image
in the group of formal diffeomorphisms as discussed in Section
\ref{difgdiff}, gets upgraded to actual diffeomorphisms analytic in
sectors.
In this section we discuss briefly some issues related to the wild
fundamental group and the non-perturbative effects.

\smallskip

  The aspect of the Riemann--Hilbert problem,
 which is relevant to the non-perturbative case, is related to
 methods of ``summation'' of divergent series modulo functions with
 exponential decrease of a certain order, namely Borel summability,
 or more generally multisummability (a good reference is \eg
 \cite{3Ramis2}.)

 \smallskip

  In this case, the
 local wild fundamental group is obtained via the following procedure
 (\cf \cite{3Ramis}). The way to pass from formal to
 actual solutions consists of applying a
 suitable process of summability to formal solutions \eqref{formalSol}.

 \smallskip

  The method of Borel summability is
 derived from the well known fact that, if a formal series
 \begin{equation}\label{formalser}
  \hat f (z)= \sum_{n=0}^\infty f_n \, z^n
 \end{equation}
 is convergent on some disk, with $f(z)=S \hat f (z)$ the sum of the
 series \eqref{formalser} defining a holomorphic function, then the formal
 Borel transform
 \begin{equation}\label{formalborel}
 \hat B \hat f (w) = \sum_{n=1}^\infty \frac{f_n}{(n-1)!} w^n
 \end{equation}
 has infinite radius of convergence and the sum $b(w):= S\hat B \hat
 f(w)$ has the property that its Laplace transform recovers the
 original function $f$,
 \begin{equation}\label{LaplaceB}
 f(z)= \Lc (b)(z)= \int_0^\infty b(w) \, e^{-w/z} \, dw,
 \end{equation}
 that is, $S \hat f(z)=(\Lc \circ S \circ \hat B) \hat f (z)$.
 The advantage of this procedure is that it continues to make sense for
 a class of (Borel summable) divergent series, for which a ``sum'' can
 be defined by the procedure
 \begin{equation}\label{Bsummation}
  f(z):=(\Lc \circ S \circ \hat B) \hat f (z).
 \end{equation}
 Very useful generalizations of \eqref{Bsummation} include replacing
 integration along the positive real axis in \eqref{LaplaceB} with
 another oriented half line $h$ in $\C$,
 $$ \Lc_h\, (b)(z)= \int_h \, b(w) \, e^{-w/z} \, dw, $$
 as well as a more refined notion of Borel summability that involves
 ramification
 \begin{equation}\label{ramif}
 \rho_k(f)\, (z) = f(z^{1/k}),
 \end{equation}
 with $\hat B_k=\rho_k^{-1} \, \hat B \, \rho_k$ and
 $\Lc_{k,h}=\rho_k^{-1} \, \Lc_{h^k}\, \rho_k$, with corresponding
 summation operators $S_h:= \Lc_h \circ S \circ \hat B$ and
 $S_{k,h}:=\Lc_{k,h} \circ S \circ \hat B_k$.
 A formal series \eqref{formalser} is Borel $k$-summable in the
 direction $h$ if $\hat B_k \hat f$ is a convergent series such that
 $S \hat B_k\hat f$ can be continued analytically on an angular sector
 at the origin bisected by $h$ to a holomorphic function exponentially
 of order at most $k$.

 \smallskip

  The condition of $k$-summability can be more conveniently expressed in
 terms of an estimate on the remainder of the series
 \begin{equation}\label{ksumestimate}
 \left| f(x) - \sum_{n< N} a_n x^n \right| \leq c\,\, A^n \Gamma(1+N/k) \,
 |x|^n ,
 \end{equation}
 on sectors of opening at least $\pi/k$. This corresponds to the case
 where the Newton polygon has one edge of slope $k$.

 \smallskip

  There are formal series that fail to be Borel $k$-summable for any
 $k>0$. Typically the lack of summability arises from the fact that the
 formal series is a combination of parts that are summable, but for
 different values of $k$ (\cf \cite{3Ramis2}). This is taken care of by
 a suitable notion of {\em multisummability} that involves
 iterating the Borel summation process. This way, one can sum a
 formal series $\hat f$ that is $(k_1,\ldots,k_r)$-multisummable in the
 direction $h$ by
 \begin{equation}\label{multisum}
 f(x):=S_{k_1,\ldots, k_r;h}\, \hat f,
 \end{equation}
 with the summation operator
 \begin{equation}\label{sumopmult}
 S_{k_1,\ldots,k_r;h}= \Lc_{\kappa_1,d}\cdots \Lc_{\kappa_r,d} S \hat
 B_{\kappa_r}\cdots \hat B_{\kappa_1},
 \end{equation}
 for $1/k_i = 1/\kappa_1 + \cdots +1/\kappa_i$ and $i=1,\ldots, r$.

 \smallskip

 Actual solutions of a differential system \eqref{ODE} with
 \eqref{germA} can then be obtained from formal solutions of the
 form \eqref{formalSol}, in the form
 \begin{equation}\label{actualSol}
 F_h (x)= H_d(u) u^{\nu L} e^{Q(1/u)}, 
 \end{equation}
 with $u^\nu =z$, for some $\nu\in\N^\times$,
 by applying summation operators $S_{k_1,\ldots, k_r;h}$
 to $\hat H$, indexed by the positive slopes $k_1>k_2>\ldots >k_r>0$ of
 the Newton polygon of the equation, and with the half line $h$ varying
 among all but a finite number of directions in $\C$. The singular
 directions are the jumps between different determinations on angular
 sectors, and correspond to the Stokes phenomenon. This further
 contributes to the divergence/ambiguity principle already illustrated
 in \eqref{divamb}.

 \smallskip

 We have corresponding summation operators
 \begin{equation}\label{determinations}
 f^\pm_\epsilon (x) = S_{k_1,\ldots, k_r;h_\epsilon^\pm} \, \hat f(x),
 \end{equation}
 along directions $h_\epsilon^\pm$ close to $h$,
 and a corresponding Stokes operator
 $$ {\rm St}_{h}  = (S_{k_1,\ldots, k_r;h}^+)^{-1}\,
 S_{k_1,\ldots, k_r;h}^- \,\,. $$
 These operators can be interpreted as monodromies associated to the
 singular directions. They are unipotent, hence they admit a
 logarithm. These $\log {\rm St}_h$ are related to Ecalle's alien
 derivations (\cf \eg \cite{3CNP}, \cite{3Ecalle}).

 \smallskip

The wild fundamental group (\cf \cite{3Ramis}) is then obtained by
considering a semidirect product of an affine group scheme $\Nc$,
which contains the affine group scheme generated by the Stokes
operators ${\rm St}_{h}$, by the affine group scheme $\cG$ of the
formal case,
 \begin{equation}\label{piwild}
 \pi_1^{wild}(\Delta^*)= \Nc\rtimes \cG.
 \end{equation}

\smallskip

At the Lie algebra level, one considers 
a free Lie algebra ${\mathcal R}$
(the ``resurgent Lie algebra'') generated by symbols $\delta_{(q,h)}$ with
$q\in \Ec$ and $h\in\R$ such that $re^{ih}$ is a direction of maximal
decrease of $\exp(\int q \frac{dz}{z})$ (these 
correspond to the alien derivations). There are
compatible actions of the exponential torus ${\mathcal T}$ and of the
formal monodromy $\gamma$ on ${\mathcal R}$ by
 \begin{equation}\label{actTZ}
 \begin{array}{l}
 \tau \exp(\delta_{(q,h)}) \tau^{-1} = \exp(\tau(q) \delta_{(q,h)}),
 \\[2mm]
 \gamma \exp(\delta_{(q,h)}) \gamma^{-1}=
 \exp(\delta_{(q,h -2\pi i)}). \end{array}
 \end{equation}
The Lie algebra $\Lie \,\Nc$ is isomorphic to a certain completion of
${\mathcal R}$ as a projective limit (\cf Theorem 6.3 of \cite{vdp}). 

 \smallskip

  The structure \eqref{piwild} of the wild fundamental group
 reflects the fact that, while the algebraic hull $\bar\Z$ corresponds to the
 formal monodromy along a nontrivial loop in an infinitesimal
 punctured disk around the 
 origin, due to the presence of singularities that accumulate at the
 origin, when considering Borel transforms, the monodromy along a loop
 in a finite disk also picks up monodromies around all the singular
 points near the origin. The logarithms of these monodromies correspond
 to the alien derivations.

 \smallskip

  The main result of \cite{3Ramis} on the wild Riemann--Hilbert
 correspondence is that again there is an equivalence of
 categories,
 this time between germs of meromorphic connections at the
 origin (without the regular singular assumption) and finite
 dimensional linear representations of the wild fundamental group
 \eqref{piwild}. 

\medskip

Even though we have seen in Section \ref{SGalois} that only an analog
of the exponential torus part of the wild fundamental group appears in
the Riemann--Hilbert correspondence underlying perturbative
renormalization, still the Stokes part will play a role when
non-perturbative effects are taken into account. In fact,
 already in its simplest form \eqref{Bsummation},
 the method of Borel summation is well known in
 QFT, as a method for evaluating divergent formal series $\hat f(g)$ in
 the coupling constants. In certain theories (super-renormalizable
 $g\phi^4$ and Yukawa
 theories) the formal series $\hat f(g)$ has the property that
 its formal Borel transform $\hat B \hat f(g)$ is convergent, while
 in more general situations one may have to use other $k$-summabilities
 or multisummability. Already in the cases with $\hat B \hat f(g)$
 convergent, however, one can see  that $\hat f(g)$ need
 not be Borel summable in the direction $h=[0,\infty)$, due to the fact
 that the function $S\hat B \hat
 f(g)$ acquires singularities on the positive real axis. Such
 singularities reflect the presence of {\em nonperturbative effects},
 for instance in the presence of tunneling between different
 vacua, or when the perturbative vacuum is really a metastable state
 (\cf \eg \cite{3Parisi1}).

 \smallskip

  In many cases of physical interest (\cf \eg
 \cite{3Parisi1}--\cite{3Parisi4}), singularities in the Borel plane
 appear along the positive real axis, namely $h=\R_+$ is a
 Stokes line. For physical reasons one wants a summation method that
 yields a real valued sum, hence it is necessary to sum ``through''
 the infinitely near singularities on the real line.
 In the linear case, by the method of Martinet--Ramis \cite{3Ramis},
 one can sum along directions near the Stokes line, and correct the
 result using the square root of the Stokes operator. In the
 nonlinear case, however, the procedure of summing along Stokes
 directions becomes much more delicate (\cf \eg \cite{3Costin}).

\medskip

In the setting of renormalization, in addition to the perturbative
case analyzed in CK \cite{3CK1}--\cite{3cknew}, there are two possible
ways to proceed, in order to account for the nonperturbative effects
and still obtain a geometric description for the nonperturbative
theory. These are illustrated in the diagram: 
 {\small
 \begin{eqnarray}
 \diagram
 \text{Unrenormalized perturbative} \rto^{g_{{\rm eff}}(z)}
 \dto^{\text{Birkhoff}} &
 \text{Unrenormalized nonperturbative} \dto^{\text{Birkhoff}} \\
 \text{Renormalized perturbative} \rto^{g_{{\rm eff}}^+(0)} &
 \text{Renormalized nonperturbative}
 \enddiagram
 \label{pertdiagr}
 \end{eqnarray} }
On the left hand side, the vertical arrow corresponds to the result of
 CK expressing perturbative renormalization in terms of
 the Birkhoff decomposition \eqref{Birkgeff}, where $g_{{\rm
 eff}}^+(0)$ is the effective coupling of the renormalized perturbative
 theory. The bottom horizontal arrow introduces the nonperturbative
 effects by applying Borel summation techniques to the formal series
 $g_{{\rm eff}}^+(0)$. On the other hand, the upper horizontal arrow
 corresponds to applying a suitable process of summability to the
 unrenormalized effective coupling constant $g_{\rm eff}(z)$, viewed
 as a power series in $g$, hence replacing formal diffeomorphisms by
 germs of actual diffeomorphisms analytic in sectors. The right
 vertical arrow then yields the renormalized nonperturbative theory by
 applying a Birkhoff decomposition in the group of germs of analytic
 diffeomorphisms. This type of Birkhoff decomposition was investigated
 by Menous \cite{3Menous}, who proved its existence in the non-formal case
for several classes of diffeomorphisms, relevant to non-perturbative renormalization.

\bigskip
\section{Questions and directions}
\bigskip

In this section we discuss some possible further directions that
complement and continue along the lines of the results presented in
this paper. Some of these questions lead naturally to other topics,
like noncommutative geometry at the archimedean primes, which will be
treated elsewhere. Other questions are more closely related to the
issue of renormalization, like incorporating nonperturbative effects,
or the crucial question of the relation to noncommutative geometry via
the local index formula, which leads to the idea of an underlying
renormalization of the geometry by effect of the divergences of
quantum field theory.

\medskip
\subsection{Renormalization of geometries}
\medskip

In this paper we have shown that there is a universal affine group
scheme $U^*$, the ``cosmic Galois group", that maps to the group of diffeographisms ${\rm
Difg}(\sT)$ of a given physical theory $\sT$, hence acting on the 
set of physical constants, with the renormalization group action
determined by a canonical one-parameter subgroup of $U^*$. We
illustrated explicitly how all this happens in the sufficiently
generic case of $\sT=\phi^3_6$, the $\phi^3$ theory in dimension
$D=6$. 

\smallskip

Some delicate issues arise, however, when one wishes to apply a
similar setting to gauge theories. First of all a gauge theory may
appear to be non-renormalizable, unless one handles the gauge degrees
of freedom by passing to a suitable BRS cohomology. This means that a
reformulation of the main result is needed, where the Hopf algebra of
the theory is replaced by a suitable cohomological version. 

\smallskip

Another important point in trying to extend our results to a gauge
theoretic setting, regards the chiral case, where one faces the
technical issue of how to treat the $\gamma_5$ within the dimensional
regularization and minimal subtraction scheme. In fact, in dimension
$D=4$, the symbol $\gamma_5$ indicates the product
\begin{equation}\label{gamma5}
 \gamma_5= i \gamma^0\gamma^1\gamma^2\gamma^3, 
\end{equation}
where the $\gamma^\mu$ satisfy the Clifford relations 
\begin{equation}\label{Cliff}
\{ \gamma^\mu,\gamma^\nu \} = 2 g^{\mu\nu}\, I, \ \ \text{ with } \ \
\Tr(I)=4, 
\end{equation}
and $\gamma_5$
anticommutes with them, 
\begin{equation}\label{anticomm}
\{ \gamma_5, \gamma^\mu \} =0.
\end{equation}

\smallskip

It is well known that, when one complexifies the dimension around a
critical dimension $D$, the naive prescription which formally sets 
$\gamma_5$ to still anticommute with symbols $\gamma^\mu$ while
keeping the cyclicity of the trace is not
consistent and produces contradictions (\cite{Collins}, \S 13.2). Even
the very optimistic but unproven  
claim that the ambiguities introduced by this naive prescription 
should be always proportional to the coefficient of the chiral gauge 
anomaly would restrict the validity of the naive approach to
theories with cancellation of anomalies.   

\smallskip

There are better strategies that allow one to handle the $\gamma_5$
within the Dim-Reg scheme (see \cite{MSR} for a recent detailed
treatment of this issue). One approach (\cf Collins \cite{Collins} \S
4.6 and \S 13) consists of providing an explicit construction of
an infinite family of gamma matrices $\gamma^\mu$, $\mu\in \N$,
satisfying \eqref{Cliff}. These are given by infinite rank matrices.
The definition of $\gamma_5$, for complex
dimension $d\neq 4$, is then still given through the
product \eqref{gamma5} of the first four gamma matrices.
Up to dropping the anticommutativity relation \eqref{anticomm} (\cf 't
Hooft--Veltman \cite{HoVe}) it can be shown that this definition is
consistent, though not fully Lorentz invariant, due to the
preferred choice of these spacetime dimensions.  
The Breitenlohner--Maison approach (\cf \cite{BM}, \cite{MSR}) does
not give an explicit expression for the gamma matrices in complexified
dimension, but defines them (and the $\gamma_5$ given by
\eqref{gamma5}) through their formal properties.  
Finally D. Kreimer in \cite{dirkano} produces a scheme in which 
$\gamma_5$ still anticommutes with $\gamma^\mu$ but the trace
is no longer cyclic. His scheme is presumably equivalent to the 
BM-scheme (\cf \cite{dirkano} section 5).

\smallskip

The issue of treating the gamma matrices in the Dim-Reg and
minimal subtraction scheme is also related to the important question
of the relation between our results on perturbative renormalization
and noncommutative geometry, especially through the local index
formula. 

\smallskip

The explicit computation in Proposition \ref{univsin} of the
coefficients of the universal singular frame is a concrete starting
point for understanding this relation. The next necessary step is
how to include the Dirac operator, hence the problem of the gamma
matrices. 
In this respect, it should also be mentioned that the local index
formula of \cite{cmindex} is closely related to anomalies (\cf
\eg \cite{Paycha}). 
From a more conceptual standpoint, the connection to
the local index formula seems to suggest that the procedure of
renormalization in quantum field theory should in fact be thought of
as a ``renormalization of the geometry''. The formulation of
Riemannian spin geometry in the setting of noncommutative geometry, in
fact, has already built in the possibility of considering a geometric
space at dimensions that are complex numbers rather than
integers. This is seen through the dimension spectrum, which is the
set of points in the complex plane at which a space manifests itself
with a nontrivial geometry. There are examples where the dimension
spectrum contains points off the real lines (\eg the case of Cantor
sets), but here one is rather looking for something like a deformation
of the geometry in a small neighborhood of a point of the dimension
spectrum, which would reflect the Dim-Reg procedure. 

\smallskip

The possibility of recasting the Dim-Reg procedure in such setting is
intriguing, due to the possibility of extending the results to curved
spacetimes as well as to actual noncommutative spaces, such as those
underlying a geometric interpretation of the Standard Model
(\cite{CoSM}, \cite{ChCo}).  

\smallskip

There is another, completely different, source of inspiration for the 
idea of deforming geometric spaces to complex dimension. In arithmetic
geometry, the Beilinson conjectures relate the values and orders of
vanishing at integer points of the motivic $L$-functions of algebraic
varieties to periods, namely numbers obtained by integration of
algebraic differential forms on algebraic varieties (\cf \eg
\cite{KoZa}). It is at least 
extremely suggestive to imagine that the values at non-integer points
may correspond to a dimensional regularization of algebraic
varieties and periods.

\medskip
\subsection{Nonperturbative effects}
\medskip

In the passage from the perturbative to the nonperturbative theory
 described by the two horizontal arrows of diagram \eqref{pertdiagr},
 it is crucial to understand 
 the Stokes' phenomena associated to the formal series $g_{\rm
 eff}(g,z)$ and $g_{{\rm eff}}^+(g,0)$. In particular, it is possible
 to apply Ecalle's ``alien calculus'' to the formal diffeomorphisms
 $$ g_{\rm eff}(g,z) =  \left( g + \quad
 \sum_{\hbox{\psfig{figure=Fig1aphy.eps}}}
 \quad g^{2\ell + 1} \,
 \frac{U(\G)}{S(\G)} \right) \left( 1 - \quad
 \sum_{\hbox{\psfig{figure=Fig1bphy.eps}}} \quad g^{2\ell} \,
 \frac{U(\G)}{S(\G)} \right)^{-3/2}. $$
 
 \smallskip

  There is, in fact, a way of constructing a set of invariants $\{
 A_{\omega}(z) \}$ of the formal diffeomorphism
 $g_{\rm eff}(\cdot,z)$ up to conjugacy by analytic diffeomorphisms
 tangent to the identity. This can be achieved by considering a
 formal solution of the difference equation 
\begin{equation}\label{ueq}
 x_z(u+1)= g_{{\rm eff}}(x_z(u),z),  
\end{equation}
 defined after a change of variables $u\sim 1/g$. Equation \eqref{ueq}
has the effect of conjugating $g_{{\rm eff}}$ to
a homographic transformation.
 The solution $x_z$ satisfies the {\em
 bridge equation} (\cf \cite{3Ecalle} \cite{3Fauvet})
 \begin{equation}\label{bridge}
  \dot\Delta_{\omega} x_z = A_{\omega}(z)\, \partial_u \,
 x_z,
 \end{equation}
 which relates alien derivations $\dot\Delta_{\omega}$
 and ordinary derivatives and provides the
 invariants $\{ A_{\omega}(z) \}$, where $\omega$ parameterizes 
the Stokes directions. Via the analysis of the
bridge equation 
 \eqref{bridge}, one can investigate the persistence at $z=0$
 of Stokes' phenomena induced by $z\neq 0$ (\cf \cite{3Fauvet}),
 similarly to what happens already at the perturbative level in the
 case  of the renormalization group
 $F_t=\exp(t\beta)$ at $z=0$, induced via the limit formula
 \eqref{Ren9} by ``instantonic effects'' (\cf \eqref{Ren14}) at
 $z\neq 0$. In this respect, Fr\'ed\'eric
 Fauvet noticed a formal analogy between  the bridge equation
 \eqref{bridge} and the action on \eqref{Ren16}
 of the derivations $\partial_\Gamma$, for $\Gamma$ a 1PI graph with
 two or three external legs, given by
 $$ \partial_\Gamma \, g_{{\rm eff}} = \rho_\Gamma\, g^{2\ell +1}
 \frac{\partial}{\partial g}  \, g_{{\rm eff}}, $$
 where $\rho_\Gamma=3/2$ for 2-point graphs, $\rho_\Gamma=1$ for
 3-point graphs and $\ell=L(\Gamma)$ is the loop number (\cf
 \cite{3cknew} eq.(34)).

 \smallskip

  Moreover, if the formal series $g_{\rm eff}(g,z)$ is
 multisummable, for some multi-index $(k_1,\ldots, k_r)$ with $k_1 >
 \cdots > k_r >0$, then the corresponding sums \eqref{multisum}
 are defined for almost all the directions $h$ in the plane of the
 complexified coupling constant. At the critical directions there are
 corresponding Stokes operators ${\rm St}_h$
 $$ {\rm St}_h(z) : g_{{\rm eff}}(g,z) \mapsto
 \sigma_h(g,z) \, g_{{\rm eff}}(g,z). $$
 These can be used to obtain representations $\rho_z$ of (a
 suitable completion of) the wild fundamental group
 $\pi_1^{wild}(\Delta^*)$ in the group of analytic diffeomorphisms
 tangent to the identity.
  Under the wild Riemann--Hilbert
 correspondence, these data acquire a geometric interpretation in the
 form of a nonlinear principal bundle over the open set $\C^*$ in the
 plane of the complexified
 coupling constant,
 with local trivializations over sectors and transition functions given
 by the $\sigma_h(g,z)$, with a meromorphic connection
 locally of the form
 $\sigma_h^{-1} \, A \, \sigma_h +
 \sigma_h^{-1}\,\,d\sigma_h$. This should be understood as
 a microbundle connection. In fact, in passing from the case of finite
 dimensional linear representations to local diffeomorphisms, it is
 necessary to work with a suitable completion of the wild fundamental
 group, corresponding to the fact that there are infinitely many
 alien derivations in a direction $h$.

\medskip
\subsection{The field of physical constants}
\medskip

 The computations ordinarily performed by physicists show that 
 many of the ``constants" that occur in quantum field theory, such as
 the coupling constants $g$ of the fundamental interactions
(electromagnetic, weak and strong), are in fact not at all
 ``constant''. They really depend on the energy scale $\mu$ at which
 the experiments are realized and are therefore functions $g(\mu)$.
 Thus, high energy physics implicitly extends the ``field of
 constants'', passing from the field of scalars $\C$ to a field of
 functions containing the $g(\mu)$. The generator of the
 renormalization group is simply $\mu\,\partial/\partial \mu$.

\smallskip

 It is well known to physicists that the renormalization group plays
 the role of a group of ambiguity. One cannot distinguish between two
 physical theories that belong to the same orbit of this group. In
 this paper we have given a precise mathematical content to a Galois
 interpretation of the renormalization group via the canonical
 homomorphism \eqref{rgU}.
 The fixed points of the renormalization group are ordinary scalars,
 but it can very well be that quantum physics conspires to prevent us
 from hoping to obtain a theory that includes all of particle physics
 and is constructed as a fixed point of the renormalization group.
 Strong interactions are asymptotically free and one can analyse them
 at very high energy using fixed points of the renormalization group,
 but the presence of the electrodynamical sector shows that it is
 hopeless to stick to the fixed points to describe a theory that
 includes all observed forces. The problem is the same in the infrared,
 where the role of strong and weak interactions is reversed.
 
\smallskip

 One can describe the simpler case of the elliptic function field
 $K_q$ in the same form, as a field of functions $g(\mu)$  with a
 scaling action generated by $\mu\,\partial/\partial \mu$. This is
 achieved by passing to loxodromic functions, that is,
 setting $\mu = e^{2 \pi i z }$, so that the first periodicity (that in
 $z\mapsto z+1$) is automatic and the second is written as
 $g( q \, \mu)=g( \mu) $. The group of automorphisms of an elliptic
 curve is then also generated by $\mu\,\partial/\partial \mu$.

\smallskip

In this setup, the equation
$\mu\,\partial_\mu\, f = \beta \, f$,
relating the scaling of the mass parameter $\mu$ to the beta
function (\cf \eqref{renaction}), can be seen as a regular
singular Riemann--Hilbert problem on a punctured disk $\Delta^*$,
with $\beta$ the generator of the local monodromy
$\rho(\ell)=\exp(2 \pi i\, \ell \beta)$. This interpretation of
$\beta$ as log of the monodromy appears in \cite{mk} in the context of
arithmetic geometry \cite{3CK}, \cite{3KM}.

\smallskip

The field $K_q$ of elliptic functions plays an important role in the
recent work of Connes--Dubois Violette on noncommutative spherical
manifolds (\cite{3CDV1} \cite{3CDV2}). There the Sklyanin algebra (\cf
\cite{3Skl}) 
 appeared as solutions in dimension three of a classification problem
 formulated in \cite{3CoLa}. The {\em regular} representation of such
 algebra generates a von Neumann algebra, direct integral of
 approximately finite type II$_1$ factors, all isomorphic to the
 hyperfinite factor $R$. The corresponding homomorphisms of the
 Sklyanin algebra to the factor $R$ miraculously factorizes
 through the crossed product of the field $K_q$ of elliptic functions,
 where the module $q=e^{2 \pi i \tau }$ is real, by the automorphism of
 translation by a real number (in general irrational).
One obtains this way the factor $R$ as a crossed product of
 the field $K_q$ by a subgroup of the Galois group. 
The results of \cite{CoMa} on the quantum statistical mechanics of
2-dimensional $\Q$-lattices suggests that an analogous construction  
for the type ${\rm III}_1$ case should be possible using the 
to modular field. 

\smallskip

This type of results are related to the question of an interpretation
of arithmetic geometry at the archimedean places in terms of
noncommutative geometry, which will be treated in \cite{CoMa2}.
In fact, it was shown in \cite{3CoZ} that the classification of approximately
finite factors provides a nontrivial Brauer theory for central simple
algebras over $\C$. This provides an analog, in the archimedean case,
of the module of central simple algebras over a nonarchimedean local
field. In Brauer theory the relation to
the Galois group is obtained via the construction of central simple
algebras as crossed products of a field by a group of automorphisms.
Thus, finding natural examples of constructions of
factors as crossed product of a field $F$, which is a
transcendental extension of $\C$, by a group of automorphisms is the
next step in this direction.

\medskip
\subsection{Birkhoff decomposition and integrable systems}
\medskip

The Birkhoff decomposition of loops with values in a complex Lie group
$G$ is closely related to the geometric theory of solitons developed
by Drinfel'd and Sokolov (\cf \eg \cite{DS}) and to the corresponding 
hierarchies of integrable systems. 

\smallskip

This naturally poses the question of whether there may be interesting
connections between the mathematical formulation of perturbative
renormalization in terms of Birkhoff decomposition of \cite{3cknew} and 
integrable systems. Some results in this direction were obtained in
\cite{Saka}. 

\smallskip

In the Drinfel'd--Sokolov approach, one assigns to a pair $(\fg,X)$ of
a simple Lie algebra $\fg=Lie\, G$ and an element in a Cartan subalgebra
$\fh$ a hierarchy of integrable systems parameterized by data
$(Y,k)$, with $Y\in \fh$ and $k\in \N$. These have the form of a Lax
equation $U_t - V_x +[U,V]=0$,
which can be seen as the vanishing curvature condition for a connection
\begin{equation}\label{F0eq}
\nabla =\left( \frac{\partial}{\partial x} - U(x,t;z),
\frac{\partial}{\partial t} - V(x,t;z) \right).
\end{equation}
This geometic formulation in terms of connections proves to be a
convenient point of view. In fact, it immediately shows that the
system has a large group of symmetries given by gauge
transformations 
$U \mapsto \gamma^{-1} U \gamma + \gamma^{-1} \, \partial_x\, \gamma$
and $V \mapsto \gamma^{-1} V \gamma + \gamma^{-1} \,\partial_t\, \gamma$. 
The system associated to the data $(Y,k)$ is specified by a ``bare''
potential 
\begin{equation}\label{bareconnection}
\nabla_0 =\left( \frac{\partial}{\partial x} - X\, z,
\frac{\partial}{\partial t} - \tilde X \, z^k \right),
\end{equation}
with $[X,\tilde X]=0$ so that $\nabla_0$ is flat, and solutions are then
obtained by acting on $\nabla_0$ with the 
``dressing'' action of the loop group $\Lc G$ by gauge
transformations preserving the type of singularities of $\nabla_0$. 
This is done by the Zakharov--Shabat method \cite{ZaSha}. Namely, one
first looks for functions $(x,t) \mapsto \gamma_{(x,t)}(z)$,
where $\gamma_{(x,t)} \in \Lc G$, such that
$\gamma^{-1} \nabla_0 \gamma = \nabla_0$.
One sees that these will be of the form
\begin{equation}\label{fixnabla0}
\gamma_{(x,t)}(z) = \exp(x X z + t \tilde X z^k)\, \gamma(z) \, \exp(-x X z
- t \tilde X z^k),
\end{equation} 
where $\gamma(z)$ is a $G$-valued loop. If $\gamma$ is
contained in the ``big cell'' where one has Birkhoff decomposition
$\gamma(z) = \gamma^-(z)^{-1} \gamma^+(z)$,
one obtains a corresponding Birkhoff decomposition for
$\gamma_{(x,t)}$ and a connection
\begin{equation}\label{nabla}
\nabla = \gamma_{(x,t)}^-(z)^{-1} \, \nabla_0\,  \gamma_{(x,t)}^-(z) =
\gamma_{(x,t)}^+(z)\, \nabla_0\,  \gamma_{(x,t)}^+(z)^{-1},
\end{equation}
which has again the same type of singularities as $\nabla_0$. The
new local gauge potentials are of the form
$U= X z + u(x,t)$ and $V= \tilde X z^k + \sum_{i=1}^{k-1}
v_i(x,t) z^i$.  
Here $u(x,t)$ is $u=[X,{\rm Res}\gamma]$. For $u(x,t)=\sum_\alpha
u_\alpha(x,t)\, e_\alpha$, where $\fg=\oplus_\alpha \C e_\alpha \oplus
\fh$, one obtains nonlinear soliton equations $\partial_t u_\alpha =
F_\alpha(u_\beta)$ by expressing the $v_i(x,t)$
as some universal local expressions in the $u_\alpha$.

\smallskip

Even though the Lie algebra of renormalization does not fit directly
into this general setup, this setting suggests the possibility of
considering similar connections (recall, for instance, that 
$[Z_0,{\rm Res}\gamma]=Y{\rm Res}\gamma =\beta$), and working with the
doubly infinite Lie algebra of insertion and elimination defined in
\cite{CK-Lie}, with the Birkhoff decomposition provided by
renormalization.

\bigskip
 \section{Further developments}\label{Sdevel}
\bigskip

  The presence of subtle algebraic structures related to the
 calculation of Feynman diagrams is acquiring an increasingly
 important role in experimental physics. In fact, it is well known
 that the standard model of elementary particle physics gives
 extremely accurate predictions, which have been tested
 experimentally to a high order of precision. This means that, in
 order to investigate the existence of new physics, within the
 energy range currently available to experimental technology, it is
 important to stretch the computational power of the theoretical
 prediction to higher loop perturbative corrections, in the hope to
 detect discrepancies from the observed data large enough to
 justify the introduction of physics beyond the standard model. The
 huge number of terms involved in any such calculation requires
 developing an effective computational way of handling them. This
 requires the development of efficient algorithms for the expansion
 of higher transcendental functions to a very high order. The 
 interesting fact is that abstract algebraic and number theoretic
 objects -- Hopf algebras, Euler--Zagier sums, multiple
 polylogarithms -- appear very naturally in this context.

 \smallskip

  Much work has been done recently by physicists (\cf the work of
  Moch, Uwer, and Weinzierl
 \cite{3MoUwWein}, \cite{3Wein}) in developing such algorithms for nested
 sums based on Hopf algebras. They produce explicit recursive
 algorithms treating expansions of nested finite or infinite sums
 involving ratios of Gamma functions and $Z$--sums, which naturally
 generalize multiple polylogarithms \cite{3sasha2},
 Euler--Zagier sums, and multiple $\zeta$-values. Such sums typically
 arise in the calculation of multi-scale multi-loop integrals. The
 algorithms are designed to recursively reduce the $Z$-sums
 involved to simpler ones with lower weight or depth, and are based on
 the fact that $Z$-sums form a Hopf algebra, whose co-algebra structure
 is the same as that of the CK Hopf algebra. Other interesting
explicit algorithmic calculations of QFT based on the CK Hopf algebra
of Feynman graphs can be found in the work of Bierenbaum, Kreckel, and
Kreimer \cite{BiKr2}. Hopf algebra
structures based on rooted trees, that encode the combinatorics of
Epstein--Glaser renormalization were developed by
Bergbauer and Kreimer \cite{BerKr}.

 \smallskip

Kreimer developed an approach to the
Dyson--Schwinger equation via a method of factorization in primitive
graphs based on the Hochschild cohomology of the CK Hopf algebras of
Feynman graphs (\cite{Kr3}, \cite{Kr4}, \cite{Kr2}, \cf also
\cite{BrKr}). 

 \smallskip

Work of Ebrahimi-Fard, Guo, and Kreimer
(\cite{EGK}, \cite{EGK2}, \cite{EGK3}) recasts 
the Birkhoff decomposition that appears in the CK theory of
perturbative renormalization in terms of the formalism of 
Rota--Baxter relations. Berg and Cartier \cite{BeCar} related
the Lie algebra of Feynman graphs to a matrix Lie algebra
and the insertion product to a Ihara bracket.
Using the fact that the Lie algebra of 
Feynman graphs has two natural representations (by creating or
eliminating subgraphs) as derivations on the Hopf algebra of Feynman
graphs, Connes and Kreimer introduced in \cite{CK-Lie} 
a larger Lie algebra of derivations which accounts for both operations.
Work of Mencattini and Kreimer further relates 
this Lie algebra (in the ladder case) 
to a classical infinite dimensional Lie algebra.

 \smallskip

 Connections between the operadic formalism and the CK
 Hopf algebra have been considered by van der Laan and Moerdijk
 \cite{3Laan}, \cite{3Moerd}. The CK Hopf algebra
 also appears in relation to a conjecture of Deligne on the existence
 of an action of a chain model of the little disks operad on the Hochschild
 cochains of an associative algebra (\cf Kaufmann \cite{3Kau}).

 \end{document}